\documentclass[11pt]{article}

\usepackage{hyperref,cite}
\hypersetup{colorlinks,bookmarksopen,bookmarksnumbered,citecolor=blue,linkcolor=[rgb]{0.75,0,0},pdfstartview=FitH,unicode}

\newcommand{\be}{\begin{equation}}
\newcommand{\ee}{\end{equation}}
\usepackage{amsmath}% 
\usepackage{amssymb}
\usepackage{latexsym}
\usepackage{comment}

\newcommand{\half}{{\textstyle ü}}

\newcommand{\RRR}{{\hbox{\rm R\kern-2.35mm R}}}
\newcommand{\p}{\partial}

\newcommand{\sectiono}[1]{\section{#1}\setcounter{equation}{0}}

\def\itemitem{\par\indent \hangindent2\parindent \textindent}
\def\textindent#1{\indent\llap{#1\enspace}\ignorespaces}

\def\frac{\f}

\catcode`@=11

\def\matrix#1{\null\,\vcenter{\normalbaselines\m@th
    \ialign{\hfil$##$\hfil&&\quad\hfil$##$\hfil\crcr
      \mathstrut\crcr\noalign{\kern-\baselineskip}
      #1\crcr\mathstrut\crcr\noalign{\kern-\baselineskip}}}\,}
\def\pmatrix#1{\left( \matrix{#1} \right)}

{\catcode`\'=\active \def'{{}^\bgroup\prim@s}}

\catcode`@=12

\def\rgboo#1{\pdfliteral{#1 rg #1 RG}}
\def\pdfklink#1#2{%
	\noindent\pdfstartlink user
		{/Subtype /Link
		/Border [ 0 0 0 ]
		/A << /S /URI /URI (#2) >>}{\rgb{1 0 0}{#1}}%
	\pdfendlink}
\def\rgbo#1#2{\rgboo{#1}#2\rgboo{0 0 0}}
\def\rgb#1#2{\mark{#1}\rgbo{#1}{#2}\mark{0 0 0}}
\def\pdflink#1{\pdfklink{#1}{#1}}
\def\xxxlink#1{\pdfklink{[arXiv:#1]}{http://arXiv.org/abs/#1}}

\def\on#1#2{{\buildrel{\mkern2.5mu#1\mkern-2.5mu}\over{#2}}}
\def\oonoo#1#2#3{\vbox{\ialign{##\crcr
	\hfil\hfil\hfil{$#3{#1}$}\hfil\crcr\noalign{\kern1pt\nointerlineskip}
	$#3{#2}$\crcr}}}
\def\oon#1#2{\mathchoice{\oonoo{#1}{#2}{\displaystyle}}
	{\oonoo{#1}{#2}{\textstyle}}{\oonoo{#1}{#2}{\scriptstyle}}
	{\oonoo{#1}{#2}{\scriptscriptstyle}}}
\def\dt#1{\oon{\hbox{\bf .}}{#1}}  
\def\ddt#1{\oon{\hbox{\bf .\kern-1pt.}}#1}    % À À   (see below)
\def\slap#1#2{\setbox0=\hbox{$#1{#2}$}
	#2\kern-\wd0{\hfuzz=1pt\hbox to\wd0{\hfil$#1{/}$\hfil}}}
\def\sla#1{\mathpalette\slap{#1}}                % slash: see Ö   below
\def\bop#1{\setbox0=\hbox{$#1M$}\mkern1.5mu
	\lower.02\ht0\vbox{\hrule height0pt depth.06\ht0
	\hbox{\vrule width.06\ht0 height.9\ht0 \kern.9\ht0
	\vrule width.06\ht0}\hrule height.06\ht0}\mkern1.5mu}
\def\bo{{\mathpalette\bop{}}}                        % box: see õ below
\mathcode`\*="702A                  % * now always complex conjugate
\def\in{\relax\ifmmode\mathchar"3232\else{\refit in\/}\fi} % ã below 
\def\f#1#2{{\textstyle{#1\over#2}}}	   % fraction
\def\half{{\textstyle{1\over{\raise.1ex\hbox{$\scriptstyle{2}$}}}}}

\def\Gamma{\mathchar"0100}
\def\Delta{\mathchar"0101}
\def\Theta{\mathchar"0102}
\def\Lambda{\mathchar"0103}
\def\Xi{\mathchar"0104}
\def\Pi{\mathchar"0105}
\def\Sigma{\mathchar"0106}
\def\Upsilon{\mathchar"0107}
\def\Phi{\mathchar"0108}
\def\Psi{\mathchar"0109}
\def\Omega{\mathchar"010A}

\catcode128=13 \def €{\"A}                 % Option-u A
\catcode129=13 \def {\AA}                 % Option-A
\catcode130=13 \def '{\c}           	   % Option-C (cedilla)
\catcode131=13 \def ƒ{\'E}                   % Option-e E
\catcode132=13 \def "{\~N}                   % Option-n N
\catcode133=13 \def …{\"O}                 % Option-u O
\catcode134=13 \def †{\"U}                  % Option-u U
\catcode135=13 \def ‡{\'a}                  % Option-e a
\catcode136=13 \def ˆ{\`a}                   % Option-`  a
\catcode137=13 \def ‰{\^a}                 % Option-i a
\catcode138=13 \def Š{\"a}                 % Option-u a
\catcode139=13 \def ‹{\~a}                   % Option-n a
\catcode140=13 \def Œ{\alpha}            % Option-a
\catcode141=13 \def {\chi}                % Option-c
\catcode142=13 \def Ž{\'e}                   % Option-e e
\catcode143=13 \def {\`e}                    % Option-`  e
\catcode144=13 \def {\^e}                  % Option-i e
\catcode145=13 \def '{\"e}                % Option-u e
\catcode146=13 \def '{\'\i}                 % Option-e i
\catcode147=13 \def "{\`\i}                  % Option-`  i
\catcode148=13 \def "{\^\i}                % Option-i i
\catcode149=13 \def •{\"\i}                % Option-u i
\catcode150=13 \def –{\~n}                  % Option-n n
\catcode151=13 \def —{\'o}                 % Option-e o
\catcode152=13 \def ˜{\`o}                  % Option-`  o
\catcode153=13 \def ™{\^o}                % Option-i o
\catcode154=13 \def š{\"o}                 % Option-u o
\catcode155=13 \def ›{\~o}                  % Option-n o
\catcode156=13 \def œ{\'u}                  % Option-e u
\catcode157=13 \def {\`u}                  % Option-`  u
\catcode158=13 \def ž{\^u}                % Option-i u
\catcode159=13 \def Ÿ{\"u}                % Option-u u
\catcode160=13 \def  {\tau}               % Option-t
\catcode162=13 \def ¢{\oplus}           % Option-4
\catcode163=13 \def £{\relax\ifmmode\to\else\itemize\fi} % Option-3
\catcode164=13 \def ¤{\subset}	  % Option-6
\catcode165=13 \def ¥{\infty}           % Option-8
\catcode166=13 \def ¦{\mp}                % Option-7
\catcode167=13 \def §{\sigma}           % Option-s
\catcode168=13 \def ¨{\rho}               % Option-r
\catcode169=13 \def ©{\gamma}         % Option-g
\catcode170=13 \def ª{\leftrightarrow} % Option-2 ; Option-E (acute) :
\catcode171=13 \def «{\relax\ifmmode\expandafter\acute\else\expandafter\'\fi}
\catcode172=13 \def ¬{\relax\ifmmode\expandafter\ddt\else\expandafter\"\fi}
\catcode173=13 \def ­{\equiv}            % Option-= ; ^ Option-U (umlaudt)
\catcode174=13 \def ®{\approx}          % Option-"
\catcode175=13 \def ¯{\Omega}          % Option-O
\catcode176=13 \def °{\otimes}          % Option-5
\catcode177=13 \def ±{\ne}                 % Option-+
\catcode178=13 \def ²{\le}                   % Option-,
\catcode179=13 \def ³{\ge}                  % Option-.
\catcode180=13 \def ´{\upsilon}          % Option-y
\catcode181=13 \def µ{\mu}                % Option-m
\catcode182=13 \def ¶{\delta}             % Option-d
\catcode183=13 \def ·{\epsilon}          % Option-w
\catcode184=13 \def ¸{\Pi}                  % Option-P
\catcode185=13 \def ¹{\pi}                  % Option-p
\catcode186=13 \def º{\beta}               % Option-b
\catcode187=13 \def »{\partial}           % Option-9
\catcode188=13 \def ¼{\nobreak\ }       % Option-0
\catcode189=13 \def ½{\zeta}               % Option-z
\catcode190=13 \def ¾{\sim}                 % Option-'
\catcode191=13 \def ¿{\omega}           % Option-o
\catcode192=13 \def À{\dt}                     % Option-?
\catcode193=13 \def Á{\gets}                % Option-1
\catcode194=13 \def Â{\lambda}           % Option-l
\catcode195=13 \def Ã{\nu}                   % Option-v
\catcode196=13 \def Ä{\phi}                  % Option-f
\catcode197=13 \def Å{\xi}                     % Option-x
\catcode198=13 \def Æ{\psi}                  % Option-j
\catcode199=13 \def Ç{\int}                    % Option-\
\catcode200=13 \def È{\oint}                 % Option-|
\catcode201=13 \def É{\relax\ifmmode\cdot\else\vol\fi}    % Option-;
\catcode202=13 \def Ê{\relax\ifmmode\,\else\thinspace\fi}
\catcode203=13 \def Ë{\`A}                      % Option-`  A ; ^ Option-space
\catcode204=13 \def Ì{\~A}                      % Option-n A
\catcode205=13 \def Í{\~O}                      % Option-n O
\catcode206=13 \def Î{\Theta}              % Option-Q
\catcode207=13 \def Ï{\theta}               % Option-q; Option-- :
\catcode208=13 \def Ð{\relax\ifmmode\expandafter\bar\else\expandafter\=\fi}
\catcode209=13 \def Ñ{\overline}             % Option-_
\catcode210=13 \def Ò{\langle}               % Option-[
\catcode211=13 \def Ó{\relax\ifmmode\{\else\ital\fi}      % Option-{
\catcode212=13 \def Ô{\rangle}               % Option-]
\catcode213=13 \def Õ{\}}                        % Option-}
\catcode214=13 \def Ö{\sla}                      % Option-/; Option-V :
\catcode215=13 \def ×{\relax\ifmmode\expandafter\check\else\expandafter\v\fi}
\catcode216=13 \def Ø{\"y}                     % Option-u y
\catcode217=13 \def Ù{\"Y}  		    % Option-u Y
\catcode218=13 \def Ú{\Leftarrow}       % Option-!
\catcode219=13 \def Û{\Leftrightarrow}       % Option-@ ; Option-# :
\catcode220=13 \def Ü{\relax\ifmmode\Rightarrow\else\sect\fi}
\catcode221=13 \def Ý{\sum}                  % Option-$
\catcode222=13 \def Þ{\prod}                 % Option-%
\catcode223=13 \def ß{\widehat}              % Option-^
\catcode224=13 \def à{\pm}                     % Option-&
\catcode225=13 \def á{\nabla}                % Option-(
\catcode226=13 \def â{\quad}                 % Option-)
\catcode227=13 \def ã{\in}               	% Option-W
\catcode228=13 \def ä{\star}      	      % Option-R
\catcode229=13 \def å{\sqrt}                   % Option-M
\catcode230=13 \def æ{\^E}			% Option-i E
\catcode231=13 \def ç{\Upsilon}              % Option-Y
\catcode232=13 \def è{\"E}    	   	 % Option-u E
\catcode233=13 \def é{\`E}               	  % Option-`  E
\catcode234=13 \def ê{\Sigma}                % Option-S
\catcode235=13 \def ë{\Delta}                 % Option-D
\catcode236=13 \def ì{\Phi}                     % Option-F
\catcode237=13 \def í{\`I}        		   % Option-`  I
\catcode238=13 \def î{\iota}        	     % Option-H
\catcode239=13 \def ï{\Psi}                     % Option-J
\catcode240=13 \def ð{\times}                  % Option-K
\catcode241=13 \def ñ{\Lambda}             % Option-L
\catcode242=13 \def ò{\cdots}                % Option-:
\catcode243=13 \def ó{\^U}			% Option-i U
\catcode244=13 \def ô{\`U}    	              % Option-`  U
\catcode245=13 \def õ{\bo}                       % Option-B ; Option-I :
\catcode246=13 \def ö{\relax\ifmmode\expandafter\hat\else\expandafter\^\fi}
\catcode247=13 \def÷{\relax\ifmmode\expandafter\tilde\else\expandafter\~\fi}
\catcode248=13 \def ø{\ll}                         % Option-< ; ^ Option-N
\catcode249=13 \def ù{\gg}                       % Option-> 
\catcode250=13 \def ú{\eta}                      % Option-h 
\catcode251=13 \def û{\kappa}                  % Option-k 
\catcode252=13 \def ü{\half}     		 % Option-Z 
\catcode253=13 \def ý{\Gamma} 		% Option-G 
\catcode254=13 \def þ{\Xi}   			% Option-X ; Option-T : 
\catcode255=13 \def ÿ{\relax\ifmmode{}^{\dagger}{}\else\dag\fi}

\def\A{{\cal A}}      
\def\E{{\cal E}}      \def\H{{\cal H}}   
    \def\L{{\cal L}}  \def\M{{\cal M}}   \def\N{{\cal N}}  
\def\O{{\cal O}}      
\def\S{{\cal S}}  \def\T{{\cal T}}  \def\V{{\cal V}}  
    
\def\Ä{\varphi}  \def\¿{\varpi}	\def\Ï{\vartheta}

\def\Ç{{\textstyle{Ç}}}
\def\cvrule{\rgbo{0 .5 1}{\vrule}}
\def\chrule{\rgbo{0 .5 1}{\hrule}}
\def\boxit#1{\leavevmode\thinspace\hbox{\cvrule\vtop{\vbox{\chrule%
	\vskip3pt\kern1pt\hbox{\vphantom{\bf/}\thinspace\thinspace%
	{\bf#1}\thinspace\thinspace}}\kern1pt\vskip3pt\chrule}\cvrule}%
	\thinspace}
\def\Boxit#1{\noindent\vbox{\chrule\hbox{\cvrule\kern3pt\vbox{
	\advance\hsize-7pt\vskip-\parskip\kern3pt\bf#1
	\hbox{\vrule height0pt depth\dp\strutbox width0pt}
	\kern3pt}\kern3pt\cvrule}\chrule}}

\def\boxeq#1{\boxit{${\displaystyle #1 }$}}          % inside $$'s
\begin{document}

\rightline{June 12, 2013} 
\rightline{\tt LMU-ASC 39/13}
\rightline{\tt MIT-CTP-4467}    
\rightline{\tt YITP-SB-13-16}      
\begin{center}

\vskip0.7in

\centerline{\rgb{0 .5 0.5}{\huge \bf Doubled $\alpha'$-Geometry }}

\bigskip
\bigskip\bigskip

{\Large Olaf Hohm${}^1$, Warren Siegel${}^2$, and Barton Zwiebach${}^3$}
\bigskip\bigskip

${}^1$ {\it Arnold Sommerfeld Center for Theoretical Physics\\
Theresienstrasse 37\\
D-80333 Munich, Germany\\}
\pdflink{mailto:olaf.hohm@physik.uni-muenchen.de}
\bigskip

${}^2$ {\it C. N. Yang Institute for Theoretical Physics\\
State University of New York, Stony Brook, NY 11794-3840\\}
\pdflink{mailto:siegel@insti.physics.sunysb.edu}\\
	\pdfklink{http://insti.physics.sunysb.edu/\~{}siegel/plan.html}
	{http://insti.physics.sunysb.edu/\noexpand~siegel/plan.html}
\bigskip

${}^3$ {\it Center for Theoretical Physics\\
Massachusetts Institute of Technology\\
Cambridge, MA 02139\\}
\pdflink{mailto:zwiebach@mit.edu}

\bigskip
\end{center}

\bigskip  
\begin{abstract}
\normalsize
\bigskip

We develop  doubled-coordinate field theory to determine the $\alpha'$ 
corrections to the massless sector of oriented bosonic closed string theory. 
Our key tool is a string current algebra of free left-handed bosons that 
makes O(D,D) T-duality manifest.  While T-dualities are unchanged, 
diffeomorphisms and $b$-field gauge transformations  receive corrections, 
with  a gauge algebra given by an $\alpha'$-deformation of the 
duality-covariantized Courant bracket. 
The action is cubic in a double metric field, an unconstrained extension
of the generalized metric that encodes the gravitational fields. 
Our approach provides a consistent 
truncation of string theory to massless fields with corrections 
that close at finite order in~$\alpha'$.

\end{abstract}

\newpage

{\hypersetup{linkcolor=[rgb]{.7,0,.7}}
\tableofcontents }

%\tableofcontents

%\newpage

\baselineskip=16pt
\parskip=\medskipamount			% space between paragraphs (TeX)

\section{Introduction}

\subsection{History}

The massless sector of the oriented, bosonic, closed string consists of the graviton (metric), ``axion" or ``notoph"
 (2-form gauge field), and dilaton.  
The T-duality symmetry  
of the D-dimensional theory
with d compactified dimensions is O(d,d), over the integers for the full string theory but over the reals for the massless sector.  
The string theory, of course, lives in D=26, but the massless theory we are studying exists for all values of D.   
Duality 
transforms  
the (D$-$d)-dimensional scalars resulting from dimensional reduction of the metric and 2-form, but leaves invariant the dilaton, when defined as a scalar density.  
This is the usual treatment of T-duality when winding modes are ignored and  dimensional reduction is described in the language of Killing vectors that
imply the independence of the background from the d compact coordinates.  

This O(d,d) can be represented on these scalars in terms of a nonlinear $§$-model for the coset O(d,d)/O(d)$\times$O(d)\cite{Duff:1989tf}. 
 But this approach can be generalized \cite{Siegel:1993xq,Siegel:1993th,Siegel:1993bj} in a way that:
\itemitem{a) }requires no dimensional reduction,
\itemitem{b) }includes the full set of massless fields, 
\itemitem{c) }includes all gauge invariances, 
\itemitem{d) }defines covariant derivatives (connections, torsions, curvatures, Bianchi identities), and
\itemitem{e) }manifests a full O(D,D) symmetry on the fields, gauge invariances, and action.

\noindent This procedure doubles the coordinates on which all fields depend.  The reduction to D dimensions is achieved by a set of constraints that preserves the manifest O(D,D), but any solution of the constraints ``spontaneously breaks" this symmetry down to the usual O(D$-$1,1) Lorentz symmetry, reproducing the standard D-dimensional fields, gauge invariances, and action.  The O(d,d) can then be restored manifestly by compactification, which weakens the constraints.  

The left- and right-handed worldsheet currents (affine Lie algebra) form the defining representation of this O(D,D).  Through coupling quadratically  
 to these currents, the metric and 2-form combine to form the coset O(D,D)/O(D$-$1,1)${}^2$.  The action can be expressed in a manifestly O(D,D) covariant form in terms of this field and the dilaton, which acts as the spacetime integration measure.

In more recent developments the construction of such a double field theory~\cite{Hull:2009mi}
was based on closed string field theory~\cite{Zwiebach:1992ie,Kugo:1992md}.  This work identified the constraints
mentioned above as the strong version of the $L_0 - \bar L_0 = 0$ level-matching
condition of closed string fields.  In its standard and seemingly 
unavoidable (weak) form, it applies to all fields and gauge parameters.  
In the strong version, which demands that
all {\it products} of fields are also killed by $L_0 - \bar L_0$, it provides the reduction
to D dimensions.  While the construction  could
be carried to cubic order in fluctuations without imposing
the strong constraint, the full construction became tractable only once 
this constraint is imposed. 
In this case the string field gauge algebra is governed by
the bracket anticipated in~\cite{Siegel:1993th}:  The C-bracket, 
which was shown~\cite{Hull:2009zb} to be the duality-covariantized 
version of the Courant bracket of generalized geometry.  This suggested 
the possibility of a compact explicit form of the doubled
action in terms of the generalized metric $\H_{MN}$ and the duality 
invariant dilaton, a construction provided in~\cite{Hohm:2010pp}.  The simplicity of the
action allows quick confirmation
that any solution of the strong constraint gives precisely the two-derivative
action for the massless sector of closed string theory.  An alternative
form of the action in terms of the field $\E_{ij} =(g+ b)_{ij}$ was given earlier in~\cite{Hohm:2010jy}.

There is also a compelling 
generalization of Riemannian geometry for this duality-covariant framework. 
A complete formulation has been given in \cite{Siegel:1993th} 
in a frame-like formalism, including torsions, curvatures (Riemann tensor), 
differential 
Bianchi identities, and a discussion of the ambiguity of some Lorentz connections and curvatures.
In~\cite{Hohm:2010xe} this formalism has been 
related to the double field theory actions of \cite{Hohm:2010pp,Hohm:2010jy}  
and  to a metric-like 
formulation. 
The  metric-like approach  
has been examined in more detail in~\cite{Jeon:2010rw} (in a
``semi-covariant" approach that truncates connections)  and in~\cite{Hohm:2011si}. 
The fully ``invariant" formulation in~\cite{Hohm:2012mf} provides a 
unifying framework for the metric- and frame-like formalisms. This includes an    
index-free 
definition of the torsion and Riemann tensor, a 
complete algebraic Bianchi identity with torsion, and a discussion of the absence of 
an uncontracted differential Bianchi identity. 
This geometry is related to (and an extension of)    
the ``generalized geometry" of  
Hitchin and Gualtieri \cite{Hitchin:2004ut,Gualtieri:2003dx,Gualtieri:2007bq}. 

Formulations including the coupling to vector multiplets, relevant for 
heterotic and type I strings, were also 
given in \cite{Siegel:1993th,Siegel:1993bj} and worked out in the 
generalized metric formulation in \cite{Hohm:2011ex}. The $\N=1$
supersymmetric form is contained in the superspace results 
of \cite{Siegel:1993th,Siegel:1993bj}
and was worked out independently in explicit component form 
in~\cite{Hohm:2011nu}.  (See also \cite{Jeon:2011sq} for 
supersymmetric double field theory without vector multiplets.)
The Ramond-Ramond
sector of type-II superstrings is given in~\cite{Hohm:2011dv}, and its 
supersymmetric extension in~\cite{Coimbra:2011nw}.   

Double field theory formulations where the strong constraint is
somewhat relaxed have been given  for massive IIA supergravity
in~\cite{Hohm:2011cp},  for flux compactifications in~\cite{Andriot:2011uh}, and
explored in some generality in~\cite{grana-marques,Geissbuhler:2013uka}.  
See also \cite{Andriot:2012wx}
for the geometric role of non-geometric fluxes in double field theory.   
Global aspects of double field theory are discussed
in~\cite{Hohm:2012gk} where a formula for 
large gauge transformations was proposed and examined in detail.   
There are numerous other developments in double field theory and 
the closely related M-theory (see \cite{Hillmann:2009ci,Berman:2010is,Coimbra:2011ky}); 
for a recent review with further references see~\cite{Aldazabal:2013sca}.

One of the most intriguing features of the theory is the 
absence of a satisfactory duality-covariant generalized Riemann tensor.
In the geometric formalism the covariant constraints do not suffice to determine all components of the connections 
in terms of physical fields, resulting in a Riemann tensor with some
undetermined components. 
In fact, the undetermined components of the 
generalized Riemann tensor are such that this tensor encodes 
nothing more than the 
Ricci curvature and scalar curvature~\cite{Siegel:1993th,Hohm:2011si,Hohm:2012mf}.    

It has been known for some time that $\alpha'$ corrections to the massless
effective field theory preserve the T-duality symmetry of the two-derivative
action~\cite{Sen:1991zi}.  This has been verified explicitly in~\cite{Meissner:1996sa} 
to first order in $\alpha'$ in a reduction down to just one dimension.  
The $\alpha'$ corrections
to the action of bosonic closed strings include Riemann-squared terms.  
In the absence of a duality-covariant Riemann tensor, it has been hard to imagine
how one could describe $\alpha'$ corrections in a manifestly duality invariant way.
For example, 
it was seen in\cite{Hohm:2011si} that certain structures in Riemann-squared
cannot be written in terms of the generalized metric.
This lack of a suitable duality-covariant Riemann tensor
 is a clear indication that some symmetries of the
theory must receive $\alpha'$ corrections. 
While the one-dimensional results
of~\cite{Meissner:1996sa} suggest that gauge symmetries could be corrected, the
more accepted viewpoint has been that $\alpha'$ corrections to the T-duality
transformations are required.  These, however, have been
hard to determine, even for the case of compactifications over a single circle
and to first order in $\alpha'$~\cite{Kaloper:1997ux}. 

On the other hand, the string field theory based analysis~\cite{Hull:2009mi} is by construction duality covariant (although background dependent), suggesting again that
duality need not be corrected.  It was noted 
in~\cite{Hohm:2012mf}, moreover, that the gauge symmetry brackets calculated  
to lowest order in derivatives in\cite{Hull:2009mi} receive computable $\alpha'$ corrections.  These corrections have been determined, appear to agree with the
results to be presented here,  and will be considered elsewhere as supporting evidence
for the connection to string theory.  
It was simpler, however, to approach the construction by extending the current algebra methods developed in~\cite{Siegel:1993xq,Siegel:1993th,Siegel:1993bj} and this is what we will do in this paper.

\subsection{Outline and summary}

In this paper the main technical tool is a modified
 worldsheet theory that amounts to a certain consistent truncation
of string theory. 
We will have D+D bosonic worldsheet fields 
$X^M$ ($M= 1, 2 , \ldots , 2D$) of one handedness, instead of the
familiar fields $X^i(z)$ and $X^i(\bar z)$, with $i=1 , \ldots , D$.   
In this formulation there is a chirality condition
setting momenta equal to $z$-derivatives of coordinates:   
$P^M = {X'}^M \equiv Z^M$.
There is also a constraint --- the strong constraint ---  
 that must be satisfied by the functions of  
$X^M$ that are used to describe background fields. These fields
and their products must be annihilated by the differential operator
$\eta^{MN}\p_M \p_N$, where $\p_M = \partial/\partial X^M$ and 
$\eta$ is the O(D,D) metric.  This simplified version of the string
truncates the $\alpha'$ corrections of the full string theory,
which is evident from the fact that all operator products terminate.
We will see this as we obtain the equations of motion for the background fields.
Analysis indicates that this truncation duplicates string theory to cubic order in fields.  

In this paper we use the quantum mechanical approach to string theory, 
not the quantum field theory approach.  Hence ``quantum" in this context will 
always refer to the JWKB approximation in orders of $Œ'$.  
 Our main goal, of course,  is the construction of a classical double field theory, a space-time
field theory which includes $\alpha'$ corrections to the two-derivative theory. 
Perhaps this double field theory is the string field theory
that results from the modified worldsheet theory.

We extend the current algebra methods of~\cite{Siegel:1993xq,Siegel:1993th,Siegel:1993bj}
to a full-fledged discussion of the worldsheet conformal field theory, including propagators 
\be
\langle X^M (z_1) X^N(z_2) \rangle \, = \, \eta^{MN}  \, \ln (z_1 - z_2)\,,
\ee
and the associated operator product expansions
in section 2.  
Note the appearance in the
above right-hand side of the O(D,D) metric, at the place where the familiar theory
uses the space-time metric.  A major simplification is that the strong constraint
implies that there are no singular terms in the operator product $A(X) B(X)$ of any two 
$X$-dependent fields.   In this world-sheet theory we consider explicitly three
kinds of operators: scalars $f$, vectors $V$, and tensors $T$
\be
\label{gentensors99}\begin{split}
&âf = f(X) \cr
	&â V = V^M(X)Z_M \cr
	&âT = üT^{MN}(X)Z_M Z_N - ü(öT^M Z_M)' .
\end{split}\ee
The above are operators of conformal weight zero, one, and two, respectively. 
The tensor requires the two terms shown for the closure of the algebra
of operator products.  We refer to them as the two-index component and the
one-index component 
(or pseudovector part) of the tensor.

We find it useful to treat operator products systematically in section~\ref{diffdoubgeo}.
Given two operators $\O_1$ and $\O_2$, the product $\O_1 \circ_w \O_2$,
with $w \geq 0$ an integer,  
is an operator of weight $w$ that appears in the operator product 
of $\O_1(z_1)\O_2(z_2)$ as follows 
\be
\label{opemex}
\O_1 (1) \O_2 (2) \ = \  \ Ý_{w=0}^{\infty} {1\over (z_{12})^{w_1 + w_2- w}} ( \O_1 \circ_{w} \O_2) (2) \,. 
\ee
Here and in the following we use the short-hand notation $z_1\equiv 1$, etc. 
The product $\O_1 \circ_0 \O_2$ is a scalar and will be written as 
the inner product $\langle \O_1 | \O_2
\rangle$.
 We examine various infinite classes of identities 
satisfied by these products.  In general the  products do not have definite symmetry
properties, but there are symmetry relations. 

Vector operators $\Xi = \xi^M (X) Z_M$
 generate gauge transformations
 (section~\ref{vecgausym}).  
 The components $\xi^M(X)$ 
of the operator comprise
D+D  gauge parameters $\xi^i$ and  $\tilde \xi_i$.  The operator product expansion 
$\Xi_1 (z_1)\Xi_2(z_2)$ of 
two such vector operators, with parameters
$\xi_1^M$ and $\xi_2^M$,  defines fundamental structures of the theory. 
We get the inner product $Òþ_1|þ_2Ô$ as the residue of the second-order pole.
This is a symmetric, bilinear scalar operator that takes the form
  \be
  \label{vinn} 
 ~~~ Òþ_1|þ_2Ô ¼= ¼Å_1^M Å_2^N ú_{MN} - (»_N Å_1^M)(»_M Å_2^N)\,. 
 \ee
The first term is familiar from the 
classical theory and  the second term is the $\alpha'$ correction, 
arising from a quantum contribution in the OPE.   Since we do not write 
explicitly $\alpha'$ factors, 
corrections are recognized by the increased number of space-time derivatives. 
We get a vector operator $[þ_1,þ_2]_{{}_C}$ as the residue of the first-order pole. Its components take the  form
\be
\label{c-bracket-corr-vm}
[ \Xi_1 , \Xi_2 ]_{{}_C}^M  \ = \ Å_{[1}^N »_N Å_{2]}^M \, - \, \f12  \,
Å_1^K\onª»{}^M Å_{2K}\, + \, \f12 
\, (»_K Å_1^L)\onª»{}^M(»_L Å_2^K)\,. 
\ee 
(In this paper we use the (anti)symmetrization convention $[ab]=ab-ba$,  
and $A\onª» B  = A \partial B - (\partial A) B$.)
The first term on the right-hand side is the Lie bracket of vector fields.
Together with the next term it defines the ``classical" C -bracket, the duality covariantized
version of the Courant bracket of generalized geometry.   The last term, with three derivatives, is the new nontrivial correction.  The 
strong constraint implies that no higher derivative correction to the bracket can be written that is, 
as required, linear in each of the gauge parameters. 
This correction is therefore unique. 
Moreover, the bracket is fully consistent: Its Jacobiator is a trivial gauge parameter,
just as it was for the classical C bracket.
A trivial gauge parameter does not generate gauge transformations and takes the form of the $z$-derivative $f'$ of a scalar operator $f$.  The quantum C-bracket given above defines the algebra of gauge transformations in the theory we construct here.  

Associated to Courant structures of generalized geometry there are  Dorfman
structures that are often more convenient. For us,  C-type operators  have D-type counterparts.  Amusingly,  C operators arise by presenting the operator product expansion symmetrically in $z_1$ and $z_2$,  while their D counterparts arise by presenting
 the expansion with operators based at $z_2$. 
 The vector operator $[þ_1,þ_2]_{{}_D}$ is the quantum D bracket, whose classical version is
 the duality covariantized Dorfman bracket. 

Very nontrivially, the above corrections do not vanish upon
reduction from D+D to D dimensions, as done by setting $\tilde \p^i$ derivatives to zero.  For the inner product we get
\be
Òþ_1|þ_2Ô \ = \ \xi_1^i \tilde\xi_{2i} + \xi_2^i \tilde\xi_{1i}  \ - \  \p_i \xi_1^j  \, \p_j \xi_2^i \,. 
\ee
The last term is the quantum correction.  For the C bracket the vector part is not corrected,
but the one-form part is:
\be
\left( [ \Xi_1 , \Xi_2 ]_{{}_C}\right)_i  \ = \  \ldots \ + \  \f12 
\, (»_k Å_1^\ell)\onª»{}_i(»_\ell Å_2^k)  \,,
\ee
where the dots denote the contributions from the ``classical terms".
Therefore our results go beyond generalized geometry in that the
familiar inner product and the Courant bracket are deformed.

Gauge transformations $\delta_\xi \O$ of any operator $\O$ are defined by the
commutator
$\delta_\xi \O  =  \bigl[ \Ç \Xi \,, \O \bigr]$, and
 are readily evaluated with the use of operator products. For a vector
operator $V$, for example,    
\be
\delta_\xi  V^M  \ = \ \xi^{P}\partial_{P}V^{M}
  +(\partial^{M}\xi_{P}\,    -\partial_{P}\xi^{M})  V^{P}
  -\,   (\p^M »_K Å^L  ) »_L V^K \,. 
\ee
The last term is the quantum correction. In D dimensions, the quantum correction vanishes for the transformation $\delta_\xi V^i$ of a vector but does not vanish for the transformation $\delta_\xi V_i$
of the one-form (see (\ref{corrliegen})). 
In mathematical language this represents a  deformation of generalized Lie derivatives. 
 
With the gauge structure  defined, the fields of the theory are introduced
using a pair of tensor operators.   We start with $\f12 Z^2\equiv \f12 \eta^{MN}Z_M Z_N$, 
the analog of the Virasoro operator $T_\sigma$ that in the undoubled
flat-space theory is
proportional to ${X'}^iP_i$.  We then introduce
in section~\ref{dildoubvol}
the dilaton in a tensor $\S$ defined to be 
\be
\S \ \equiv \ \f12  (Z^2 - \phi'') \,.
\ee
The second term is  consistent with the general form in (\ref{gentensors99}) since 
$\phi'' = (Z^M\partial_M \phi )'$ (recall $Z= X'$).  This dilaton improvement is needed for consistency of gauge transformations.  As it turns out, 
the gauge transformation of the dilaton receives no quantum corrections (see (\ref{dil-quantum})).

The products also satisfy useful distributive type identities.
Products of the dilaton-based tensor $\S$ with a tensor $T$ lead to convenient definitions
\be
\label{112}
\S \circ_0 T  =  \f12 \hbox{tr}\, T  \,, \qquad  \S \circ_1 T =  \hbox{div} \, T \,.  
\ee
The trace of a tensor is
a scalar with leading term $\eta^{MN} T_{MN}$.
The divergence of a tensor is a vector  with leading
term $\p_N T^{MN}$.  Both have
nontrivial $\alpha'$ contributions that can be seen in~(\ref{tr-div}). 

A second tensor operator $\T$ is used  in section~\ref{doubmetsec}
to introduce the gravitational fields, metric
and two-form.  This operator is the analog of the  Virasoro operator 
$T_\tau$ that in the undoubled flat-space theory is
proportional to $(P_i)^2 + ({X'}^i)^2$.  In toroidal backgrounds, this operator
is a quadratic form on  currents with the generalized metric used to contract indices.
In our formulation we start with a {\it double metric} $\M^{MN}$ that will turn out to be related but {\it not} equal to the generalized metric $\H^{MN}$. While off-shell the latter squares to one, the former is unconstrained. The tensor operator $\T$ takes the form
\be
\T \ \equiv \ \f12  \M^{MN} Z_M Z_N  -  \f12 (\widehat \M^M Z_M)' \,.
\ee
The second term, needed for consistency with gauge
transformations, contains a field $\widehat \M^M$,  to be determined in terms of 
the double metric and the dilaton.  
The gauge transformation of the double metric $\M^{MN}$ receives $\alpha'$
and ${\alpha'}^2$ corrections (see (\ref{gtacfin})).  

Having introduced the dilaton and the double metric on the weight-two tensor
operators $\S$ and $\T$, we make the usual assumption
that the equations of motion of these fields 
are the  conditions that $\S$ and $\T$ form  the Virasoro algebra:
\be  
\label{summary_virasoro}
\begin{split}
\S (1) \S (2) \ = \ & \  {D\over z_{12}^4}  + {2 \S (2) \over z_{12}^2}  + \, {\S'  (2) \over z_{12}} 
+ \hbox{finite} \,,\\[0.7ex]
\S (1) \T (2) \  = \ & \    {2\T (2) \over z_{12}^2} +  {\T' (2) \over z_{12}} 
+  \hbox{finite}\,, \\[0.5ex]
\T (1) \T (2) \ = \ & \  {D\over z_{12}^4} \, + \,   {2\S (2) \over z_{12}^2} + 
 {\S' (2) \over z_{12}} +  \hbox{finite} \,.
\end{split}
\ee
Remarkably, the operator product $\S\S$ (first line) works out
automatically  
without imposing any condition on the dilaton.  This is required, since the dilaton equation
of motion  involves the double metric, which does not appear in $\S$.   
For the $\S\T$ operator product (second line) 
the terms on the right-hand side  appear as expected,
but the vanishing of the quartic and cubic poles give nontrivial conditions.  In the
notation  of (\ref{112}) these correspond to  
\be
\hbox{tr}\, \T \ = \ 0  \quad  \hbox{and} \quad  \hbox{div}\,  \T \ = \ 0\,.
\ee
The first equation  is the %bz fully 
$\alpha'$-corrected equation of motion of the dilaton.  The second equation
determines the auxiliary field $\widehat \M^M$ 
in terms of double metric 
and the dilaton.  For the $\T\T$ operator product  (third line),
we prove  that the only nontrivial conditions are  getting
a constant quartic pole and the  correct value for the quadratic pole.   In terms of products, 
\be
\label{ksvm}
\langle \T | \T \rangle \ = \ \hbox{constant} \;, ~~~~      \T \circ_2 \T  \ = \ 2 \S \,.  
\ee
The second condition is a tensor equation and its two-index part is the double
metric equation of motion.  In terms of the matrix $\M^{MN}$ it takes the form
$\M^2  =  1 + 2 {\cal V}$, 
where $\V$ is  quadratic in $\M$ and contains from two up to six derivatives.
 While the generalized metric squares to the identity, the double metric  squares to the identity plus
 higher derivatives terms.  We view this as a most significant departure
 from the classical theory, forced by $\alpha'$ corrections. 
 We prove that the first equation in (\ref{ksvm}) as well 
as the  one-index part of the
second equation are redundant. 

The construction of the action is done in terms
of the tensor operators 
$\S$ and $\T$, with the latter constrained to have
zero divergence.    These operators encode the double metric $\M^{MN}$ 
and the dilaton. We examine the properties 
of   divergence-free tensors and  introduce an
``overline" projector that
acting on a weight-two tensor $T$  gives  a tensor $\overline T$ with $\hbox{div}\,  \overline T =0$. 
Using this projection we define a $\star$-product mapping into 2-tensors such that $T_1 \star T_2 = T_2 \star T_1$ is  divergenceless.   
We are then able to write a manifestly gauge invariant and O(D,D) invariant
 action
\be S = Çe^Ä \bigl[ \,  Ò\T|\S\rangle \,  - \, \f16\langle \T | \Tä\TÔ \bigr] . \ee
This action is cubic in the double metric (with no quadratic term!) 
and contains up to six derivatives.  
We show by variation that the
expected equations of motion arise. 
This uses a key property of the star product: The 
complete symmetry of  $Çe^Ä \langle \overline T_1 | \overline T_2ä \overline T_3Ô$  under the exchange of any pair of $\overline T$'s. 
The dilaton equation of motion also emerges correctly, 
but takes a bit more effort 
since dilaton variations affect
the overline projection and thus  $\delta_\phi \T$ is not divergence free.

We work out explicitly the above action
in section~\ref{reltogenmetfor}, including all terms with up to two derivatives
and confirm that the generalized metric form of the two-derivative action emerges.  
This reassuring confirmation provides an 
explicit test for many of our formulae. 
The above action almost certainly encodes Riemann-squared and 
Riemann-cubed corrections to the two-derivative action, 
but we will leave a direct verification of this 
for future work. 
%bz  added for referee... 
For sure, we have constructed a completely consistent and 
{\em exactly} gauge invariant $\alpha'$-deformation of the low-energy effective action. The action
contains bounded powers of $\alpha'$, at least when written
in terms of the gravitational
variable $\M$ and the dilaton.  
It thus seems unlikely that this is the full string effective field 
theory of the massless sector.  
We believe, instead, that this theory is a consistent
truncation of string theory in which some of the 
stringy non-locality has been eliminated.

Our paper concludes with some perspectives
on the results and discussion of open questions.

\sectiono{Doubled conformal field theory}  

\subsection{Double dimensions}\label{doubledimensions}

We first describe the construction in double dimensions, then show how it reduces to the usual D dimensions.  We introduce the gauge-invariant constant metric $ú_{MN}$ of O(D,D), which we use implicitly to raise/lower and contract O(D,D) indices $M,N$. 
We also have $2D$ chiral fields $X^M (z)$ representing the doubled coordinates.
Then the D+D dimensional formalism is described by the constraints
\be\begin{split}
	\hbox{Halving:âââstrong:} &â(»^M A)(»_M B) = »^M »_M A = 0 \,,\\
	\hbox{chirality:} &âP^M = X'^M ­ Z^M¼Ü¼A(X)'¼=¼Z^M »_M A \,.
\end{split}\ee
The first line is the duality-invariant  
strong constraint on fields or gauge parameters $A, B$, which are by
definition functions of $X^M$.  
 The constraint states that they, 
as well as their products, must be annihilated by $\p^M \p_M$.
The chirality condition halves the number of oscillators in the theory 
by setting $P_M$, the canonical conjugate to $X^M$, equal to $X_M'$.  This
current is denoted as $Z_M$ and appears each time we take $z$-derivatives
(denoted by prime) of $X$-dependent operators. 

We will also have Virasoro operators $\S$ and $\T$ that must have zero
expectation values on physical states:   
\be\begin{split}
	\hbox{``Virasoro":}âââ\S : &âüZ^2 + \O(Œ') = 0 \\
	\T : &âü\M_{MN}(X) Z^M Z^N + \O(Œ') = 0  \,.
\end{split}\ee
The explicit construction of these operators will be discussed later,
and only leading terms have been shown above. 
The background field $\M$ is the double metric, 
an extension of the generalized metric,  and will play an important role in our theory. 
It should be emphasized that neither the strong constraint nor the chirality condition
acquire $\alpha'$ corrections.  %bz 

We use the Hamiltonian formalism:  The above constraints can be imposed at fixed $ $ (but will be preserved at all $ $).  The halving constraints will be used immediately for reduction to the usual D $X$'s.  The Virasoro constraints will have the usual interpretation in D dimensions, but not in D+D:  
Because of the chirality constraint, only half of the energy-momentum tensor 
should survive, yet we still impose two sets of similar constraints.

(Note that by  ``chiral", as referring to the $X^M$,  we mean left-handed only,
i.e., no ``antichiral".
Chiral bosons were described in Lagrangian language in \cite{Siegel:1983es,Hull:1988dp}.  In nonunitary gauges, such actions can be reduced to the usual $ÄõÄ$ \cite{Siegel:1983es} or to $Ä»_§(»_§-»_ )Ä$ \cite{Floreanini:1987as}, resulting in a second nonchiral set of modes that must be removed as usual by the first-class constraints implied by the original gauge invariance, which must be preserved by the interactions.  Bosons of both chiralities, D left + D right, were used in \cite{Tseytlin:1990nb,Tseytlin:1990va}, but T-duality was considered only for constant backgrounds, i.e., d = D, and thus all fields were compactification scalars.)

As will be elaborated in section~\ref{vecgausym},  
gauge transformations of an operator $T$ (inducing the transformation
of the fields contained in $T$)  
 are to be computed by the commutator 
\be ¶_Å T = [\Ç þ, T] ,ââþ = Å_M (X) Z^M \,,\ee
where $\xi_M$ are the gauge parameters. 
(Here ``$Ç$" means ``$Çdz/2¹i$".  
This is essentially an integral over all $§$ for constant $ $.
In radial quantization it's an integral enclosing the origin.  We'll use ``$È$" for closed contours not enclosing the origin.)

In previous work the focus was on  
equal-``time" ($ $) commutation relations and only Poisson brackets were used.    
Here we find it convenient to introduce operator products, 
and therefore time dependence.  
We therefore choose the Hamiltonian
\be H \, = \, \Ç d\sigma \, \S \, = \, \Ç d\sigma \, üZ^2 \,.\ee
We'll see later that the quantum corrections to $\S$ are a total derivative, so $H$ has no
corrections.  This Hamiltonian is background independent, thus very different from
the familiar background-dependent D-dimensional Hamiltonian. 
We also have the equal-$\tau$ commutation relations
\be
\label{commutatorZ}
[ Z^M (\tau, \sigma_1)\,  , Z^N (\tau, \sigma_2) ] \ = \ - i \,\eta^{MN} \delta' (\sigma_2 - \sigma_1) \,. 
 \ee
 The Heisenberg equation of motion for the operators $Z_M$ then takes the form
\be  i \partial_\tau Z^M (\tau, \sigma) \ = \ \bigl[  Z^M (\tau, \sigma) \,,  H \, \bigr] \ = \  i \,\partial_\sigma Z^M (\tau, \sigma) 
\ee
so that $Z^M$ is a chiral field:
\be
\label{chiralZ}
(\p_\sigma - \p_\tau )  Z^M \, = \, 0 \,. 
\ee
The $X^M$ are thus chiral fields as well.  
We therefore have the propagator (back in the complex plane)  
\be
\label{propagator}
 Ò X^M (1) X^N (2) Ô = ú^{MN}\ln z_{12} \,,\ee
where $z_{12}=z_1-z_2$.   
An $Œ'$ is needed on the right-hand side for proper dimensions. 
For simplicity, however, we will set $\alpha'= 1$.  Note that the sign of $Œ'$ is arbitrary:  We can freely replace $ú£-ú$, since it's the indefinite metric of O(D,D) anyway.
From the above propagator  
and the identification of $Z$ with $X'$ follow the operator products
\be
\label{opeformulae}
 \begin{split}  
 Z_M (1) Z_N (2)  \ = \ & \    
 {1\over z_{12}^2} ú_{MN}  + \hbox{finite}\,, \\
Z_M (1) A(X(2)) \ = \ & \  
{1\over z_{12}} »_M A(2)  + \hbox{finite}\,.
\end{split} \ee
A remarkable simplification occurs due to the strong constraint: There are no singular
terms in the OPE of fields. Indeed, on general grounds 
\be 
\label{gengrounds}
A(1) B(2)\ = \ \,  : A(1)¼e^{\onÁ»_M\langle X^M(1) X^N(2) \rangle\, »_N} B(2) :  \;,
\ee
as seen, e.g., by Fourier transformation of the fields 
\be A(X(1)) ­ Çdk_1¼e^{ik_1ÉX(1)}÷A(k_1) ,ââB(X(2)) ­ Çdk_2¼e^{ik_2ÉX(2)}÷B(k_2)\,,
\ee
and using the identity
\be
e^{ik_1ÉX(1)} e^{ik_2ÉX(2)} ¼=â : e^{ik_1ÉX(1)} e^{ik_2ÉX(2)} :   e^{-k_{1M} k_{2N} \langle X^M(1) X^N(2) \rangle  } \,.    
 \ee
Using the propagator (\ref{propagator}) and then the strong constraint,  (\ref{gengrounds}) gives 
\be 
\label{nocontraction}
A(1) B(2)\ = \ \,  : A(1)¼e^{\onÁ»{}^MÊ\ln (z_{12})Ê»_M} B(2)
 :â=â: A(1) B(2) : 
 \ee
The result is conceptually clear:  The propagator couples coordinates to their 
duals and strongly constrained fields never depend on both a coordinate and
its dual. Since $X$'s without derivatives occur only as arguments of fields, it follows from (\ref{nocontraction})  that no explicit $(\ln z)$'s 
will appear in our contractions.
This situation is similar to the treatment of the twistor superstring formalism for N=4 super Yang-Mills as a closed string with chiral worldsheet fields \cite{Siegel:2004dj}.  There the absence of $\ln$'s corresponds to the fact that the theory describes only particles and not true strings.

\subsection{Halving}

For reduction to D dimensions, we use the strong constraint   
to reduce the dependence of fields to half the coordinates, thus essentially eliminating half the zero-modes.  (For this paper we do not compactify, so these constraints eliminate winding modes.)  We then use the chirality constraint to eliminate half the oscillator modes:  Writing the metric as
\be ú_{MN} = \pmatrix{ 0 & ¶_m^n \cr \noalign{\smallskip} ¶^m_n & 0 \cr} \,,\ee
in terms of  the usual 
D-valued spacetime indices $m$, we have
\be\begin{split}
	\hbox{chirality:}âZ^M & £¼(X'^m, P_m) \\
	\hbox{strong:}âX^M & £¼(X^m, 0)
\end{split}\ee
where the latter refers to the arguments of fields, the only place $X$ doesn't appear as $Z$.  Solving %bz Having 
the halving constraints %bz solved 
in terms of the usual D coordinates, the Virasoro constraints can then be recognized as the usual (in Hamiltonian formalism).  

With the above conditions, we have  $H = \int d\sigma X'^m P_m$ 
and the associated action $S_H$  in Hamiltonian form is given by 
\be
\label{?!}
S_H  = \  \int d^2 \sigma  ¼P_m (\p_\tau - \p_\sigma)  X^m \,, \ee
whose counterpart in Lagrangian language has the singular form
\be S_L¼¾¼\lim_{·£0}¼{1\over ·}Çd^2 §¼ü[(\p_\tau - \p_\sigma)X]^2 \,.\ee
Using the worldsheet metric, the usual string action in D dimensions (without dilaton)
takes the Hamiltonian form 
\be S_H¼¾ Çd^2 §¼\left( \,  ÀX^m P_m - {å{-g}\over g_{11}}ü \M_{MN} Z^M Z^N - {g_{01}\over g_{11}} üZ^2 \right)\,. \ee
Then the action (\ref{?!})  
corresponds to the singular gauge
\be {å{-g}\over g_{11}} = 0â,â{g_{01}\over g_{11}} = 1 \,,\ee
a fact that %bz
may eventually be used to explain that the theory is
some kind of $\alpha'$ truncation of the full string theory (as stated at the
end of the introduction).

\sectiono{Differential double geometry}\label{diffdoubgeo}

\subsection{Operators and contour integrals}  

Although we will focus in the following sections on operators of lower conformal weight, we  provide
here  a general pedagogical discussion, relating the different applications and 
their future use.  We thus consider general operators consisting of functions of $X$ (evaluated at some value of $z$), carrying arbitrary numbers of (D+D)-valued indices, all contracted with $Z$'s and their $z$-derivatives.  We define the conformal weight ``$w$" of such an operator (eigenvalue of ``$÷w$") as the number of $Z$'s plus the number of primes ($'$):
\be ÷w(Z) = ÷w({}') = 1 ,ââ÷w(X) = 0  \,. \ee
Any of the operators to be  
considered has definite weight, but 
may consist of general linear combinations of terms of that weight, which differ by how that weight comes from $Z$'s vs.¼primes.  
(This definition of weight agrees with the conformal field theory definition of weight when operators are on-shell.) 
The lowest weight operators, which play a central role in the rest of this paper, are
\be
\label{threeactors} 
\begin{split}
w = 0,¼\, \hbox{scalars:}&âf = f(X)\,, \cr
	w = 1,¼\hbox{vectors:}&â V = V^M(X)Z_M \,,\cr
	w = 2,¼\hbox{tensors:}&âT = üT^{MN}(X)Z_M Z_N - ü(öT^M Z_M)' \,.
\end{split}\ee
(In oscillator language, these correspond to 1, $a_1ÿ$, and $(a_1ÿ)^2¢a_2ÿ$, respectively.)

We now examine identities for commutators that follow directly from consideration of contour integrals for operator products.  The basic identity is that the commutator of an integrated operator (over all $§$ for fixed $ $) with another operator equals the integral of the former over a contour enclosing the latter in the operator product:
\be [ \Ç A , B (1)  ] = È_1 d2¼A (2) B (1) \,, \ee
where $A$ and $B$ are arbitrary operators, expressed in terms of the currents $Z$ and functions of $X$ (fields).  In the following sections we'll examine relevant special cases; for now we look at general properties.

The charge $ÇA$ generates symmetry transformations $¶_A$ on ``covariant" operators $B$ as
\be
\label{general_symmetries}
 ¶_A B = [ \ÇA, B ] \,,\ee
for symmetry parameters and fields appearing as functions in the operators $A$ and $B$, respectively.  As always, {\it symmetry transformations define a Lie derivative}:  In particular, in the case of quantum mechanics the representation of the Lie derivative/infinitesimal symmetry transformation on a field (denoted by $¶_A$) can be obtained by an operator commutator (with the field represented by an operator $B$).  
Of course, the operator $B$ must contain enough terms so that these transformations close on the fields contained therein, and the operators $A$ must contain enough terms so that their algebra closes.  

In the following sections we will evaluate operator products for the relevant fields.
We will focus on two particular symmetries, to be analyzed in detail in the following sections:
\itemitem{1) }When the symmetry parameter is an O(D,D) vector, multiplying a single current $Z$, 
it describes ``gauge" symmetries, specifically those that reduce to D-dimensional coordinate transformations and the gauge transformations of the 2-form field \cite{Siegel:1993xq}.
\itemitem{2) }When the parameter is a symmetric second-rank O(D,D) tensor, 
multiplying two currents $Z$, it describes worldsheet conformal (coordinate) transformations.  
It is then natural to multiply the second rank
tensor by a single (scalar) world sheet parameter $\lambda (z)$.

\subsection{Bilinear operator products}  

In this section 
we will introduce families of bilinear (quadratic) products of operators starting
from the operator product expansion of two operators. 
Consider operators $\O_1$
and $\O_2$ of weights $w_1$ and $w_2$ respectively:
\be w_1 = ÷w (\O_1),âw_2 = ÷w (\O_2) \,.
\ee
Their OPE is now written as 
\be 
\label{opedorf}
\O_1 (1) \O_2 (2) ¼=¼ Ý_{w=-¥}^{w_1+w_2} {1\over z_{12}^w} ( \O_1 \circ_{w_1+w_2-w} \O_2) (2) \,. \ee
The above expansion defines products $\circ_w$ with $w$ an integer
greater than or equal to zero.  The subscript on the product indicates the weight of the operator, independently
of the weights of $\O_1$ and $\O_2$: 
\be \tilde w(\O_1 \circ_w \O_2) = w \,.  \ee
Note that the expansion in conformal weight is associated with the change in power of $z$, as follows from Taylor expansion and the propagators of the previous section.  
We can write the above OPE as
\be
\label{opeme}
\begin{split}
\O_1 (1) \O_2 (2) \ = \ & \ Ý_{w=0}^{\infty} {1\over (z_{12})^{w_1 + w_2- w}} ( \O_1 \circ_{w} \O_2) (2) \\
 \ = \ & \ {1\over (z_{12})^{w_1 + w_2}} \,  \O_1 \circ_{0} \O_2 (2) +
  \ {1\over (z_{12})^{w_1 + w_2-1}} \,  \O_1 \circ_{1} \O_2 (2) + \ldots 
\end{split}
\ee
In practice, the explicit forms of all these products are evaluated by use of the 
free propagators introduced in the previous section, the various terms coming from the possible combinations and permutations of these propagators.

Of particular interest is the scalar product $\circ_0$ of weight zero, which
we write as a bracket:
\be Ò\O_1 | \O_2Ô ­  \O_1 \circ_0 \O_2  \,.  
\ee
Note that this product is defined even when the operators have different weight.
In an explicit computation, the leading term in $Œ'$ contracts as many indices on the fields as possible with $ú$'s, the rest with derivatives:
\be\begin{split}
& \qquad \O  ¼=¼ \f1{w_\O!}\O^{M_1...M_{w_\O}}Z_{M_1}òZ_{M_{w_\O}} + ...â \\[1.0ex]
	& ÜâÒ\O_>|\O_<Ô = \f1{w_<!(w_>-w_<)!}(\O_>)^{M_1...M_{w_>}}»_{M_1}ò»_{M_{w_>-w_<}}(\O_<)_{M_{w_>-w_<+1}...M_{w_>}} +...  
\end{split}\ee
where $>$ and $<$ refer to the higher and lower weights.

The products satisfy a couple of useful identities associated
with differentiation:  
\be\label{derivative-identity}\boxeq{\begin{aligned}
\hbox{derivative:ââ}\O_1' \circ_w \O_2 ¼=â & (w-w_1-w_2) \,\O_1 \circ_w \O_2\,, 
\\[0.5ex]
( \O_1 \circ_w \O_2 )' ¼=â & \O_1' \circ_{w+1} \O_2 + \O_1 \circ_{w+1} \O_2' \,. ~~
\end{aligned}}\ee
The first follows by differentiating  (\ref{opedorf}) or (\ref{opeme}) with respect to $z_1$ and  recalling that 
$÷w(\O')=÷w(\O)+1$.  The second follows by differentiation with respect to $z_2$
and use of the first identity.

Since all our operators
are Grassmann even 
we have the equality 
$\O_1 (1) \O_2 (2) =   \O_2 (2)\O_1 (1)$ of operator products, and therefore the products satisfy
certain symmetry properties.  For the weight zero product, it follows from 
(\ref{opeme}) that 
\be Ò\O_1 | \O_2Ô \  = \  (-1)^{w_1+w_2}Ò\O_2 | \O_1Ô \,. \ee
More systematically,  we can compare OPE's about $z_1$ and about $z_2$ using
Taylor expansion with the relation $z_1 = z_2 + z_{12}$.  The result is that the 
symmetry property of the products
takes the form
\be
\label{symmetry-identity} \boxeq{~~\hbox{symmetry:â \ }\phantom{\Biggl(} 
\O_2 \circ_w \O_1 = (-1)^{w_1+w_2-w} e^{-\L} \O_1 \circ_w \O_2 ~\,, } \ee
where we have defined   
a (linear)  operator $\L$ that acts on products to give products:
\be
\L  (\O_1 \circ_w \O_2 ) ­ ( \O_1 \circ_{w-1} \O_2 )' \,.
\ee
The right-hand side is indeed a sum of products because of
 the second derivative identity.
One can then verify that the iterated action of this operator gives
\be (\L)^{w'} ( \O_1 \circ_w \O_2 ) ­ ( \O_1 \circ_{w-w'} \O_2 )^{(w')}\,, 
\quad \hbox{and} \quad  \circ_w =0â\hbox{for}¼w<0 \,.\ee
The superscript $(w')$ means $z$-differentiation $w'$ times. 
We have, for example  
\be
\O_2 \circ_2 \O_1 = (-1)^{w_1+w_2} \Bigl(  \O_1 \circ_2 \O_2 - 
(\O_1 \circ_1 \O_2)'  + \f12 (\O_1 \circ_0 \O_2)'' \Bigr)\,.  %bz corrected
\ee
We say that this product has exchange parity $(-1)^{w_1+w_2}$, up to $z$-derivatives.

For higher-weight products, it is 
useful to define truly symmetric products.  This
can be done by explicit symmetrization or antisymmetrization, as appropriate,
and modified further by adding 
 lower-weight products of the same exchange symmetry,
acted by $z$-derivatives to raise the weight.
 Since $\circ_1$, like $\circ_0$, does not include lower-weight products of the same symmetry, their definitions 
 are unambiguous:
\be
\label{low-cases}\begin{split}
\O_1\bullet_0\O_2 ¼­â& \O_1\circ_0\O_2 ¼=¼ Ò\O_1|\O_2Ô\,, \\[0.5ex]
\O_1\bullet_1\O_2 ¼­â& ü[ \O_1\circ_1\O_2 - (-1)^{w_1+w_2} \O_2\circ_1\O_1 ]\,.
\end{split}\ee
For the rest, several alternative possibilites suggest themselves:
\be\begin{split}
(1) &â\O_1 (1) \O_2 (2) ¼=¼ Ý_{w=-¥}^{w_1+w_2} {1\over z_{12}^w} ( \O_1 \bullet_{w_1+w_2-w} \O_2 ) ü((1)+(2)) \,,\\
(2) &â\O_1 (1) \O_2 (2) ¼=¼ Ý_{w=-¥}^{w_1+w_2} {1\over z_{12}^w} ( \O_1 \bullet_{w_1+w_2-w} \O_2) (ü(z_1+z_2))\,, \\[0.8ex]
(3) &â\O_1\bullet_w\O_2¼­¼ü[ \O_1\circ_w\O_2 + (-1)^{w_1+w_2-w} \O_2\circ_w\O_1 ] \,,
\end{split}\ee
where we use the notation
\be {\cal O}\,  ü((1) + (2))  \ \equiv  \   \, ü({\cal O}(z_1) +{\cal O} (z_2)) \,.\ee
All of these have definite exchange symmetry and satisfy a derivative identity   
\be
\begin{split}
 \O_1 \bullet_w \O_2 \ = \ & \ (-1)^{w_1+w_2-w} \O_2 \bullet_w \O_1 \,, \\[0.3ex]
 ( \O_1 \bullet_w \O_2 )' \ = \ & \ \O_1' \bullet_{w+1} \O_2 + \O_1 \bullet_{w+1} \O_2' \,.
 \end{split}
 \ee
Moreover, the three versions agree with the  definitions of $\bullet_0$ and $\bullet_1$
in (\ref{low-cases}).  The $\circ$ products can be expressed in terms of the $\bullet$ products
as follows:  
\be \O_1\circ\O_2 = [1+\hbox{tanh} (\f12 \L )\,  ]f(\L)\, \O_1\bullet\O_2\,, \ee
where $f(\L) = f(-\L)$ and $f(0) = 1$.  The function $f$ takes the following forms for our
three cases:
\be\begin{split}
(1) &âf(\L) = \hbox{cosh}^2Ê \f12 \L  \\
(2) &âf(\L) = \hbox{cosh} \f12 \L   \\
(3) &âf(\L) = 1\,, 
\end{split}\ee
as easily verified by Taylor expansion about $z_2$.  
All these (anti)symmetrized products differ from the asymmetric 
ones only by total $z$-derivative terms, which   
play an auxiliary role.  A particularly convenient choice of them
will lead to a unique symmetric product,  the star-product $\star$, defined 
with the help of the dilaton in section \ref{starproduct-sec}.

In the following we will make extensive use of the symmetry and derivative identities
(\ref{symmetry-identity}) and (\ref{derivative-identity}),  
 usually without reference, except for a few early examples and some exceptional cases.  This should be obvious:  For any expression $A'\circ B$ we use the derivative identity to remove the prime; for any expression where we wish to reorder a product we use the symmetry identity.  For convenient reference, we have collected the most frequently used identities in the Appendix.

The operator product expansion in terms of $\circ$ products can be used
to evaluate commutators, such as $[ \ÇÂ \O_1 , \O_2 ]$.  Here $\lambda(z)$ 
is a worldsheet parameter that depends on $z$, but not on $X(z)$,  so it does not
contribute propagators.  We then find 
\be \bigl[ \ÇÂ \O_1 , \O_2\,  \bigr] = Ý_{w=1}^{w_1+w_2} {1\over (w-1)!} Â^{(w-1)} \O_1 \circ_{w_1+w_2-w} \O_2 \,,\ee
where  the integration around the position $z_2$ of the second operator
picks out just the singular part of the operator product.
In here we used the integration identity
\be È_{z_2}{dz_1\over 2¹i}Ê{1\over z_{12}^{n+1}}ÊA(z_1)¼=¼{1\over n!}A^{(n)}(z_2) 
\,, \ \ n\geq 0\,. \ee  
For example, we have
\be 
\begin{split}  
È_{z_2}{dz_1\over 2¹i}Ê{1\over z_{12}^3}Ê2A(z_1) =& \  A''(z_2) 
= (Z^M »_M A)' = (Z^M)'»_M A + Z^M(»_M A)'\\
	= & \  Z'^M »_M A + Z^M Z^N »_N »_M A \,.
	\end{split}\ee

\subsection{Cubic relations}  

Although we have used operator product expansions in place of commutators, commutators are equivalent to just the singular parts of OPE's.  However, OPE's of more than 
two  operators can be unwieldy.  In particular, Jacobi identities are easier than 
associativity identities, 
which require keeping finite terms after the first product, contributing to infinite sums.

Two important identities are the distributivity identity
\be  \hbox{\bf distributivity:â} [ \Ç A , B (1) C (2) ] = [ \Ç A , B (1) ] C (2) + B (1)  [ \Ç A , C (2) ]
\,,  \ee
which follows from
\be\begin{split}
\left[ ÇA, B (1) C (2) \right] & = È_{1,2} d3¼A (3) B (1) C (2) = È_1 d3¼A (3) B (1) C (2) + È_2 d3¼A (3) B (1) C (2) \\[.1in]
	& = \left[\ÇA, B (1) \right] C (2) + B (1) \left[\ÇA, C (2) \right]\,, 
\end{split}\ee
and the Jacobi identity
\be \hbox{\bf Jacobi:â} [ \Ç A_{[1} , [ \Ç A_{2]} , B ] ] = [ [ \Ç A_1 , \Ç A_2 ] , B ] \,,\ee
which follows from distributivity upon integrating $B(1)$ about $z_2$.

The distributivity identity proves the symmetry invariance of field equations, since we will derive the field equations through preservation of the Virasoro operator algebra.
The Jacobi identity proves the closure of the symmetry transformations of the background fields we will introduce below:
\be [\ÇA_1,\ÇA_2] = \ÇA_{12}âÜâ[¶_{A_1},¶_{A_2}] = -¶_{A_{12}} \,. \ee
The explicit action of $¶_A$ on various fields, and the explicit form of $A_{12}$ in terms of $A_1$ and $A_2$, as evaluated by the above operator commutators, is a subject of the following sections.  

We can derive various identities for these infinite classes of products by applying these identities, and expanding in powers of $z$, including the implicit ones now appearing as derivatives on $Â$.  For the distributivity identity, which we write as
\be [\ÇÂ\O_1 , \O_2 \O_3 ] - \O_2 [\ÇÂ \O_1 , \O_3 ] = [\ÇÂ \O_1 , \O_2 ] \O_3 \,,\ee
we find
\be
\label{didn}\begin{split}
[\ÇÂ\O_1 , \O_2 \O_3 ] ¼=â& Ý_{w=-¥}^{w_2+w_3} Ý_{w'=1}^{Ýw_i-w} {1\over z^w} {Â^{(w'-1)} \over (w'-1)!} \
	\O_1 \circ_{öw} ( \O_2 \circ_{w_2+w_3-w} \O_3 ) \\
\O_2 [\ÇÂ \O_1 , \O_3 ] ¼=â& Ý_{w'=1}^{w_1+w_3} Ý_{w=-¥}^{Ýw_i-w'} {1\over z^w} {Â^{(w'-1)} \over (w'-1)!} \
	\O_2 \circ_{öw} ( \O_1 \circ_{w_1+w_3-w'} \O_3 ) \\
[\ÇÂ \O_1 , \O_2 ] \O_3 ¼=â& Ý_{w''=0}^¥  Ý_{w'=w''+1}^{w_1+w_2+w''} {w'-1\choose w''}Ý_{w=-¥}^{Ýw_i-w'} {1\over z^w} {Â^{(w'-1)}\over (w'-1)!}  \\
	& ð ( \O_1 \circ_{w_1+w_2+w''-w'} \O_2 ) \circ_{öw} \O_3 \,, \\
\end{split}\ee
where $z = z_{23}$, all operators and $\lambda$'s are evaluated at $z_3$,  and 
\be öw ­ Ýw_i - w - w' ,âÝw_i=w_1+w_2+w_3 \,. \ee
(For the last line in (\ref{didn})  the third sum arises because we need to Taylor expand $Â(2)$ about $z_3$.)

We then compare terms of fixed order $w$ and $w'$ in derivatives of $Â$ and powers of $z$.  Paying attention to the limits of summation we find
\be\label{Tom} \boxeq{
\ \O_1 \circ_{öw} ( \O_2 \circ_{w_2+w_3-w} \O_3 )  -
	\O_2 \circ_{öw} ( \O_1 \circ_{w_1+w_3-w'} \O_3 )  =
	Ý_{w''=1}^{w'} \hskip-3pt{w'\hskip-1pt -1\choose w''\hskip-1pt -1}( \O_1 \circ_{w_1+w_2-w''} \O_2 ) \circ_{öw} \O_3 \ }
\ee
where always
\be -¥ ² w + w' ² Ýw_i , ââ 1 ² w'\,,\ee
which means that we get identities for any $w$ satisfying
\be
w \leq  -1 +  Ýw_i  \, = \, w_{max} \,. 
\ee
We get one identity for $w= w_{max}$, two identities for $w = w_{max} -1$,
three identities for $w = w_{max}-2$ and so forth and so on. 

For Jacobi, we examine
\be [ \Ç Â_1 \O_1 , [ \Ç Â_2 \O_2 , \O_3 ] ] - [ \Ç Â_2 \O_2 , [ \Ç Â_1 \O_1 , \O_3 ] ] = [ [ \Ç Â_1 \O_1 , \Ç Â_2 \O_2 ] , \O_3 ] \,.  \ee
The manipulations and results are almost the same as for distributivity, only now only singular terms contribute, so
\be 1 ² w ,¼1 ² w' \,,  \ee
and things are antisymmetric in $w$ and $w'$ (when 1's and 2's are switched).  The result is the same as for distributivity, except for the restriction on the lower limit of $w$ to singular terms.

\sectiono{Vector gauge symmetry} \label{vecgausym}

In this section we will examine the quantum corrections to generalized brackets and (Lie) derivatives introduced in previous papers for the gauge symmetries.  We begin by considering general properties of current algebra that should generalize to other string models.
For the vector operator  
\be
þ (z) = Å^M(X(z))Z_M(z) \,, 
\ee
with gauge parameter $Å^M(X)$, we define the gauge transformation $\delta_\xi B$ 
of the arbitrary operator $B$ of weight $\tilde w (B)$ by  
\be ¶_Å B \equiv  [\Çþ , B ] = þ \circ_B B\,,  
\ \  \  \circ_B ­ \circ_{÷w(B)} \,. 
\ee
The equality after the definition follows by using (\ref{opedorf}) for $\Xi$ and $B$:
\be
\Xi (1) B(2) \ = \  \hbox{regular} \, + \,  {1\over z_{12}}  (\Xi \circ_B  B) (2)   \, + \,  
 {1\over z_{12}^2}  (\Xi \circ_{B-1}  B)(2)   + \ldots \,.  
\ee 
The gauge transformation $\delta_\xi$ vanishes if the gauge parameter
  is ``gauge for gauge", $Å^M = »^M ½$, since
\be \Çþ \  =  \ÇZ^M »_M ½ = \ǽ' = 0 \,.  \ee
It is a fundamental property 
 that {\it all} products $\circ$ are gauge covariant:   
 \be
 \label{covariance-products} 
 \boxeq{\phantom{\Bigl(} ¶_Å(A\circ_w B) = (¶_ÅA)\circ_w B + A\circ_w(¶_ÅB) \,. \ \ } 
 \ee 
This follows from the distributive identity  
\be [\Çþ, \, A(1) B(2)] = [\Çþ,A](1) B(2)  + A(1) [\Çþ,B](2) \,,\ee
and use of (\ref{opedorf}) for each term to find
\be [\Çþ, \, A\circ_w B] = [\Çþ,A]\circ_w B + A\circ_w [\Çþ,B]\,,\ee
which is equivalent to (\ref{covariance-products}).  
Taking a $z$-derivative is also a covariant operation, 
\be
\label{zder-cov}
\boxeq{ \phantom{\Bigl(} \ ¶_Å(A') = (¶_Å A)' \,, \ \ } 
\ee
as we check using the
derivative identities: 
\be (þ\circ_A A)' = þ'\circ_{A+1} A + þ\circ_{A+1}A' =\,  0 \, + \, þ\circ_{A'}A' \,.\ee
All versions of $\bullet$ products are also gauge covariant:  They are 
 built from $\circ$ products and $z$-derivatives of $\circ$ products.
So is the $\L$ operator, in the sense that $\delta_\xi \L =  \L\, \delta_\xi$ holds
when acting on (sums of) bilinear products of operators.  

In the following we will discuss the operator product of currents
that give us inner brackets,  C and D brackets,  and outer products,
all of them with $\alpha'$ corrections.  They will be evaluated explicitly,
and the C bracket Jacobiator will be shown to be a trivial vector. 
We also evaluate the gauge transformations of scalars, vectors, and tensors.

\subsection{Inner and outer products, brackets}

The key ingredients for the theory we are to construct arise in a simple manner
from the OPE expansion $\Xi_1(1) \Xi_2(2)$ of two currents $\Xi_1 = \xi^M_1 Z_M$ and $\Xi_2 = \xi^M_2 Z_M$.
Indeed, the operators in this expansion define the inner product, the various brackets, and a set of useful products.  

The brackets
in generalized geometry come in Courant and Dorfman varieties.  Their 
double field theory versions,
without $\alpha'$ corrections, are the C bracket of \cite{Siegel:1993th} and the D-bracket~\cite{Hull:2009zb}.  The C bracket
when restricted from D+D dimensions to D dimensions becomes the
Courant bracket \cite{Hull:2009zb}.  Similarly, upon reduction, the D bracket 
becomes the Dorfman bracket. 

The C and D varieties of brackets arise by doing the OPE of two currents 
in slightly different ways.  In the C case the normal ordered operators
are averaged over the two points, while in the D case the operators are
located at the position of the second current.
In the following, the ``quantum" contributions to the OPE give the 
$\alpha'$ corrected brackets, as well as corrected inner products 
and other products.  We call these the new brackets and products. 
Upon reduction to D dimensions they give new versions of the
Courant and Dorfman brackets, as well as a new inner product.

We thus have two forms of the OPE:
\be\begin{split}
\label{keyexp}
þ_1(1) þ_2(2) ¼­â& \left[ {1\over z_{12}^2} Òþ_1|þ_2Ô
 + {1\over z_{12}} [þ_1,þ_2]_{{}_C} ¼ + :þ_1 þ_2:_{{}_C} \right] ü((1) + (2)) 
 + \O (z_{12}) \,,  \\[0.4ex]
 þ_1(1) þ_2(2) ¼­â& \left[ {1\over z_{12}^2} Òþ_1|þ_2Ô
 + {1\over z_{12}} [þ_1,þ_2]_{{}_D} ¼ + :þ_1 þ_2:_{{}_D} \right] (2) + \O (z_{12})  \,. 
\end{split}\ee
In our previous notation, we thus have
\be [V_1,V_2]_D ­ V_1\circ_1 V_2,â \ \,  [V_1,V_2]_C ­ V_1\bullet_1 V_2,â
	\ \,  :V_1 V_2:_{{}_D} ­ V_1\circ_2 V_2,â\ \,  :V_1 V_2:_{{}_C} ­ V_1\bullet_2 V_2 \,,\ee
as well as the previously defined $Ò¼|¼Ô­\circ_0=\bullet_0$, where we have made a particular choice of the ambiguous $\bullet_2$.

The two above expansions are simply related by
\be
 A \, \f12 ( (1) + (2))  \ = \ \left(  A  + \f12  z_{12} A'  + \f14 \, z_{12}^2  A'' \right) (2)   \,    + {\cal O}(z_{12}^3) \,,
\ee
and thus
the inner product, at the second-order pole, is the same for the two, while the other terms are related by
\be
\label{cvd}
\begin{split}
[þ_1,þ_2]_{{}_D}  ¼=â& [þ_1,þ_2]_{{}_C} +\,  ü Òþ_1|þ_2Ô'  \\[0.3ex]
:þ_1 þ_2:_{{}_D} ¼= â\ & \hskip-5pt :þ_1 þ_2:_{{}_C}  +¼ ü  [þ_1,þ_2]_{{}D} ' \,. 
\end{split}\ee
The C form is more useful for symmetry:  Clearly
\be Òþ_1|þ_2Ô = Òþ_2|þ_1Ô ,ââ[þ_1,þ_2]_{{}_C} = - [þ_2,þ_1]_{{}_C} ,â¼: þ_1 þ_2 :_{{}_C} 
¼=¼ : þ_2 þ_1 :_{{}_C} \,. \ee
The D bracket, as opposed to the C bracket, is not antisymmetric in its 
inputs. One readily sees that the C bracket is obtained by antisymmetrization
of the D bracket
\be
\label{casantid}
 [þ_1,þ_2]_{{}C} ­ ü[þ_{[1},þ_{2]}]_{{}_D} \,. \ee

The brackets can be also viewed as  current algebra commutators.
For example, consider the single commutator $[\Çþ_1,þ_2]$.  We can use
the OPE in (\ref{keyexp}) to see that this selects the D bracket
\be
\label{dder}
 [\Çþ_1,þ_2] \ = \ [þ_1,þ_2]_{{}_D} \ = \  [þ_1,þ_2]_{{}_D}^M Z_M\,  .
\ee
Clearly the D-bracket then defines a distributive ``D-derivative",
\be
\label{btfvc}
 [\Çþ_3,þ_1 (1) þ_2 (2) ]  
= [\Çþ_3,þ_1 (1) ] þ_2 (2) + þ_1 (1) [\Çþ_3,þ_2 (2) ]\,, \ee
as follows from the distributivity identity of the previous subsection.  This is a special case of the distributivity of the Lie derivative/gauge transformation $¶_Å$.

Of course, the algebra of integrated currents, and thus gauge transformations, closes.
We can now express this algebra in terms of the new brackets:
From (\ref{dder})
\be [\Çþ_1,\Çþ_2] = \Ç[þ_1,þ_2]_{{}_D} = \Ç[þ_1,þ_2]_{{}_C} \,,  \ee
using the fact that the 2 brackets differ only by a total derivative.
We can thus identify
\be [\Çþ_1,\Çþ_2] = \Çþ_{12}âÜâþ_{12} = [þ_1,þ_2]_{{}_C} \,, \ee
without loss of generality, so that $þ_{12}$ preserves the antisymmetry of $Çþ_{12}$.
This defines the algebra of gauge transformations:
\be [¶_{Å_1},¶_{Å_2}]\, = \, -[¶_{Å_2},¶_{Å_1}] \, = \,  -¶_{Å_{12}},â \ Å_{12}^M 
= [þ_1,þ_2]_{{}_C}^M\,. \ee
All these objects will be computed explicitly in the following subsection.
   
\subsection{Evaluation}

We now evaluate the OPE of two currents for the theory under consideration.  
The possible contractions give:	
\be\begin{split}
  þ_1(1) þ_2(2) ¼=¼ & : Å_1^M (1) Z_M (1) Å_2^N (2) Z_N (2) \\[0.4ex]
	& + Å_1^M (1) ÒZ_M (1) Å_2^N (2)Ô Z_N (2) + Å_2^M (2) ÒZ_M (2) Å_1^N (1)Ô Z_N (1) \\[0.4ex]
	& + Å_1^M (1) Å_2^N (2) ÒZ_M (1) Z_N (2)Ô 
	+ ÒZ_M (2) Å_1^N (1)Ô ÒZ_N (1) Å_2^M (2)Ô : \, . 
\end{split}\ee
The last term, with a double contraction, is the quantum correction.
Using (\ref{opeformulae}) to evaluate the above contractions we find
\be
\label{89vm}
\begin{split}  
   þ_1(1) þ_2(2) ¼=¼ & \ \  : (Å_1^MZ_M) (1) (Å_2^N Z_N) (2) \\[0.5ex]
	& + {1\over z_{12}} \bigl( \, Å_1^M (1) \, (»_M Å_2^NZ_N) (2) -  Å_2^M (2) \, (»_M Å_1^N Z_N) (1) \, \bigr)  \\
	& + {1\over z_{12}^2} \bigl( \, Å_1^M (1) Å_{2M} (2) 
	- »_M Å_1^N \hskip-1pt (1) \,  »_N  Å_2^M\hskip-1pt (2) \, \bigr)  : \, \, . 
\end{split}\ee
As we will see, the second line will contribute
to  the usual Lie bracket/commutator.  
A contribution from   
the first term on the third line modifies it to the classical C- or D-bracket.  
A contribution from the last 
term  gives the quantum correction.  
We will use the following expansion of a normal-ordered product
of operators
\be
\label{expident}
 A(z_1) B(z_2) ¼= ¼\left[ AB - ü z_{12} \,  A\onª» B 
- ü z_{12}^2 A'B' + \O (z_{12}^3) \right] ü((1) + (2)) \,.
\ee
(As usual, 
 normal ordering is assumed for operators evaluated at the same point.)
Here  the $z$ 
derivative $\onª»$ is defined to act
as $A\onª» B \equiv AB' - A' B$. We now use this equation to expand 
the right-hand side of (\ref{89vm}) 
and we obtain a result that can be put in the form of 
the top equation in 
(\ref{keyexp}).

The residue of the second order pole defines a new symmetric inner product given by
\be 
\hbox{\bf Inner product:}
 ~~~ Òþ_1|þ_2Ô ¼= ¼Å_1^M Å_2^N ú_{MN} - (»_N Å_1^M)(»_M Å_2^N)\,. 
 \ee
The first term is the familiar one and the second is the $\alpha'$ correction, arising from a quantum contribution in the OPE.  This correction vanishes if 
any of the $\xi$'s is trivial ($\xi^M = \p^M \chi$) and the whole inner product
vanishes if both $\xi$'s are trivial.  Equivalently, 
$ ÒA'|B'Ô = 0$, recalling that  $(A')^M = »^M A$.
Using the strong constraint, the new inner product can also be written as
\be Òþ_1|þ_2Ô = Å_1^M Å_2^N ú_{MN} + ü K_1^{MN}K_{2MN}\,,\ee
where $K_{MN}$ is the ``field strength" of the gauge parameter:
\be K_{MN} \equiv »_{[M}Å_{N]} ­ »_M Å_N - »_N Å_M \,.\ee

Reducing to D dimensions by setting $\tilde \p^i$ derivatives to zero gives, 
with $(\tilde\xi_{1i} , \xi_1^i)$ and $(\tilde\xi_{2i} , \xi_2^i)$ the one-form
and vector components of $\xi_1^M$ and $\xi_2^M$, respectively:
\be
Òþ_1|þ_2Ô \ = \ \xi_1^i \tilde\xi_{2i} + \xi_2^i \tilde\xi_{1i}  \ - \  \p_i \xi_1^j  \, \p_j \xi_2^i \,. 
\ee
The last term is the quantum correction.

The first-order pole contains the
corrected C-bracket, a skew bracket that we write compactly as: 
\be [þ_1,þ_2]_{{}_C} ¼= ¼[þ_1,þ_2]_{{}_L} - üÒþ_1|\onª»|þ_2Ô \,. \ee
Here  $\onª»$ translates as $»=Z^M »_M$, the bracket $[¼,¼]_{{}_L}$ is the commutator/Lie bracket
\be [þ_1,þ_2]_{{}_L} ­ (Å_{[1}^N »_N Å_{2]}^M)Z_M\,, \ee
and the correction to the C-bracket is produced by the correction of the
inner product.  More explicitly the above formula reads
\be
\label{c-bracket-corr}
\hbox{\bf  C bracket:}~~[ \Xi_1 , \Xi_2 ]_{{}_C}^M  \ = \ Å_{[1}^N »_N Å_{2]}^M \, - \, \f12  \,
Å_1^K\onª»{}^M Å_{2K}\, + \, \f12 
\, (»_K Å_1^L)\onª»{}^M(»_L Å_2^K)\,. 
\ee 
The last term, with three derivatives, is the new correction.  

Upon
reduction from D+D to D dimensions
the vector part of the bracket is not corrected, but the one-form part is
\be
\begin{split}
\left( [ \Xi_1 , \Xi_2 ]_{{}_C}\right)^i  \ =\ &  \  \xi_{[1}^k \p_k \xi_{2]}^i    \,, \\[0.5ex]
\left( [ \Xi_1 , \Xi_2 ]_{{}_C}\right)_i  \ = \ &  \ \xi_{[1}^k \p_k \tilde\xi_{2]i} \, + \f12
\bigl( \xi_1^k \onª»{}_i \tilde \xi_{2k} + \tilde \xi_{1k} \onª»{}_i \xi_2^k \bigr) \ + \  \f12 
\, (»_k Å_1^\ell)\onª»{}_i(»_\ell Å_2^k)  \,.
\end{split}
\ee
The last term with three derivatives is the quantum correction.

Finally, the regular term in the OPE defines 
a tensor operator of weight two. 
The two-index part defines an outer (``star" $¡$) product constructed
from the two $\xi$'s. The one-index part defines a product $(\flat)$, built
from  the two $\xi$'s as well.  
As will be explained in the next section, 
the two-index part defines a tensor by itself, but the one-index part does not.
Thus the $(\flat)$ product, which enters the tensor as total derivative,  
is less interesting.  
We have
\be
\label{twoproducts}
 :þ_1 þ_2:_{{}_C} ¼= ¼þ_1 ¡ þ_2 - (þ_1 \flat þ_2)'_{{}_C} \, . 
 \ee
Both products are symmetric, a property they inherit from the OPE, 
\be
þ_1 ¡ þ_2 \ = \ þ_2 ¡ þ_1\,, ~~~    (þ_1 \, \flat \, þ_2)_{{}_C}  \ = \ (þ_2 \,\flat \, þ_1)_{{}_C} \,.
\ee
Explicitly,
\be\begin{split}
\hbox{\bf  Outer product:~~} þ_1 ¡ þ_2 â­ &¼ ü \left[ \, \xi_{(1}^M \, \xi_{2)}^N 
+ »^P \xi^{(M}_{(1}\, »^{N)}\xi_{2)P}\,  - ü \, »^M \xi_{(1}^P\, »^N \xi_{2)P} \right. \\[0.6ex]
	& \left.  \ \ \ + ü »^M »^P \xi_{(1}^Q\, »^N »_Q \xi_{2)P} \right] Z_M Z_N \,. 
\end{split}\ee
We also give the $\flat$-product for completeness:
\be (þ_1 \,\flat \, þ_2)_{{}_C} Ê­¼ ü\, \xi_{(1}^K »_K \xi_{2)}^M \, Z_M \,. \ee

Being bilinear and symmetric, these two products (as well as the inner product) can be written in terms
of squares. For example, 
\be
þ_1 ¡ þ_2 = ü [ (þ_1 + þ_2) ¡ (þ_1 + þ_2) - þ_1 ¡ þ_1 - þ_2 ¡ þ_2 ]\,,
\ee
so we need only define that:
\be
þ ¡ þ = [ Å^M Å^N + »^P Å^{(M}\, »^{N)} Å_P - ü \, »^M Å^P »^N Å_P + ü \, »^M »^P Å^Q
\, »^N »_Q Å_P\, ] Z_M Z_N \,. 
\ee

We now turn to the D form of the OPE.  Using the relation to the C form,
\be
[þ_1,þ_2]_{{}_D}  = [þ_1,þ_2]_{{}_C} + ü Òþ_1|þ_2Ô'  = [þ_1,þ_2]_{{}_L} + Òþ_2|»|þ_1Ô \,.
\ee
In components,
\be
\label{d-bracket-corr}
\hbox{\bf  D bracket:}~~[ \Xi_1 , \Xi_2 ]_{{}_D}^M  \ = \ 
Å_{[1}^K »_K Å_{2]}^M 
	+ \, \p^M \xi_1^K \, \xi_{2K} 
-\,   \p^M »_K Å_1^L  \,  »_L  Å_2^K  \,  
\,. 
\ee
 
Since the extra term in $:þ_1 þ_2:_{{}_D}$ is a total derivative, ``$Ê¡Ê$" is unchanged, but
we have a different bilinear, symmetric, auxiliary product:
\be
\label{twoproductsDversion}
 :þ_1 þ_2:_{{}_D} ¼\equiv ¼þ_1 ¡ þ_2 - (þ_1 \flat þ_2)'_{{}_D} \, , 
 \ee
 where
 \be
 (þ_1 \flat þ_2)_{{}_D} \ = \ (þ_1 \flat þ_2)_{{}_C} - \f12 [þ_1,þ_2]_{{}D} \,. 
 \ee
In components
\be
 (þ_1 \flat þ_2)_{{}_D} ¼\equiv ¼ \left[  \,  Å_2^K»_K Å_1^M 
	-\f12 \, \p^M\xi_1^K \, \xi_{2K} 
+ \f12\,  \p^M »_K Å_1^L   \, »_L  Å_2^K 
 \right] Z_M \,.\ee

As for any gauge transformation, one can view the D bracket as defining a (generalized) Lie derivative.
For a current $V = V^M Z_M$ 
\be
\label{vdorf}
{\bf L}_\xi  V \ \equiv  \ [ \Xi , V ]_{{}_D}  \quad\to \quad
{\bf L}_\xi  V^M  \ = \ \xi^{P}\partial_{P}V^{M}
  +(\partial^{M}\xi_{P}\,    -\partial_{P}\xi^{M})  V^{P}
  -\,   \p^M »_K Å^L  \,  »_L V^K\,,
\ee
where the last term is the $\alpha'$ correction 
to the generalized 
Lie derivative introduced in \cite{Siegel:1993th}.   
Upon reduction to D dimensions, the Lie derivative of a vector receives
no correction but the Lie derivative of a one-form does
\be
\label{corrliegen}
\begin{split}
({\bf L}_\xi  V)^i   \ = \ & \ \xi^k \p_k V^i - V^k \p_k \xi^i   \\
({\bf L}_\xi  V)_i  \ = \ & \ \xi^k \p_k V_i\, + \, \p_i \xi^p \,V_p  + (\p_i \tilde \xi_p - \p_p \tilde \xi_i) V^p \, - \, \p_i \p_k \xi^p \p_p V^k \,.
\end{split}
\ee
The last term on the second line is the correction.

\subsection{Jacobiator and N-tensor}  

The C-bracket, while antisymmetric, is not a Lie bracket, since it does not
satisfy a Jacobi identity.  The failure to satisfy a Jacobi identity is measured by
the Jacobiator $J_C (\Xi_1, \Xi_2, \Xi_3)$ defined by
\be
\begin{split}
J_C (\Xi_1, \Xi_2, \Xi_3) \ \equiv \ & \  [\, [þ_1,þ_{2}]_{{}_C} , þ_{3} ]_{{}_C} + \,
[\,  [þ_2,þ_{3}]_{{}_C} , þ_{1} ]_{{}_C} + \, [\,  [þ_3,þ_{1}]_{{}_C} , þ_{2} ]_{{}_C}  \\[1.0ex]
= \ &  - \bigl( [þ_{1},[þ_2,þ_{3}]_{{}_C}]_{{}_C}  + 
[þ_{2},[þ_3,þ_{1}]_{{}_C}]_{{}_C}  + [þ_{3},[þ_1,þ_{2}]_{{}_C}]_{{}_C} \bigr) \\[1.0ex]
= \ & - \f12  [þ_{[1},[þ_2,þ_{3]}]_{{}_C}]_{{}_C} \,,
\end{split}
\ee
where the antisymmetrization on the last line is over the three indices, making
the Jacobiator manifestly antisymmetric on the three currents $\Xi_1, \Xi_2, \Xi_3$. 
In this section we calculate this Jacobiator.  As it turns out, the above C-Jacobiator
is actually proportional to the D-Jacobiator, defined by 
\be
\begin{split}
J_D (\Xi_1, \Xi_2, \Xi_3) \ \equiv \  - \f12  [þ_{[1},[þ_2,þ_{3]}]_{{}_D}]_{{}_D} \,.
\end{split}
\ee
While the D bracket is not antisymmetric, the above Jacobiator is. 

To motivate the answer for this calculation let us consider the rewriting:
\be
\label{jacascurrcomm}
  [þ_{[1},[þ_2,þ_{3]}]_{{}_D}]_{{}_D} \ = \  [ \Çþ_{[1} , [ \Çþ_2 , þ_{3]} ] ] \,.
  \ee
The right-hand side is a current that when integrated must give zero since
$[ \Çþ_{[1} , [ \Çþ_2 , \Çþ_{3]} ] ] =0$ trivially.  Therefore this current must
be a total derivative of a scalar $N$ built from the three currents
\be
\label{anticipate}
  [þ_{[1},[þ_2,þ_{3]}]_{{}_D}]_{{}_D}  = 4\, N' = 4 Z^M »_M N\,, \ee
where the coefficient was adjusted for later convenience. 
Note that the nontriviality of the Jacobiator does not imply the violation of the usual type of Jacobi identities for operator commutators, where the same operators appear in all terms, in contrast to the right-hand side of (\ref{jacascurrcomm}), where the choice of currents to be integrated varies from term to term.  

$J_C$ and $J_D$ can be calculated conveniently at the same time.  We first relate
 $J_C$ to $J_D$.  Using twice the fact that the C-bracket is the antisymmetric part of the D-bracket (see (\ref{casantid})), we find:
\be
\label{fstep}
 [þ_{[1},[þ_2,þ_{3]}]_C]_C = [þ_{[1},[þ_2,þ_{3]}]_D]_C 
 = ü( [þ_{[1},[þ_2,þ_{3]}]_D]_D - [[þ_{[1},þ_2]_D,þ_{3]}]_D ) \,.\ee
Then using the distributivity (\ref{btfvc}) 
of the D-bracket 
\be\begin{split}
[þ_{[1},[þ_2,þ_{3]}]_D]_D & =¼ [[þ_{[1},þ_2]_D,þ_{3]}]_D + [þ_{[2},[þ_1,þ_{3]}]_D]_D \\[0.5ex]
	ܼ[[þ_{[1},þ_2]_D,þ_{3]}]_D & =¼ 2\,[þ_{[1},[þ_2,þ_{3]}]_D]_D \,,
\end{split}\ee
a curious relation, since the $(+2)$ would be replaced by a $(-1)$ for an antisymmetric bracket (it shows that similar looking definitions of the
D Jacobiator can be quite different).  Back in (\ref{fstep}) we find the anticipated relation
between Jacobiators
\be
\label{fhuu}
¼[þ_{[1},[þ_2,þ_{3]}]_C]_C  =¼ - ü [þ_{[1},[þ_2,þ_{3]}]_D]_D  \quad Ü \quad
 J_C \,= \, -\f12 J_D \,. 
\ee
We then again express the C-Jacobiator in terms of the D-Jacobiator using 
(\ref{casantid}) for the inner C-bracket and the first of (\ref{cvd}) 
for the outer C bracket: 
\be [þ_{[1},[þ_2,þ_{3]}]_C]_C = [þ_{[1},[þ_2,þ_{3]}]_D]_C 
= [þ_{[1},[þ_2,þ_{3]}]_D]_D - ü Òþ_{[1}|[þ_2,þ_{3]}]_DÔ' \,. \ee  
Again, using (\ref{casantid}) we can replace the D by a C inside the inner product, so that we have found
\be
- 2 J_C  \, = \, - 2 J_D - ü Òþ_{[1}|[þ_2,þ_{3]}]_CÔ' \quad Ü\quad 
 J_C  \, = \,  J_D + \f14 Ò[þ_{[1},þ_{2}]_C | þ_{3]}Ô' \,. 
\ee
It follows from this equation and (\ref{fhuu}) that 
\be
J_C (\Xi_1, \Xi_2, \Xi_3) \ = \ \f1{12} Ò[þ_{[1},þ_{2}]_C | þ_{3]}Ô' \ = \ N'\, ,  
\ee
where $N$ can be written as
\be
N (\Xi_1, \Xi_2, \Xi_3) \ = \ \f16 \bigl( Ò[þ_{1},þ_{2}]_C | þ_{3}Ô +
Ò[þ_{2},þ_{3}]_C | þ_{1}Ô + Ò[þ_{3},þ_{1}]_C | þ_{2}Ô \bigr) \,.
\ee
This result takes exactly the same form as that
classical C-bracket Jacobiator~\cite{Hull:2009zb}, the only change is that now
we use the $\alpha'$ corrected brackets and inner product.
More explicitly,
\be N = -\f18 \, \bigl( Å_{[1}^M Å_2^N K_{3]MN} + Å_{[1}^M K_2^{NP} »_M K_{3]NP} + \f23 K_{[1M}{}^N K_{2N}{}^P K_{3]P}{}^M \, \bigr) \,.\ee
Note also that $J_D = -2 N'$ and that is consistent with (\ref{anticipate}).  
The N-tensor was introduced in \cite{Siegel:1993th} as a field strength.  In D dimensions, it reduces to the Nijenhuis tensor, that appears in the 
computation of the Jacobiator for the Courant bracket.

\subsection{Gauge transformations}  

We have already seen examples of the three different kinds of covariant operators
listed in (\ref{threeactors}): scalars, vectors, and tensors. 
The gauge transformations of the first two have already been treated:
\be\label{gaugetwo}\begin{split}
¶_Å f ¼= &â Å^M »_M f \,,\\[0.5ex]
¶_Å V^M ¼­ &â [þ,V]_{{}_D}^M ¼=¼ 
 \xi^{P}\partial_{P}V^{M}
  +(\partial^{M}\xi_{P}\,    -\partial_{P}\xi^{M})  V^{P}
  -\,   \p^M »_K Å^L  \, »_L V^K \,. 
 \end{split}\ee
For the tensor we have $\delta_\xi T =  [\,  \Ç þ \,,  T \,   ] $ which means that
\be
\label{tgT}
 ü(\delta_\xi T^{MN}) Z_M Z_N - ü\bigl( (\delta_\xi \hat T^M) Z_M\bigr)' \ = \ 
 \Bigl[\,  \Ç þ \,,\,  ü  T^{MN} Z_M Z_N - ü (\hat T^M Z_M\bigr)'  \,   \Bigr] \,.
\ee
The computation of the first contribution on the right-hand side gives  
\be
[\,  \Ç þ \,,  TZZ \,   ] ¼=\  (¶_Å T^{MN}) Z_M Z_N \ - \ [(ë_\xi öT^M)Z_M]'   \,,
\ee
where the gauge transformation of the two index tensor gets determined to be
\be\label{gauge56}
\boxeq{
\begin{aligned}
¶_Å T^{MN} ¼= &â  Å^P »_P T^{MN} + ( »^{M}Å_{P} - »_{P}Å^{M})T^{PN}  
 + ( »^{N}Å_{P} - »_{P}Å^{N})T^{MP} \phantom{\Bigl(}
\\[0.8ex]
	&  - ü \,\bigl[ \, »^N T_Q{}^P »_P »^{[Q} Å^{M]}  + 2 \, »_Q T^{KM} »^{N}
	\hskip-1pt »_K Å^Q  
	+ (MªN) \bigr]  
   \\[0.8ex]
	&  -\,    \f14 \,  »_K »^{(M} \hskip-1pt T^{PQ}\,  »^{N)} 
	\hskip-1pt »_P »_Q Å^K \,, \phantom{\Bigl(}
\end{aligned} }\ee
and the extra piece, showing the necessity of the pseudovector part, is found to be
\be
ë_Å öT^M  \ =  \  - T_{PQ}»^P »^{[Q} Å^{M]}  \  - \ ü  »_K T^{PQ}»^M \hskip-1pt 
»_P »_Q Å^K \,.
\ee
The second contribution on the right-hand side of (\ref{tgT}) gives 
\be
[\, \Ç þ  \,,  (öT Z)'  ] ¼=\  [ \Ç þ  \,,  öT Z ]'  \ = \ \bigl( \, [þ,öT]_{{}_D}^M Z_M  \bigr)' \,,
\ee
where $[þ,öT\, ]_{{}_D}$ is the transformation the pseudovector $öT$ would have if it were a true vector.  All in all we have
\be
\delta_\xi \hat T^M \ = \ [þ,öT]_{{}_D}^M +ë_Å öT^M \,,
\ee
and therefore 
\be\label{gauge87}
¶_Å öT^M ¼= \ [ \, þ\, ,öT\,]^M_{{}_D}\, -\,  T_{PQ}»^P »^{[Q} Å^{M]}  \  - \ ü  »_K T^{PQ}\, 
»^M \hskip-1pt»_P »_Q Å^K  \,.
\ee
This completes our determination of the gauge transformation of the tensor $T$.
Note that $:VW:$ is a particular case of tensor $T$.  
Also, $¶_Å T^{MN}$ depends only on $T^{MN}$ while
$¶_Å öT^M$ depends both on $\hat T^M$ and $T^{MN}$.
This means that $T^{MN}$ and $(T^{MN},öT^M)$ are both representations, but $öT^M$ by itself is not.  $T$ is ``not fully reducible", as e.g., the adjoint representation of the Poincar«e group.

\sectiono{Dilaton and double volume}\label{dildoubvol}   
 
In this section we introduce and study the Virasoro tensor operator $\S$ 
that involves the dilaton field.
Virasoro operators are tensor operators that generate conformal symmetries.
This kind of symmetry transformations takes the form discussed earlier in (\ref{general_symmetries}) and the following paragraphs.  Thus associated with
a tensor operator $T$ we have the operator
\be ñ (1) = Â(1)T(1) = Â(z_1)ü[T^{MN}(X)Z_M Z_N - (öT^M Z_M)' ](z_1) \,, \ee
obtained by multiplying the tensor by a world sheet parameter $\lambda (z)$.
The corresponding symmetry transformation $\delta_\lambda B$ of any
operator $B$ is defined by the commutator
\be ¶_Â B = [\Çñ , B ] \,.\ee
The closure of this symmetry  algebra, with one or more tensors involved, is quite nontrivial and requires conditions
that can be interpreted as field equations for the components of the tensor operators.
This will be the subject of the next section, where we introduce a second Virasoro
operator $\T$  that encodes the gravitational variables of the theory.

\subsection{Virasoro operator \texorpdfstring{$\S$}{S}} 

As mentioned earlier, the worldsheet Hamiltonian is given by $ÇüZ^2$.  The 
 two-dimensional energy-momentum tensor $üZ^2$ can have a total-derivative ``improvement term".  Such a term is implied by the coupling of the dilaton to the worldsheet curvature and is proportional to 
 $»_à^2 Ä$ \cite{Fradkin:1984pq,Fradkin:1985ys,Banks:1986fu}. We therefore take
 the tensor operator $\S$  to be given by 
\be
\label{deftsigma}
 \S \ \equiv \  ü (Z^2 - \phi'')âÜâT^{MN} = ú^{MN},¼öT^M = »^M Ä \,,
\ee
where we indicated the tensor components to the right.

The gauge transformations calculated in the previous section allow us
to determine the dilaton gauge transformation. We see from (\ref{gauge56}) that our choice
$T^{MN}=ú^{MN}$ is consistent as the right-hand side of that equation vanishes
for such $T^{MN}$.  Equation (\ref{gauge87}) then gives us
\be
\begin{split}
\delta_\xi (\p^M \phi)\   = \  \p^M (\delta_\xi \phi) \ =  & \ 
\ [ \,\Xi, \partial \phi\,]^M_{{}_D}  + \eta_{PQ}  \p^P \p^M\xi^Q \\[0.8ex]
= \ & \  \xi^K\p_K \p^M \phi  +  \p^M \xi^K\p_K \phi  + \p^M (\p \cdot\xi)  \\[0.8ex]
= \ & \  \p^M ( \xi^K\p_K\phi  + \p \cdot\xi) \,, \end{split}
\ee
where we used (\ref{vdorf}).  We then conclude that 
\be
\label{dil-quantum}  ¶_\xiÄ\  = \  ÅÉ» Ä + »ÉÅ \,, \ee
unmodified from the classical result.
This means that $e^\phi$ transforms as the ``volume element" (measure) for spacetime integration.  
For example, as in conventional gravity, it can be used to define the divergence of a vector without using a metric.  We'll see next that this also allows $T$ to be reduced by fixing $öT$ in terms of $T^{MN}$.

(We have dropped a constant that can be added
to the right-hand side of (\ref{dil-quantum}), given that only the derivative of the gauge transformation
is determined.  The associated transformation $\delta \phi =  \gamma $, with constant
$\gamma$ and independent of $\Xi$,  leaves field equations invariant but scales 
the action through the
$e^Ä$ factor in the measure,  showing that dilaton shifts change the coupling constant.)

Without the dilaton improvement term the following OPE holds for the operator $\f12 Z^2$:
\be üZ^2 (1) \, üZ^2 (2)\  = \  {D\over z_{12}^4}  + { Z^2 (2) \over z_{12}^2} + \, {(ü Z^2)' (2) \over z_{12}} + \hbox{finite}\,. \ee
Using this result, some additional calculation
(with repeated use of the strong constraint) gives the remarkable fact
that the OPE of the improved operators $\S$ is exactly the same:
\be \S (1) \S (2) =  {D\over z_{12}^4}  + { 2\S (2)\over z_{12}^2}  + \, {\S' (2) \over z_{12}} + \hbox{finite}\,. \ee

Next we consider products $\S(1)\O(2)$, expanded about $z_2$, for arbitrary operators $\O$.  We first note that the least singular terms, $1/z_{12}$ and $1/z_{12}^2$, are completely classical:  They are determined from terms with a single propagator 
contracting with $üZ^2$.  
If we used two propagators contracting with $üZ^2$
this leaves no $z_1$ dependence 
except in the $z_{12}$'s, so nothing to expand about $z_2$.  
But then the only term less singular than $1/z_{12}^3$ is 
killed by the strong constraint.  
For the $Ä''$ term, we contract $Ä$, Taylor expand about $z_2$, and take the $»_1^2$ from $»_1^2 Ä(1)$ to act last.  This gives terms of the form
\be\begin{split} 
& »_1^2 \left[ {1\over z_{12}^n} (»_M ... »_N Ä) (1) \O^{M...N} (2) \right] \\
 =â& »_1^2 \left[ {1\over z_{12}^n} (»_M ... »_N Ä) 
\O^{M...N} (2) + {1\over z_{12}^{n-1}} (»_M ... »_N Ä)' (2) \O^{M...N} (2) + ... \right]  \,. 
\end{split}\ee
But $»_1^2$ on any negative power of $z$ will yield terms at least as singular as $1/z^3$.  This is true for any number of propagators:  $Ä$ has no classical contribution to 
the $1/z_{12}$ and $1/z_{12}^2$
terms either.  We then have
\be
\f12 Z^2 (1) \O (2) \ = \ \ldots  +  {w_\O \,\O(2) \over (z_{12})^2} 
 +  {\O'(2) \over z_{12}} + \hbox{finite} \,, 
\ee
which imply
\be \label{straightforward} \boxeq{
\phantom{\Bigl(} \S \circ_{w_\O+1} \O = \O'  , ââ \S \circ_{w_\O} \O = w_\O \O \,. } \ee 
For the two  next least divergent terms we make the definitions of the
quantum generalizations of the trace and divergence:
\be \boxeq{\phantom{\Bigl(}  \hbox{\rm div}(\O)\, ­ \,\S \circ_{w_\O-1} \O  , ââ 
\f12\, \hbox{\rm tr}(\O) \, 
­ \, \S \circ_{w_\O-2} \O \,. \ } \ee  
The divergence lowers the weight by one, the trace lowers the weight by two.  
We can apply the derivative identities to the above 
 general expressions for $\S\circ\, $.  For the latter cases we find
\be
\boxeq{  
\begin{aligned}  \phantom{\Bigl(}
\hbox{\rm tr}\, (\O') & = (\hbox{\rm tr}\,\O)' +6Ê\hbox{\rm div}\,\O\, , \\
\ \ \hbox{\rm div}\,(\O') & = (\hbox{\rm div}\,\O)' +2w_\O \O \,, \ \ \ 
\end{aligned}
}
\ee
while the $\circ_{w_\O+1}$ identity is trivial and the $\circ_{w_\O}$ identity shows that the expression for $\circ_{w_\O+1}$ is implied by that for $\circ_{w_\O}$.

For the tensor $T$ 
the trace gives a scalar. For the vector $V$ the divergence gives a scalar and the
trace gives zero. For the scalar $f$ both the trace and the divergence give zero.  
Thus
\be \boxeq{ \phantom{\Bigl(} \hbox{\rm tr}\, V = \hbox{\rm tr}\, f = \hbox{\rm div}\, f = 0\,. \ } \ee
The derivative identities then specialize: 
\be
\boxeq{
\begin{aligned}
\hbox{\rm tr}(V') ¼= & \  \  6 \,\hbox{\rm div} V  \\[0.5ex]
\hbox{\rm div}(V') ¼= & \ \  2V + (\hbox{\rm div} V)'     \\[0.3ex]
\hbox{\rm div}(f') ¼= & \ \  0 \,. 
\end{aligned}
}
\ee

We can write these products collectively as the OPE
\be 
\S(1) \O(2) ¼=¼ \, \hbox{finite} \,+\, {1\over z_{12}} \O' + {1\over z_{12}^2} w_\O \O + {1\over z_{12}^3} \hbox{div}(\O) 
	+ {1\over z_{12}^4} ü \hbox{tr}(\O) + ...\,, 
\ee
so that conformal transformations take the form   
\be \bigl[ \Ç Â\S , \O \bigr] ¼=¼ Â\O' + w_\O Â'\O +üÂ'' 
\hbox{div}(\O) + \f1{12}Â''' \hbox{tr}(\O) + ... \, , \ee
where the first two terms are the usual (free, ``on-shell") universal terms.

Straightforward calculation gives the covariants
\be
\label{tr-div}
\begin{split}
\hbox{tr}ÊT ¼=â & ú^{MN}T_{MN} -3 ( T^{MN}»_M »_N Ä + »ÉöT + öTÉ»Ä ) \,,\\[0.5ex]
(\hbox{div}ÊT)^M ¼=â &  »_N T^{MN} + T^{MN} »_N Ä -ü T^{NP}»_N »_P »^M Ä - öT^M  - ü »^M (» É öT + öTÉ»Ä ) \,, \\[0.3ex]
\hbox{div}ÊV ¼=â & »ÉV+VÉ»Ä \,, 
\end{split}\ee
as well as the trivial cases
\be \boxeq{ \phantom{\Bigl(}  \ \hbox{\rm tr}\, (\S) = 2D,ââ\hbox{\rm div}\, (\S) = 0 \,. \ \  } \ee
This explicit expression for the ``divergence" of a vector also identifies $e^Ä$ as the integration measure, taking the place of ``$å{-g}Ê$":
\be \hbox{div}\, V = e^{-Ä}»É(e^Ä V) \,.\ee
The case of a trivial tensor is of some interest.  Such tensor is the $z$ derivative of a
vector operator $V$: 
\be
T = V'  = (Z_M V^M)'  âÜâ  \hat T^M  =  -2V^M  \,, \quad  T_{MN} = 0 \,.  
\ee

\subsection{Projection to divergence free tensors} \label{prodivten}

Before introduction of the dilaton, we found that operators 
of $w>1$ were not fully reducible.  We'll see now a further 
reduction, the separation of the ``divergence" and ``divergenceless" pieces.  
In the case of weight-two tensors $T$, 
this allows us to treat $T^{MN}$ and $öT^M$ separately.

We therefore look for a solution to the constraint $\hbox{div}\, Ñ\O=0$ by projecting 
out the $\hbox{div}$ piece of $\O$.  The solution is not unique; we look for a solution by taking $z$-derivatives of iterated divergences
\be Ñ\O = \Bigl( Ý_{n=0}^{w_\O} c_n \A^n\Bigr) \O ,ââ\A^n \O 
­ (\hbox{div}^n \O)^{(n)} \,. \ee
Here, for example  $\A^2 \O =  (\hbox{div}\, \hbox{div}Ê\O )''$
and $\A^0  \O = \O$.   
Note that the sum can be taken to $¥$ since $\hbox{div}$ vanishes on a scalar.  Using the 
$\hbox{div}\, (\O')$ identity, we find by induction
\be \hbox{div}\A^n = \A^n \hbox{div} +2\,
\bigl(nw_\O - \f{n(n+1)}2\bigr)\A^{n-1}\hbox{div} \,. \ee
This allows the constraint to be solved as (using recursion or differential equation)
\be Ñ\O = Ñ{Ñ\O} = g(\A)\O ,ââg(x) = 
Ý_{n=0}^{w_\O} {[2(w_\O-1) -n]!\over n![2(w_\O-1)]!}(-x)^n\,.  \ee
These polynomials are essentially the Neumann polynomials, or the leading terms in the modified Bessel functions of the second kind:
\be \begin{split}
{4\over (a+1)!}x^{a/2+1}O_{a+1}(2åx) &¼=¼ Ý_{n=0}^{[(a+1)/2]}{(a-n)!\over n!a!}x^n \\
{2\over a!}x^{(a+1)/2}K_{a+1}(2åx) &¼=¼ Ý_{n=0}^a{(a-n)!\over n!a!}x^n +...
\end{split}\ee 
Similarly,
\be \A^n »_z \ = \  »_z \A^n +2\,\Bigl(nw_{\O'} - \f{n(n-1)}2 \Bigr) »_z \A^{n-1} \,,\ee
implies
\be Ñ{\O'} = 0 \,.\ee
The result is that for arbitrary-weight operators (except vectors, $w_\O=1$) we can initially ignore all total $z$-derivative terms, as they will be fixed in terms of the rest by the $¼Ñ{\phantom M}¼$ operation.  

In particular, this applies to all products $\O_1\circ_w\O_2$ for $w>1$.  This means we can replace all ``$\bullet$" products defined previously by a new product with nicer properties:  From the symmetry condition on $\circ$, we see that $¼Ñ{\phantom M}¼$ automatically (anti)symmetrizes it,
\be   
\label{define-bullet} 
\boxeq{\phantom{\Biggl(} \ 
\O_1\bullet_w\O_2 \ ­ \ Ñ{\O_1 \circ_w \O_2\phantom{\bigl(}} = (-1)^{w_1+w_2-w} \O_2 \bullet_w \O_1 
\, . \ \ } \ee

We then define the $\hbox{Div}$ operation through
 the following relation 
\be Ñ\O \ ­ \ \O -{1\over 2(w_\O-1)}  (\hbox{Div}\, (\O))'  \,, \ee  %bz   
which implies that
\be
\hbox{Div}\, (\O') = 2w_\O\O ,â\hbox{Div}\, (Ñ\O) = 0 \,,\ee
as well as 
\be \hbox{Div}\, \O = h(\A)\hbox{div}\, \O ,ââ
h(x) = Ý_{n=0}^{2w_\O-3} {[(2w_\O-3)-n]!\over (n+1)!(2w_\O-3)!} (-x)^n \,.\ee
Note that $h(\A)$ is an invertible  
finite polynomial.  This means that $\hbox{div}$ determines $\hbox{Div}$, and vice versa.  In particular,
\be
\label{divDiv} \hbox{div}\, \O = 0âÛâ\hbox{Div}Ê\O = 0 \ee
so the two constraints are freely interchangeable.  The advantage of $\hbox{Div}$ 
over $\hbox{div}$ is that on $\O'$, $\hbox{div}$ gives $2w_\O\O+(\hbox{div}Ê\O)'$, while $\hbox{Div}$ gives simply the first term.  This allows $\hbox{Div}Ê\O=0$ to be more easily solved than $\hbox{div}Ê\O=0$, although the solution is the same.

In particular,
\be
\label{tbart} ÑT ¼=¼ T -ü(\hbox{Div}ÊT)' ,â \hbox{Div}ÊT ¼=¼ \hbox{div}ÊT -ü(\hbox{div}^2 T)' \,,
\ee
so that  
\be
\label{Tprojection}
\boxeq{\phantom{\Biggl(} 
 ÑT ¼=¼ T -ü(\hbox{\rm div}ÊT)' + \f14 (\hbox{\rm div} ^2\, T)'' \, . ~~  } 
 \ee
Thus, since the terms subtracted affect only the pseudovector part 
\be âÑT^{MN} =¼ T^{MN} \,.\ee
Using the explicit expressions for the divergence of a tensor and of a vector we find
\be\begin{split}
(\hbox{Div}\, T)^M  ¼=¼& -öT^M +(»_N T^{MN} +T^{MN}»_N Ä) -üT^{NP}»_N »_P »^M Ä \\
	& -ü»^MÓ»_N »_P T^{NP} +T^{NP}[»_N »_P Ä +(»_N Ä)(»_P Ä)]Õ \,. 
\end{split}\ee
The expression  
 for $ß{ÑT}$ is equal to the value of $\hat T$ for which $\hbox{Div}\, T=0$.  From the
 above equation we get 
\be
\label{gintroduced}
\hbox{Div}\,T = 0âÜâöT^M \, =\,   G^M (T^{MN}, \phi ) \,,\ee
where we have introduced the vector function    
\be
\label{G-formula}\begin{split}
	G^M(T^{MN}, \phi)  ¼\equiv \ 
	 & \  (»_N T^{MN} +T^{MN}»_N Ä) -üT^{NP}»_N »_P »^M Ä \\[0.5ex]
	&  -ü»^MÓ»_N »_P T^{NP} +T^{NP}[»_N »_P Ä +(»_N Ä)(»_P Ä)]Õ\,. 
\end{split}\ee
The tensor $\overline T$ is thus given by   
\be
\label{overlineT}
\boxeq{ \phantom{\Biggl(} 
\overline T  \ = \ \f12 T^{MN} Z_M Z_N  \, - \, \f12  [ 
G^M(T^{MN}\hskip-1pt , \phi) Z_M]' \,. \  } 
\ee
Note that the pseudovector part $\hat T^M$ of $T$ has dropped out of
$\overline T$, while appearing in $\hbox{Div}\,T$ in the simplest nontrivial way.  The divergenceless tensor    $\overline T$ has a pseudovector part but it is determined in terms of $T^{MN}$ and the dilaton through the function $G$.   

Another useful evaluated expression is
\be
\label{troverlineT}
\begin{split}
\hbox{tr}(ÑT) &¼=\ \hbox{tr}ÊT -3Ê\hbox{div}^2 T \\[0.5ex]
& ¼=\ ú^{MN}T_{MN} -3\,  [ \,T^{MN} \p_M\phi \p_N\phi +»_M (»_N T^{MN} + 2T^{MN} »_N Ä)  ] \,.
\end{split}\ee
We also have the trivial case
\be \hbox{Div}\,\S = 0âÜâÑ\S = \S \,. \ee

\sectiono{Double metric}\label{doubmetsec}

Having studied the properties of the tensor operator $\S$ encoding the 
dilaton background, we now introduce the second Virasoro (tensor) operator $\T$
that encodes the gravitational background.  We take, in full generality 
\be\begin{split}
T^{MN} = \M^{MN},âöT^M = ß{\M}^M âÜâ& \T =\,  ü[\M^{MN}Z_M Z_N - (ß{\M}^M Z_M)']\,.
\end{split}\ee
In here the field $\M^{MN}$ will be called the double metric.  Nothing is assumed about it to begin.
The field $ß{\M}^M$ is an additional degree of freedom that will eventually get determined
in terms of the double metric and the dilaton.

\subsection{Field equations}

The field equations for $\M^{MN}, ß{\M}^M,$ and the dilaton appear as enforcement of the Virasoro algebra for the operators $\S$ and $\T$. 
  Since only singular terms contribute to commutators, we look 
  at the table of products $T\circ_w T$ only for $w²3$.  The Virasoro algebra requires: 
\be
\bordermatrix{ \hfill w¼= ¼& 0 & 1 & 2 & 3 \cr
	\hfill\S\circ_w\S¼= ¼& D & 0 & 2\S & \S' \cr   
	\hfill\S\circ_w\T¼= ¼& 0 & 0 & 2\T & \T' \cr
	\hfill\T\circ_w\T¼= ¼& D & 0 & 2\S & \S' \cr}
\ee
(Note that $\circ_2$ is a quantum generalization of the anticommutator when applied to tensors $T$: e.g., $\S\circ_2T=2T$.)
Ghost contributions, which we don't discuss in this paper, would cancel 
the $Ò\S|\SÔ$ and $Ò\T|\TÔ$ terms.   
The ghosts are not  
necessary for the classical
field theory we are building, presumably because they
do not couple to the background.  
The $\S\S$ equations are satisfied off shell, as discussed earlier.
The $\S \T$ equations for $\circ_3$ and $\circ_2$ also hold off-shell 
(see (\ref{straightforward})).  The ones for $\circ_1$ and
$\circ_0$ are, respectively, 
\be
\label{dieqn} \begin{split}
\hbox{div} (\T) = 0 : &âß\M^M \\
\hbox{tr}\,  (\T) = 0 : &âÄ\,. 
\end{split}
\ee
The first equation fixes $ß\M^M = G^M (\M^{MN}\hskip-1pt , \phi)$, as
defined in (\ref{gintroduced}). 
The second equation can be viewed as the dilaton field equation.   
Let us now consider the $\T \T$ equations.  
By the symmetry identity, 
 $\T\circ_3\T$ and $\T\circ_1\T$  
are derivatives of $\T\circ_2\T$ and $\T\circ_0\T$, 
so only the latter are relevant.  The $\circ_2$ condition $\T \circ_2 \T = 2 \S$ gives
two equations and the $\circ_0$ condition just one: 
\be\begin{split}
(\T\circ_2\T)^{MN} = 2ú^{MN} : &â\M^{MN} \,, \\[0.5ex]
(\T\circ_2\T)^M = 2»^M Ä : &â \hbox{redundant,} \\
Ò\T|\TÔ = D : &â \hbox{redundant.} 
\end{split}\ee
The first is a nontrivial equation for the field $\M^{MN}$; the last
two are redundant to the first and
those in (\ref{dieqn}), as we now show.

We first reorganize  a bit the equations above.  Since $\hbox{div} \,\T = 0$
we have $\overline \T = \T$.  We can then let  $\T \to \overline \T$ 
 everywhere thus taking care of the first equation in (\ref{dieqn}).  Note also that the vanishing of 
any tensor $T$ is equivalent to the vanishing of $\overline T$ and the vanishing of
$\hbox{div} \, T$ (or alternatively, the vanishing of $\overline T$ and $\hbox{Div}\, T$).  
We do this with the $\T \circ_2 \T = 2\S$ equation, recalling that
$\overline \S = \S$.  We then have
\be
\label{collect-eqns}\begin{split}
\hbox{tr}\,  (Ñ\T) = 0 : &âÄ\,, \\
Ñ{Ñ\T\circ_2Ñ\T} = 2\S : &â\M^{MN} \,,\\[0.5ex]
\hbox{div}\, (Ñ\T\circ_2Ñ\T) = 0 : &â \hbox{redundant,} \\
Ò\overline\T|\overline \TÔ = D : &â \hbox{redundant.}
\end{split}\ee

Consider again the distributivity identities, now for $\O_1=\S,\O_2=T_1,\O_3=T_2$, so $w_i=2$, and also $w=2$, but $öw=0$ or 1.  Then
\be \S\circ_{öw}(T_1\circ_2 T_2) -T_1\circ_{öw}(\S\circ_{öw} T_2) 
= Ý_{w''=1}^{4-öw}{3-öw\choose w''-1}(\S\circ_{4-w''}T_1)\circ_{öw} T_2 \,. \ee
Using the $\S\circ$ and derivative identities,
\be
\label{distr01101}
\hskip.5in{}\boxeq{
\begin{aligned}
  \hbox{\rm tr}  ( T_1  \circ_2 T_2 )  \ = \ & \ 
  \, T_1 \circ_{0} ( \hbox{\rm tr} \, T_2 )  \ +   ( \hbox{\rm tr} \,T_1 ) \circ_{0} T_2  
    \ +6 \, ( \hbox{\rm div}\, T_1 ) \circ_{0} T_2 
    \ +4 \, \langle T_1  |  T_2\rangle \,,
       ~~   \\[1.0ex]  
 \hbox{\rm div}  ( T_1  \circ_2  T_2 )  \ = \ & \ 
  T_1 \circ_{1} (\hbox{\rm div}\,  T_2 )  
    \ + ( \hbox{\rm div}\,  T_1 ) \circ_{1} T_2 \ + \ T_1  \circ_{1} T_2 \,.
   \ \ \end{aligned} }
\ee

Noting that $T \circ_1 T = üÒT|TÔ'$ (symmetry identity) and setting $T_1 = T_2 = ÑT$, we get
\be
\label{traverss}\begin{split}
\hbox{tr}\, (ÑT\circ_2ÑT) &¼=¼ 2Ò\hbox{tr}\,ÑT|ÑTÔ +4ÒÑT|ÑTÔ\,,  \\
\hbox{div}\,(ÑT\circ_2ÑT) &¼=¼ üÒÑT|ÑTÔ' \,. 
\end{split}\ee
The first can be re-expressed using (\ref{troverlineT}) and the second, to find 
\be\begin{split}
\hbox{tr}\left(Ñ{ÑT\circ_2ÑT}\right) &¼=¼ 2Ò\hbox{tr}\, ÑT|ÑTÔ +4ÒÑT|ÑTÔ \,. 
\end{split}\ee
Substituting $\T$ for $T$ and applying the $\M$ and $Ä$ field equations, 
the last two equations become 
\be\begin{split}
\hbox{div}\, (Ñ\T\circ_2Ñ\T) &¼=¼ üÒÑ\T|Ñ\TÔ'\,, \\
4D &¼=¼ 4ÒÑ\T|Ñ\TÔ\,, 
\end{split}\ee
proving, as we wanted,  that  the last two equations in (\ref{collect-eqns}) are redundant.

\subsection{\texorpdfstring{$ä$}{Star} product}  \label{starproduct-sec}

In this subsection we consider a number of properties that will
allow us to write an action and vary it to determine its field equations.
A useful star-product will be introduced.  This product yields weight-two
divergence-free tensors.  It is also symmetric, and together with the inner
product defines a scalar that is totally symmetric in its three tensor inputs.

The action will take the form $\int e^\phi L$ where $L$ is a scalar. 
As noted earlier, for an arbitrary vector $V$ integration by parts 
shows that 
\be Çe^ÄÊ\hbox{div}\, V = 0 \,. \ee
It is convenient to introduce 
 the equivalence symbol $\sim$ for objects that are the same under the integral 
\be
A \sim B    \quad  \to \quad  \int e^\phi \, A  =   \int e^\phi  B  \,. 
\ee
We thus have
\be
\hbox{div}\, V \ \sim \  0 \,. 
\ee
Since $\hbox{tr} (V') \ = \ 6\,  \hbox{div} \, V $  we also have
\be
\hbox{tr} (V') \ \sim \ 0 \,, 
\ee 
which states that the trace of a trivial tensor gives no contribution to the action.
Since $\overline T =  T + V'$ for some $V$,  we also have that
\be
\label{trTr} 
\hbox{tr} (T ) \ \sim \  \hbox{tr} (\overline T ) \,. 
\ee

We now use the distributive identity (\ref{Tom}) 
 with $\O_1=\S, \O_2 = \O, \O_3 = T$, and $öw=0,w'=3$: 
\be \hbox{div}\, (\O\circ_1 T) 
= Ò\hbox{div}Ê\O|TÔ +Ò\O|\hbox{div}ÊTÔ +(w_\O-2)Ò\O|TÔ\,,  \ee
where we also used the identifications~(\ref{saction}). 
For the cases of the scalar, vector, and tensor, we get
\be\label{divcirc1}\begin{split}
\hbox{div}(f\circ_1 T) &¼=¼ Òf|\hbox{div}ÊTÔ -2Òf|TÔ\,, \\[0.3ex]
\hbox{div}(V\circ_1 T) &¼=¼ Ò\hbox{div}ÊV|TÔ +Ò\, V|\hbox{div}ÊTÔ -ÒV|TÔ\,,  \\[0.3ex]
\hbox{div}(T_1\circ_1 T_2) &¼=¼ Ò\hbox{div}ÊT_1|T_2Ô +ÒT_1|\hbox{div}ÊT_2Ô\,.
\end{split}\ee
Applied to divergenceless tensors $\overline T$ we have
\be
\label{thedivless}\begin{split}
\hbox{div}(f\circ_1 \overline T) &¼=¼  -2Òf|\overline TÔ\,, \\[0.3ex]
\hbox{div}(V\circ_1 \overline T) &¼=¼ Ò\hbox{div}ÊV|\overline TÔ  -ÒV|\overline TÔ \,,\\[0.3ex]
\hbox{div}(\overline T_1\circ_1 \overline T_2) &¼=¼ 0\,. 
\end{split}\ee
The first equation implies that
\be
\label{prop1}
ÒÑT|f Ô \,  \sim  \, 0  \,. 
\ee
The second equation, using the first, can be written as
\be ÒV|\overline TÔ  \ = \ - \hbox{div} \, [ V \circ_1 \overline T  
+ \f12 (\hbox{div}ÊV) \circ_1 \overline T ] \,,\ee
which implies that $ÒÑT|VÔ \ \sim \ 0$.  Thus, all in all,  
\be
\label{prop2} 
\boxeq{\phantom{\Bigl(} \ 
ÒÑT|f Ô \,  \sim  \, 0\,, \qquad  ÒÑT|VÔ \ \sim \ 0 \,. \ } 
\ee
Thus divergenceless tensors have the remarkable property that their inner product
against a scalar or a vector are zero under the integral. We now note that the overlap of a projected tensor $\overline T_1$
and an unprojected tensor $T_2$
picks up its projected part: 
\be
\label{orthogonal_projection}
~ÒÑT_1|T_2 Ô =  ÒÑT_1|\overline T_2  + V' Ô \sim  ÒÑT_1|\overline T_2Ô \,, 
\ee
where we used 
$ \langle \, Ñ T\, |   V' \rangle  =  \langle \,V'  |   Ñ T\, \rangle
= -3 \langle V  |  Ñ T\, \rangle \sim 0$. 
The overline projection is  an orthogonal projection.

We now show that there are two equivalent ways of forming a scalar in order to
use it in the action.
From (\ref{distr01101}) we have
\be \hbox{tr} \, (ÑT_1\circ_2 ÑT_2) = ÒÑT_{(1}|\, \hbox{tr} ÊÑT_{2)}Ô +4\, ÒÑT_1|ÑT_2Ô \,.  \ee
The first term on the right-hand side is equivalent to zero under the integral
on account of (\ref{prop1}) so that 
\be
\label{twosim}
  ÒÑT_1|ÑT_2Ô \ \sim \  \f14 \, \hbox{tr} \, (ÑT_1\circ_2 ÑT_2) \,.\ee

Recall now our definition of symmetric products $\bullet_w$  
in (\ref{define-bullet}).  The case $w=2$, for which the output (regardless of the inputs)
is a tensor, will be particularly useful.   We will call this product a ``star" product:  
$ä \ ­  \  \bullet_2 $.  We thus have:
\be
\label{definestar}    
\boxeq{ \phantom{\Bigl(} \O_1 \star \O_2  \ \, \equiv \,  \ \overline{\O_1 \phantom{\hat i}
\hskip-2pt \circ_2 \O_2} \,. \ \ }
\ee 
Using this notation and recalling (\ref{trTr}), we see that (\ref{twosim}) takes the form
\be 
\label{trs}
 ÒÑT_1|ÑT_2Ô \ \sim \  \f14 \, \hbox{tr} \, (ÑT_1\star ÑT_2) \,.\ee

To perform the variation of the action we need to show that under
the integral
$ÒT_1| T_2ä T_3Ô$  is totally symmetric when $\hbox{div}\, (T_i) = 0$.  
So we look at distributivity identities for three tensors ($w_i=2$) with an inner product outside ($öw=0$) and a $\circ_2$ inside.  All these identities have a term with $\circ_3$ also; the ones with only 
one such term, appearing with the same coefficient, (and no $\circ_4$) are those with $(w,w')=(4,2),(3,3),(2,4)$.  The former two are the simplest; taking their difference, we find
\be
\label{thesavior}
 ÒT_{2}|T_1\circ_2 T_{3}Ô - ÒT_{3}|T_1\circ_2 T_{2}Ô \, =    \, 
  ÒT_{2}|T_1\circ_1 T_{3}Ô -  ÒT_{3}|T_1\circ_1 T_{2}Ô - ÒT_1|T_2\circ_1 T_3Ô \, \,   + ÒT_1|ÒT_2|T_3ÔÔ \,.
\ee
Applying this identity to divergenceless tensors, all terms on the right-hand side
are equivalent to zero on account of (\ref{prop2}) and therefore 
\be
\label{scalar-withcirc2}
 \langle \, Ñ T_1\, | \,  Ñ T_2 \circ_{2} Ñ T_3\, \rangle  
  \ \sim  \  \langle \, Ñ T_3 \, | \, Ñ T_2 \circ_{2} Ñ T_1 \, \rangle \,.
\ee
Now note that 
\be
T_1 \star T_2 \ = \ Ñ{T_1 \circ_2 T_2}  \ = \ T_1 \circ_2 T_2 + V'   \,,
\ee
for some vector $V$.   Since   
$ \langle \, Ñ T\, | \,  V' \, \rangle  
\sim 0$,  replacing $\circ_2$ with $ä$ has no 
effect on the above symmetry:
\be
\boxeq{\phantom{\Biggl(} 
 \langle \, Ñ T_1\, | \,  Ñ T_2 \star Ñ T_3\, \rangle  
  \ \sim  \  \langle \, Ñ T_3 \, | \, Ñ T_2 \star Ñ T_1 \, \rangle \,.  ~~}
  \ee
Since the product $\star$ is symmetric, this shows that
$\langle \, Ñ T_1\, | \,  Ñ T_2 \star Ñ T_3\, \rangle$
is totally symmetric. Note that the form in (\ref{scalar-withcirc2}) is also totally  
symmetric because the product $\circ_2$ is symmetric up to
$z$-derivatives.

As a useful exercise we consider the explicit form of the star product of two
projected tensors:
\be
\overline T_1 \star  \overline  T_2 \ = \ \overline T_1 \circ_2 \overline T_2 
- \f12  (\hbox{div} \,(\overline T_1 \circ_2 \overline T_2))'
+ \f14  (\hbox{div}\,\hbox{div}\, (\overline T_1 \circ_2 \overline T_2))'' \,.
\ee
We then note that $\hbox{div}(\overline T_1 \circ_2 \overline T_2) 
=  \overline T_1 \circ_1 \overline T_2$, because of  (\ref{distr01101}), 
and that the last term above drops out by the last of (\ref{thedivless}).  As a result, 
we have the simplified form
\be
\overline T_1 \star  \overline  T_2 \ = \ \overline T_1 \circ_2 \overline T_2 
- \f12 (\overline T_1 \circ_1 \overline T_2)' \,.
\ee
When the two tensors are the same, further simplification is possible using the symmetry property, 
\be
\label{ttstar}
\overline T \star  \overline  T \ = \ \overline T \circ_2 \overline T 
- \f14\,  \langle\overline T | \overline T\rangle '' \,.
\ee

\subsection{Action}

We use the double metric defined previously,
\be
\T =\,  ü[\M^{MN}Z_M Z_N - (ß{\M}^M Z_M)']\,,
\ee
with the condition that
\be
\hbox{div} \, \T  \ = \ 0  \quad Ü \quad  \overline\T \ = \ \T \,,
\ee
so that the identities for barred tensors can be used for $\T$. 
The tensor field $\T$ so constrained is only a function of $\M^{MN}$ and the dilaton. 
This means that $ß{\M}^M = G^M (\M^{MN}\hskip-1pt , \phi) $ is determined in terms of $\M^{MN}$ and the dilaton.

We now claim that the action is given by 
\be 
\label{theaction}
\boxeq{ S = Çe^Ä L ,ââL \ =\  Ò\T|\S  -\f16\Tä\TÔ \; . }\ee
Using (\ref{trs}) we also have the alternative form, equivalent
up to total derivatives 
\be
 L \ =\  ü\, \hbox{tr} \, [ \, \T -\, \f1{12} \, \T\star (\Tä\T) \, ] \,.
\ee
This action is gauge invariant because the dilaton 
provides a measure and $L$ is a gauge scalar. This is clear by construction since 
we begin with tensors under gauge transformations and all our operations are covariant: the products,  projections, inner products.  The gauge transformations
are simply
\be
\begin{split}
\delta_\xi \T \ = \ & \  \Xi \circ_2  \T \,, \\
\delta_\xi  \S \, \ = \ & \ \Xi \circ_2 \S \,.
\end{split}
\ee
The gauge transformed $\T$ is divergence free with respect to the divergence
operator 
that uses the gauge transformed dilaton.   The explicit form
of the gauge transformations
can be read from (\ref{gauge56}) and (\ref{dil-quantum}), and for completeness we give them here:
\be
\label{gtacfin}
\begin{split}
¶_Å \M^{MN} ¼= &â Å^P »_P \M^{MN} + ( »^{M}Å_{P} - »_{P}Å^{M})\M^{PN}  
 + ( »^{N}Å_{P} - »_{P}Å^{N})\M^{MP}  
\\[0.8ex]
	& - ü\, \bigl[ \, »^M \hskip-2pt\M^{PQ}\, »_P (»_Q Å^N -»^N Å_Q )  + 2\,»_Q \M^{KM}\, »^{N} »_K Å^Q
	+ (MªN)  \bigr] 
	\ \   \\[0.8ex]
	&  -\,    \f14 \,  »_K »^{(M} \hskip-2pt \M^{PQ}\,  »^{N)} 
	\hskip-1pt »_P »_Q Å^K \,, \\[0.8ex]
¶_Å \phi \  ¼= &â	 \xi\cdot \p\phi  + \p\cdot \xi \,. 
\end{split}
\ee

We now vary the action
to derive the equations of motion.  Consider first the variation $¶_\M$   
of double metric $\M_{MN}$.  
The only field to vary is $\T$ and  $\delta_\M \T$ is still projected. 
The result is
\be ¶_\M S = Çe^Ä Ò¶_\M\T|\S -ü\Tä\TÔ \,,\ee
using the total symmetry of $Çe^ÄÒÑT_1|ÑT_2äÑT_3Ô$.
Now note that $Ò\delta T_1|T_2Ô$ gives $AT_2=0$ for some operator $A$ of the form
\be A = I +Œ'A_1 +Œ'^2A_2 \,, \ee 
where $A_1$ is second-order in spacetime derivatives and $A_2$ is fourth-order.  There are no higher-order terms because of the strong constraint:  Derivatives can be contracted only with the indices on $T$.  (This includes derivatives acting on $Ä$.)  For the same reason, $A$ is easily invertible, also terminating at fourth-order.  Thus $AT_2=0$ implies $T_2=0$.  
In our case we have $\langle {\delta_\M  \T}| T_2\rangle$, and since the bra is
projected, the pseudovector part of $T_2$ drops out  giving us the equation
$(T_2)_{MN} = 0$. 
 If the tensor $T_2$ on the ket is itself divergenceless, 
 then it also follows that $\hat T_2 =0$,
and thus simply that $T_2 =0$.  This is the case for us, so the field equations following from the $\delta_\M $ variation is
\be   
\Tä\T = 2\S \,. \ee
This implies our OPE field equation $(\T\circ_2 \T-2\S)^{MN}=0$, since all 
products are the same for the two-index part of the tensor.

Now we consider the $Ä$ variation.  This is more subtle because the projection uses $\S$, which contains $Ä$, so the projection itself is varied.
Consider an arbitrary tensor $T$ and its constrained projection $\overline{T}$.
Using (\ref{Tprojection}) we have that the variation of the projection is 
not projected
\be
\label{Tprojection-var}
\delta  ÑT ¼=¼ \overline{\delta T}
 -ü(\delta_\phi\hbox{div}ÊT)' + \f14 (\delta_\phi\hbox{div} (\,\hbox{div} \, T) )''
  + \f14 (\hbox{div} (\delta_\phi\hbox{div} \, T) )''  \, .
\ee
While we could proceed without calculating the doubly primed terms (which will
drop out) it is of interest to obtain a general formula for the variation of the
projection.  Taking dilaton variations of the products
 $\S\circ T$ and $\S\circ V$
one quickly shows that  
\be
\label{deltaphitraces}
\delta_\phi \,  \hbox{tr} \,T  \ = \ -6 \langle \delta \phi | T \rangle \,, \qquad
\delta_\phi  \, \hbox{div}\,  T \ = \  - \delta\phi \circ_1 T  \,, \qquad
 \delta_\phi\, \hbox{div}\, V \ = \ 
\langle V | \delta \phi \rangle \,,
\ee
giving us
 \be
\label{Tprojection-var-pho}
\delta  ÑT ¼=¼ \overline{\delta T}
 +ü(\delta \phi \circ_1 T)' + \f14 \bigl( \,\langle \hbox{div}\,  T | \delta \phi \rangle
 - \hbox{div}(\delta \phi \circ_1 T)  \bigr)'' \, .
 \ee
Since $\hat T_M$ drops out of $\overline T$, and thus from the full variation $\delta \overline T$, it is
possible to rewrite the above right-hand side solely in terms of $\overline T$. 
For this we note that
\be
\delta \phi \circ_1 T \ = \ \delta \phi \circ_1 \overline T
\, + \, \f12 \delta \phi \circ_1 (\hbox{div} \, T)'  \, - \, \f14 
\delta \phi \circ_1 (\hbox{div}\, \hbox{div} \, T)''\,. 
\ee
The last term vanishes by repeated use of the symmetry and 
derivative identities ($f \circ_1 V' = - \langle V | f\rangle ' $ which then implies
$f_1 \circ_1 f'' = 0$), and we get
\be
\delta \phi \circ_1 T \ = \ \delta \phi \circ_1 \overline T
\, - \, \f12 \langle \hbox{div} \, T | \delta \phi\rangle '  \,. 
\ee
Using this identity twice in (\ref{Tprojection-var-pho}) as well as (\ref{divcirc1})
 we finally find the desired variation formula:
\be ¶ÑT \ = \  Ñ{¶T} \, + \,  ü(¶Ä\circ_1ÑT)' \, +\, üÒ¶Ä|ÑTÔ'' \,.  \ee
Applied to our divergence free $\T$ it reads
\be ¶\T  \ = \  Ñ{¶\T} \, + \,  ü(¶Ä\circ_1\T)' \, +\, üÒ¶Ä|\T Ô'' \,,   \ee
The first term is the $¶_\M$ we have already evaluated; we now consider only the latter two terms that comprise the $\delta_\phi\T$ variation   
\be
\delta_\phi \T \ = \  ü(¶Ä\circ_1\T)' \, +\, üÒ¶Ä|\T Ô'' \,.
\ee 

The action is of the form
\be Çe^ÄÒÑT_1|ÑT_2Ô ,ââ ÑT_1 =\T ,ââ ÑT_2 = \S -\f16 \Tä\T  \,. \ee
From the above variation we see, since $ÒV'|ÑTÔ=ÒÑT|V'Ô¾0$ in the integral, that the variation of the projection operator on $T_1$ or $T_2$ will vanish, as the integral of a total derivative.  (In particular, $¶\S=-ü(¶Ä)''$.)  However, $\Tä\T$ is itself the projection of $\T\circ_2\T$:  So we can ignore the projection to get $ä$ from $\circ_2$, but not the projections implicit in the $\T$'s inside the product:  In fact, these are the only $¶_Ä$'s that contribute to the action, other than that of the measure $e^Ä$.  
We thus find   
\be \delta_\phi S \ = \ Çe^Ä[ (¶Ä)L  -\f13 Ò\T|¶_\phi\Tä\TÔ  ] \,, \ee
where we have noted that the variations $\delta_\phi \T$ on bras 
are total derivatives,
and applied the symmetry of the $\star$ product.  
The result is then
\be
\label{dilvaract} ¶_Ä S = Çe^Ä[ (¶Ä)L + \f16Ò\T|(¶Ä\circ_1\T)ä\TÔ] \, , 
\ee
where the double derivative term does not contribute because of the derivative identity $f''\circ_2 T = 0$.

From the distributive identity with $\O_1 = V, \O_2 = \overline T, \O_3 = \overline T$, and $w=2, w'=3$ we find 
\be
\langle \overline T | V\circ_2 \overline T \rangle \ = \ \langle V | \overline T\circ_2 
\overline T\rangle - 2 \langle \overline T | V\circ_1 \overline T\rangle 
- 2 \langle \overline T | \langle V | \overline T\rangle \rangle  \,.
\ee
Because both vectors or scalars contracted with projected tensors are total derivatives
under $\int e^\phi$, the last two terms in the above right-hand side can be dropped:
\be
\langle \overline T | V\circ_2 \overline T \rangle \ \sim  \ \langle V | \overline T\circ_2 
\overline T\rangle \,.
\ee
Using (\ref{ttstar}) for the above right-hand side, we then have
\be
\langle \overline T | V\circ_2 \overline T \rangle \ \sim  \ \f14 \langle V | \langle
\overline T |\overline T\rangle'' \rangle  \sim \  \f12 \langle V | \langle
\overline T |\overline T\rangle \rangle  \,, 
\ee
where the last step used the
derivative identity and the symmetry property  
twice.  Applying this to $ÑT=\T$
we get 
\be
\label{yuih}
\langle \T | V\circ_2 \T \rangle \ \sim   \  \f12 \langle V | \langle\, 
\T |\T\rangle \rangle \,. 
\ee
For further simplification, we use the first equation in (\ref{traverss}), 
applied to $\T$  to get  
\be
\label{travers}
\hbox{tr}\left( \T \star \T\right) \, =\  2Ò\hbox{tr} \T | \T Ô +4Ò\T | \T Ô \,. \ee
By the double metric equation of motion, the left hand side is
a constant.  Under a derivative
we can therefore replace $Ò\T | \T Ô$ by $ -\f12 Ò\hbox{tr} \T | \T Ô$.  Indeed, 
using $\langle V|f\rangle=V^M\p_Mf$, we see that 
$Ò\T | \T Ô$ appears under a derivative in (\ref{yuih}) and therefore, 
\be
\label{yuihx}
\langle \T | V\circ_2 \T \rangle \ \sim   \  -\f14 \langle V | \langle
\T |\hbox{tr}\,\T\rangle \rangle \,.
\ee
With this result we can return to our variation (\ref{dilvaract}), identify
$V = \delta \phi \circ_1 \T$ and then obtain
\be
\label{dilvaractx} ¶_Ä S = Çe^Ä \, \bigl[ (¶Ä)\f13 \hbox{tr} \T \, 
 - \, \f1{24} Ò¶Ä\circ_1\T|
Ò\T|\hbox{tr}\, \TÔÔ \, \bigr]\,,   \ee
where we simplified $L$ by using the double metric equation of motion.

In order to isolate the $\delta \phi$ factor in the second term, we consider another distributive identity 
($\O_1 = \S, \O_2 = f_2, \O_3 = V$
with $\hat w =0, w=0, w'=3$)
\be 
\hbox{div}\, (f_2\circ_1 V) = Òf_2|\hbox{div}\, VÔ -Òf_2|VÔ
\quad \to \quad  ÒV|f_2Ô \, \sim \, - Òf_2|\hbox{div}\, VÔ\, . \ee  
Taking now $V = f_1 \circ_1 \overline T$ and noting that $\hbox{div}\, (f_1 \circ_1 \overline T) =
-2 \langle f_1 | \overline T\rangle $ (see (\ref{thedivless})), we have
\be
\label{xys}
Ò f_1 \circ_1 \overline T|f_2Ô \, \sim \, 2 \, Òf_2\,|\,\langle f_1 | \overline T\rangleÔ\,. 
\ee
The distributive identity ($\O_1 = f_1, \O_2 = f_2, 
\O_3 = T$ with $\hat w =0, 
w = w' =1$) 
\be Òf_1|f_2\circ_1 TÔ - Òf_2|f_1\circ_1 TÔ = 0 \,,\ee
informs us that the left-hand side of (\ref{xys}) is symmetric under the exchange
of the two functions, so that we have
\be
Ò f_1 \circ_1 \overline T|f_2Ô \, 
\sim \ 2\, Òf_1\,|\,\langle f_2 | \overline T\rangleÔ\,  =  \, 2 f_1  \, 
\langle f_2 | \overline TÔ  \,,\ee
where in the last equality we noted that  the inner product 
of two functions is equal to their ordinary
product (by the strong constraint there are no contractions 
in the operator product of two functions).
This is our desired result.  With the relevant choices of $f$'s and tensor it reads:
\be
\label{xysx}
Ò \delta \phi \circ_1 \T|\, \langle \T | \hbox{tr}\, \T \rangleÔ \, 
  =  \, 2 \delta \phi  \, \langle \T | \langle \T | \hbox{tr}\, \T \rangle\, Ô  \,. \ee
Back to the action variation (\ref{dilvaractx}), we
can finally rewrite the second term with $\delta \phi$ separated out:  
\be
\label{dilvaractxx} ¶_Ä S = \f13Çe^Ä\,(¶Ä) \bigl[  \,\hbox{tr} \, \T \, -\, 
 \f14  \langle \T | Ò\T|\hbox{tr}\,\TÔ\, Ô\, \bigr] \,.
 \ee
The result is that variation of $Ä$ gives an operator of the 
form $I+...$ acting on $\hbox{tr}\, \T$, so the $Ä$ field equation is the expected
$\hbox{tr}\, \T = \, 0.$

\subsection{Field equation evaluation}

We'll need some explicit evaluations of operator products of tensors.
For two equal tensors the OPE gives
 \be\label{TTPROducts}\begin{split}
 ÒT|TÔ¼=â& \f12T^{PQ}T_{PQ}  -
\p_PT^{LK}  \p_LT_K{}^P  
+ \f14\,  \p_P\p_Q T^{KL}  \p_K \p_L T^{PQ}  \\[0.5ex]
 &\hskip-8pt  -\f32  
 ÒöT|öTÔ \,-3\,\p_PöT^KT_K{}^P 
 + \f32 \p_P\p_Q öT^K \p_KT^{PQ}  \\[1.0ex]
 (T\circ_2 T)_{MN}¼=â& \{ T, T \}_{MN}\,
 - \f12  \p_MT^{PQ} \p_NT_{PQ}  
 +\,  T^{PQ} \p_P\p_QT_{MN} 
 \\[0.5ex]
& +2\, \p_{(N}T^{LK} \, \p_LT _{M)K} 
- 2\p_QT_M{}^P   \p_P T_N{}^Q  \\[0.5ex]
&+ \p_M\p_PT^{LK}\,  \p_N\p_LT_K{}^P
- \, \p_{(N}\p_KT^{PQ} \p_P\p_QT^K{}_{M)} \,  \\[0.5ex]
& - \f14\p_M\p_P\p_Q T^{KL}  \,\p_N\p_K \p_L T^{PQ}  \\[0.5ex]
&+ öT^K\p_K T_{MN}\, 
 + \,  (\p_{(N} öT^K- \p^K öT_{(N} )T_{M)K} \
   - \p_{(N}\p_PöT^K \p_K T^P{}_{M)}  \\
&  +  \frac{1}{2} \p_P (\p_{(M}öT_Q - \p_Q öT_{(M}) \p_{N)} T^{PQ} 
 - \frac{1}{4}  \p_{(M}\p_P\p_Q öT^K \p_{N)} \p_KT^{PQ}  \,, 
  \end{split}\ee
where $ÒöT_1|öT_2Ô$ means the inner product of 
two pseudovectors $öT^M Z_M$ treated as if they were vectors.

Note that the above imply the corresponding results for two
different tensors, since for any bilinear product $\diamond$ we have  
\be \O_{(1}\diamond\O_{2)} = (\O_1+\O_2)\diamond(\O_1+\O_2) - \O_1\diamond\O_1 -\O_2\diamond\O_2 \,.\ee
In practice, this means to just substitute $T_1$ and $T_2$ for the 
two  $T$'s in each term in the above equations in the two possible ways, 
then average to get $ÒT_1|T_2Ô$ and $(T_1\circ_2 T_2)_{MN}$.
Note that $(T_1\circ_2 T_2)_{MN}$ is symmetric under $1 \leftrightarrow 2$ 
because the lack of symmetry in $T_1 \circ_2 T_2$ only affects the pseudovector part.

Our full double-metric field equation $(\M \circ_2 \M)_{MN} = 2\eta_{MN}$ is therefore  
 \be\label{dmfeq}\begin{split}  
 (\M^2)_{MN} ¼=â& \, \eta_{MN}
 + \f14  \p_M\M^{PQ} \p_N\M_{PQ}  
 - \f12\,  \M^{PQ} \p_P\p_Q\M_{MN} 
 \\[0.5ex]
& -\, \p_{(N}\M^{LK} \, \p_L\M_{M)K} 
+\p_Q\M_M{}^P   \p_P \M_N{}^Q  \\[0.5ex]
&-\f12 \p_M\p_P\M^{LK}\,  \p_N\p_L\M_K{}^P
+\f12 \, \p_{(N}\p_K\M^{PQ} \p_P\p_Q\M^K{}_{M)} \,  \\[0.5ex]
& + \f18\p_M\p_P\p_Q \M^{KL}  \,\p_N\p_K \p_L \M^{PQ}  \\[0.5ex]
&-\f12 G^K\p_K \M_{MN}\, 
 -\f12 \,  (\p_{(N} G^K- \p^K G_{(N} )\M_{M)K} \
   +\f12 \p_{(N}\p_PG^K \p_K \M^P{}_{M)}  \\
&  -  \frac{1}{4} \p_P (\p_{(M}G_Q - \p_Q G_{(M}) \p_{N)} \M^{PQ} 
 + \frac{1}{8}  \p_{(M}\p_P\p_Q G^K \p_{N)} \p_K\M^{PQ}  \,.
  \end{split}\ee
where $G^M = G^M (\M , \phi)$, as defined in (\ref{G-formula}).  While $G^M$ has  
terms with one derivative and terms with three derivatives, the latter
carry the index on a derivative $\p^M$ and cannot contribute in the last term. 
The equation of motion has terms with zero, two, four, and six derivatives.  There
cannot be terms with more than six derivatives since the strong constraint does
not allow one to write any such terms.

\sectiono{Relation to generalized metric formulation} \label{reltogenmetfor}

In this section we will relate our formalism to the generalized metric. 
In particular, we confirm that  for the two-derivative 
approximation the field equations reduce to the known double field theory equations in terms of the 
generalized metric ${\cal H}_{MN}$ and the dilaton~\cite{Hohm:2010pp}. We  review the ${\cal H}$
equation  and show that it arises from  
the $\M$ field equation. Then we show that ${\rm tr}(\T)$ 
reproduces the generalized curvature scalar ${\cal R} (\H, \phi) $, which encodes the dilaton 
equation.

\subsection{Classical action}

Consider the action $S = \int e^\phi L$  with $L(\M)$ a Lagrangian for an arbitrary matrix 
$\M_{MN}$, whose indices are raised and lowered
with $\eta^{MN}$:
\be\label{origLagr}
L \ = \ 
 \f18   \M^{MN} \p_M \M^{KL} \p_N \M_{KL} -  
\frac{1}{2}\M^{MN}  \p_N \M^{KL}  \p_L \M_{MK}   - \M^{MN} \p_M \p_N \phi  
\;. 
\ee
If we were to set $\M$ equal to the (constrained) generalized metric $\H$, the resulting
$L$  is the 
simplest form of the double field theory 
Lagrangian of \cite{Hohm:2010pp}.  
This connection requires the identification 
\be
\phi =  - 2\,  d \,. 
\ee
Varying with respect to the unconstrained $\M$ we find 
\be  
\label{var-L-dft}
\delta_\M  S  = \int e^\phi \delta_\M L\,, \, \qquad 
\delta_{\M} L \ = \  
\delta \M^{MN}  {\cal K}_{MN} (\M)  \,, 
\ee
where
\be
\label{kisVM}
\begin{split}  
{\cal K} _{MN} (\M) \ \equiv 
~&~
\frac{1}{8}\, \partial_{M}{\M}^{KL}
  \,\partial_{N}{\M}_{KL}
  -\f14  (\partial_L + \partial_L \phi ) 
  ({\M}^{LK} \partial_K {\M}_{MN})
  - \,\partial_{M}\partial_N \phi\,  
    \\[1.0ex]
  & \hskip-10pt -\frac{1}{4} \partial_{(M}{\cal M}^{KL}\,\partial_{L}
  {\cal M}_{N)K}
  + \f14  (\partial_L + \partial_L \phi )  ({\M}^{KL} \partial_{(M}
   {\M}_{N)K}
  + {\M}^K{}_{(M}  \partial_K {\M}^L{}_{N)}  ) \,.
   \end{split}
\ee
It is  convenient to rewrite 
this expression in terms of  the 
pseudovector part of $\T$,
 \be
 \ G^M(\M,\phi) \ = \ \partial_L{\M}^{LK} + \partial_L \phi \, {\M}^{LK}+\cdots\;, 
 \ee 
leaving out higher-derivative terms in (\ref{G-formula}) that are irrelevant 
for our present purposes. One finds 
\be
\label{kis}
\begin{split}  
{\cal K} _{MN}\equiv 
~&~
\frac{1}{8}\, \partial_{M}{\M}^{KL}
  \,\partial_{N}{\M}_{KL}
  -\f14  
  {\M}^{LK} \partial_K \partial_L{\M}_{MN} 
   -\f14  
  G^{K} \partial_K {\M}_{MN}
  - \,\partial_{M}\partial_N \phi\,  
    \\[1.0ex]
  & \hskip-10pt -\frac{1}{4} \partial_{(M}{\M}^{KL}\,\partial_{L}
  {\M}_{N)K}
  + \f14    {\M}^{KL} \partial_{(M}\partial_L
   {\M}_{N)K}
  +\  \f14   \partial_L  {\M}_{(M}{}^{K}  \partial_K {\M}_{N)}{}^L   \\[1.0ex]
  & \hskip-10pt  + \f14   G ^{K} \partial_{(M}
   {\M}_{N)K}
  +\ \f14  {\M}^K{}_{(M}  \partial_K G_{N)}   -  \frac{1}{2}{\M}^K{}_{M}
  {\M}^L{}_{N}  \partial _K \partial_L \phi  \,.
   \end{split}
\ee

It is useful to note that the above variation (\ref{var-L-dft}) also applies to the equivalent form 
$L'$ of the Lagrangian
that yields the same action as $L$:  
\be
\begin{split}
L'  \ = \  \,{\cal R} (\M, \phi)& \ \equiv \    
\f18 \,   \M^{MN} \p_M \M^{KL} \p_N \M_{KL} -  \frac{1}{2}\M^{MN}  \p_N \M^{KL}  \p_L \M_{MK}  \\[1.0ex]
 & \quad  - 2 \M^{MN} \p_M \p_N \phi  -  \p_M \p_N \M^{MN} 
 -  \M^{MN} \p_M\phi \, \p_N \phi  \, - 2 \p_M \M^{MN} \p_N \phi \; \, . 
\end{split} 
\ee 
  Note that ${\cal R} (\H, \phi)$
is the scalar curvature of~\cite{Hohm:2010pp}.
 The ${\cal K} (\M)$ above also coincides with
 ${\cal K}({\cal H})$  in \cite{Hohm:2010pp} (eqn.(4.49)), when $\M$ is replaced ${\cal H}$.  
 Since ${\cal H}$ is 
 a constrained field, its equation of motion is not the vanishing of ${\cal K}({\cal H})$. Rather, the field equation 
 is given by eq.~(4.57) in \cite{Hohm:2010pp} which, written out explicitly, reads   
  \be
   {\cal K}(\H) -\H\, {\cal K}(\H) \, \H \ = \ 0\;.  
  \ee 
Multiplying from the right with $\H$ and  using $\H^2=1$ gives  
 \be\label{DFTHEq}
 {\cal K}(\H)\,  {\cal H} - {\cal H} \, {\cal K}(\H) \ = \ 0 \,, 
 \ee
which is the form of the equation of motion that we will re-derive below.   

\subsection{Double metric action and field equation}

Consider the action 
 \be\label{ACtionHERE}
  S \ = \ \int e^{\phi} L \,,   \qquad   L \ = \  
  \f12 \, {\rm tr}(\T)-\frac{1}{6}\langle\T|\T\star\T\rangle \;, 
 \ee
to second order in derivatives.  The Lagrangian can be easily computed, recalling 
that the $\star$ product projects onto divergence-free tensors, so that    
the  pseudovector part of a star product is given by  
\be
(T_1 \star T_2)^M=G^M \bigl( (T_1\circ_2 T_2)^{MN}, \phi \bigr)  \,,
\ee
with the right-hand side defined in (\ref{G-formula}). 
The term tr$(\T)$ can be evaluated explicitly from the first equation in (\ref{tr-div}), 
using the determined expression $\hat{\T}^M=G^M (\M^{MN}, \phi)$.   
Similarly, the cubic term can be straightforwardly 
computed from the explicit form of the inner product and $\circ_2$ given in (\ref{TTPROducts}).  
One finds 
 \be
 \begin{split}
  L \ = \ \,&\frac{1}{2}\Big[{\rm Tr}\left( \M-\frac{1}{3}\M^3\right)-3\M^{MN}\p_M\p_N\phi
  -3 \p_NG^N(\M) -3 \p_N\phi \, G^{N}(\M) \\
  &+\frac{1}{12}\M^{MN}\p_M\M^{PQ}\p_N\M_{PQ}
   -\frac{1}{6}\M^{MN}\M^{PQ}\p_P\p_Q\M_{MN}\\
   &-\frac{2}{3}\M^{MN}\p_M\M^{LK}\p_L\M_{KN}+\frac{1}{3}\M^{MN}\p_Q\M_{M}{}^{P}\p_P\M_{N}{}^{Q}
   +\frac{2}{3}\p_P\M^{LK}\p_L(\M^2)_K{}^{P} \\
   &-\frac{1}{6}\M^{MN}G^K(\M)\p_K\M_{MN}
   +\p_PG^K(\M)(\M^2)_K{}^{P}\\
   &+G^M(\M) G_{M}(\M^2)+\p_PG^K(\M^2) \M_{K}{}^{P}\Big]\;. 
 \end{split}
 \ee
For brevity we have dropped the dilaton input from $G(\M, \phi)$ 
and $G(\M^2, \phi)$.  
It is now a straightforward though somewhat tedious 
calculation to verify that, up to total derivatives,
the corresponding action reads   
 \be\label{3TermVersion}
   \begin{split}
   S 
     \ &= \  \int e^{\phi}\Big[\frac{1}{2} ú^{MN}( \M-\frac{1}{3}\M^3)_{MN}+\frac{1}{2}(\M^2-1)^{MP} \M_{P}{}^{N}
   \p_M\p_N\phi\\
   &\qquad +\frac{1}{8}\M^{MN}\p_M\M^{PQ} \p_N\M_{PQ}-\frac{1}{2}\M^{MN} \p_N\M^{KL}\p_L\M_{KM}
   -\M^{MN}\p_M\p_N\phi\Big]\;. 
  \end{split}
 \ee  
Since the last line coincides with (\ref{origLagr}), its variation equals ${\cal K}_{MN}$ determined in (\ref{kisVM}).
Thus, the total variation with respect to $\M$ is given by  
 \be
  0 \ = \ \frac{1}{2}(\eta-\M^2)_{MN}
          +\frac{1}{2}(\M^2)_{(M}{}^{K} \p_{N)}\p_K\phi +\frac{1}{2}\M_{M}{}^{P}\M_{N}{}^{Q}\p_P\p_Q\phi
  -\frac{1}{2}\p_M\p_N\phi+{\cal K}_{MN} (\M)\;.   
 \ee 
Using the zeroth-order relation $\M^2=1$ in the second term 
we find 
 \be\label{SEMifinalMeq}
 \begin{split}
  0 \ = \ & \frac{1}{2}(\eta-\M^2)_{MN} 
  +\frac{1}{2}\M_{M}{}^{P}\M_{N}{}^{Q}\p_P\p_Q\phi
  +\frac{1}{2}\p_M\p_N\phi+{\cal K}_{MN}(\M)\;.  
\end{split}
 \ee

Let us now show that this equation coming from the
double metric action  is  
the $\T$ equation  $ (\T \star \T)_{MN}  =  2\,\eta_{MN}$ from the OPE.   We rewrite this in matrix notation as 
\be\label{maineq}
\M^2  \ = \ 1 + 2 {\cal V} (\M) \quad \to \quad  \f12 (1-\M^2) + \V(\M)  \ = \ 0\,.  
\ee
Equation (\ref{dmfeq}) allows us to identify  
the two-derivative part $\V^{(2)}$ of $\V$ as   
 \be\label{strangeV}
 \begin{split}
  {\cal V}_{MN}^{(2)}(\M)  \ = \ &\ \ \frac{1}{8}\p_M\M^{PQ}\p_N\M_{PQ}-\frac{1}{4}\M^{PQ}\p_P\p_Q\M_{MN}
  -\frac{1}{2}\p_{(M}\M^{KL}\p_L\M_{N)K} \\[0.5ex]
  &\hskip-6pt +\frac{1}{2}\p_Q\M_{M}{}^{P}\p_P \M_{N}{}^{Q}
  -\frac{1}{4}G^K\p_K\M_{MN}
  -\frac{1}{4}\left(\p_{(M}G^K-\p^KG_{(M}\right)\M_{N)K}\;, 
 \end{split}
 \ee
 where only the parts of 
 $G^K$ with one derivative are included.   
Next we have to relate $\V_{MN}^{(2)}$ to ${\cal K}_{MN}$.  Using (\ref{kis}) we find  
 \be\label{FirstVKcomp}
 \begin{split}
  \V_{MN}^{(2)} (\M) \ & = \ \ {\cal K}_{MN} (\M) -\frac{1}{4}G^K\partial_{(M}\M_{N)K}
  -\frac{1}{4} \p_{(N}G^K \M_{M)K}\\[0.9ex]
   &-\frac{1}{4}\p_{(M}\M^{KL}\p_L\M_{N)K}
  -\frac{1}{4}\M^{KL}\p_L\p_{(M}\M_{N)K}+\p_M\p_N\phi+\frac{1}{2}\M^{K}{}_{M}\M^{L}{}_{N}\p_K\p_L\phi\;. 
 \end{split}
 \ee 
We now  use $\M^2=1$ in the two-derivative terms, which implies in particular
 \be
  -\frac{1}{2}\M^{KL}\p_M\M_{NK} \ = \ \frac{1}{2}\M_{NK}\p_M\M^{KL} \;.
 \ee  
Acting here with $\p_L$ implies 
 \be
 \begin{split}
   -\frac{1}{2}\M^{KL}\p_L\p_M\M_{NK} \ &= \ \frac{1}{2}\p_L\M^{LK}\p_M\M_{NK}
   +\frac{1}{2}\p_L\M_{NK}\,\p_M\M^{KL}+\frac{1}{2}\M_{NK}\p_M\p_L\M^{LK} \\
   \ &= \ \frac{1}{2}G^K\p_M\M_{NK}-\frac{1}{2}\p_L\phi \M^{LK}\p_M\M_{NK}+\frac{1}{2}\p_M\M^{KL}\p_L\M_{NK}\\
   &\qquad +\frac{1}{2}\M_{NK}\p_MG^K-\frac{1}{2}\M_{NK}\p_M\big(\p_L\phi \,\M^{LK}\big) \\[0.5ex]
     \ &= \ \frac{1}{2}G^K\p_M\M_{NK}+\frac{1}{2}\p_M\M^{KL}\p_L\M_{NK}
   +\frac{1}{2}\M_{NK}\p_MG^K-\frac{1}{2}\p_M \p_N\phi\,  \;. 
  \end{split}  
  \ee 
 Using this in (\ref{FirstVKcomp}) many terms cancel and we finally get 
 \be\label{finalKVrel}
 \begin{split}
   \V_{MN}^{(2)}  (\M) \ = \ & \,{\cal K}_{MN} (\M)  
   +\frac{1}{2}\p_M\p_N\phi+\frac{1}{2}\M^{K}{}_{M}\M^{L}{}_{N}\p_K\p_L\phi\;. 
   \end{split}
 \ee 
 Inserting this in (\ref{maineq}) we obtain
  \be
   0 \ = \ \frac{1}{2}(\eta-\M^2)_{MN}+{\cal K}_{MN}(\M)
   +\frac{1}{2}\p_M\p_N\phi+\frac{1}{2}\M^{K}{}_{M}\M^{L}{}_{N}\p_K\p_L\phi\;, 
  \ee  
which is in perfect agreement with (\ref{SEMifinalMeq}), as we wanted to show.   
 
\medskip

We will now show that 
(\ref{maineq}) implies the equation of motion and the 
constraint for the generalized metric $\H$.
Indeed, multiplying by $\M$ from the left and subtracting the 
same equation but multiplied by $\M$ from the right we quickly
see that 
\be
\label{ghklei}
\V(\M) \, \M-\M\, \V(\M) \ = \  0\,.  
\ee
Next we do an $\alpha'$ expansion by writing
\be
\label{popi}
\M \ = \ {\cal H}  +  ñ ({\cal H})  \,, ~~~ \hbox{with} ~~ {\cal H}^2= 1\,, 
\ee
where $ñ ({\cal H})$ is first order in $\alpha'$, containing two derivatives. 
To leading order (\ref{ghklei}) gives 
\be
\label{jui}
{\cal V}^{(2)}(\H) \, {\cal H} - {\cal H}\,  {\cal V}^{(2)}(\H) \ = \ 0 \;.  
\ee
We quickly  
confirm   that the difference in (\ref{finalKVrel}) between ${\cal K}$ and $\V^{(2)}$ 
drops out from the 
above field equation. Thus, 
\be
{\cal V}^{(2)}(\H) \,  {\cal H} - {\cal H}\,  {\cal V}^{(2)}(\H) \ = \ {\cal K}(\H)\,  {\cal H} 
- {\cal H} \, {\cal K}(\H) = 0 \;.   
\ee
The last equality is the field equation (\ref{DFTHEq}) for ${\cal H}$ in DFT. 
We have reproduced it correctly from the double-metric.

We now determine $\Lambda(\H) $ in (\ref{popi}).  Using the expansion in (\ref{maineq}) gives
\be
\label{eqnsolve}
{\cal H} ñ ({\cal H})   +  ñ ({\cal H})  {\cal H}  \ = \  2 {\cal V}^{(2)} ({\cal H}) \;.   
\ee
This equation also contains the field equation (\ref{jui}): It is obtained by  multiplying the above by ${\cal H}$ from 
the left, and subtracting the equation in which we multiply by ${\cal H}$ from the
right.
On the other hand solving for $ñ$ from (\ref{eqnsolve}) looks
at first sight impossible, since it would appear to trivialize ${\cal V}^{(2)}$.  
But this is not the case if the solution involves the field equation.
Indeed we can take  
\be\label{ExplSOL}
ñ \ = \  ü \{  {\cal H} \,, {\cal V}^{(2)}(\H) \} \;.   
\ee
Then back on the left-hand side of (\ref{eqnsolve}) and using ${\cal H}^2 =1 $ we get
\be
{\cal V}^{(2)} + {\cal H} {\cal V}^{(2)} {\cal H} \ = 2{\cal V}^{(2)} 
+ {\cal H} \, [ \, {\cal V}^{(2)}\,, {\cal H}\, ] \ = \ 2 {\cal V}^{(2)}\;, 
\ee
where in the last step we used the equation of motion.  
So we can write, 
\be\label{M(H)}
\M ({\cal H})  \ = \ {\cal H} +  ü   \{ \,  {\cal H} \,, {\cal V}^{(2)} ({\cal H}) \,  \} 
+ {\cal O} (\alpha'^2) \;. 
\ee 
We note that this parameterization of $\M$ in terms of $\H$ has
assumed the equation of motion for $\H$.  It can therefore be used in the form 
$\M ({\cal H})  \ = \ {\cal H} +   {\cal H}{\cal V}^{(2)}$.

\subsection{Dilaton equation}

We now analyze the dilaton equation 
\be\label{Honcemore}
  {\rm tr}(\T)  \ = \ \eta^{MN}\M_{MN}-3\partial_M\p_N{\cal H}^{MN}
  -6{\cal H}^{MN}\p_M\p_N\phi-6\p_M{\cal H}^{MN}\p_N\phi-3{\cal H}^{MN}\p_M\phi\,\p_N\phi\,,
\ee  
where we were allowed to replace $\M={\cal H}$ in the ${\cal O}(\alpha')$ term. 
We will show that, in the two-derivative approximation, 
it gives rise to the scalar curvature of double field theory when written in terms of the generalized metric $\H$, 
and thus to the usual dilaton equation.

First, we insert (\ref{ExplSOL}),   
 \be
  \eta^{MN}\M_{MN} \ = \  \eta^{MN}ñ_{MN}({\cal H}) \ = \ 
  {\cal H}^{MN}{\cal V}_{MN}\;. 
 \ee
The tensor $\V^{(2)}$ is given in (\ref{strangeV}),  
 \be
 \begin{split}
  {\cal V}_{MN}^{(2)} \ = 
  \ &\ \ \frac{1}{8}\p_M\H^{PQ}\p_N\H_{PQ}-\frac{1}{4}\H^{PQ}\p_P\p_Q\H_{MN}
  -\frac{1}{2}\p_{(M}\H^{KL}\p_L\H_{N)K} \\[0.5ex]
  &\hskip-6pt +\frac{1}{2}\p_Q\H_{M}{}^{P}\p_P \H_{N}{}^{Q}
  -\frac{1}{4}G^K\p_K\H_{MN}
  -\frac{1}{4}\left(\p_{(M}G^K-\p^KG_{(M}\right)\H_{N)K}\;, 
 \end{split}
 \ee 
where we replaced everywhere $\M$ by $\H$. 
Thus,
 \be\label{simpHV}
  \begin{split}
   {\cal H}^{MN}{\cal V}_{MN}^{(2)} \ = \ &\frac{1}{8}{\cal H}^{MN} 
   \p_M{\cal H}^{KL}\,\p_N{\cal H}_{KL} -\frac{1}{4}\H^{MN}\H^{PQ}\p_P\p_Q\H_{MN}
  -\H^{MN} \p_{M}\H^{KL}\p_L\H_{NK} \\[0.5ex]
  &+\frac{1}{2}\H^{MN}\p_Q\H_{M}{}^{P}\p_P \H_{N}{}^{Q}\;.
 \end{split}
 \ee  
A few terms dropped out by the constraint on $\H$, in particular all the $G$ terms. 
Using some identities following from $\H^2=1$,  
 \be
  \begin{split}
      -\frac{1}{4}{\cal H}^{KL}{\cal H}^{PQ}\p_K\p_L{\cal H}_{PQ}\ &= \ \frac{1}{4}
   {\cal H}^{KL}\p_K{\cal H}^{PQ}\p_L{\cal H}_{PQ}\;,  \\
   \frac{1}{2}{\cal H}^{PQ} \p^K{\cal H}_{LP}\,\p^L{\cal H}_{KQ}\ &= \ -\frac{1}{2}{\cal H}^{PL}
   \p_L{\cal H}^{KQ}\,\p_K{\cal H}_{PQ}\;,
  \end{split}
 \ee   
one finds 
 \be
   \eta^{MN}\M_{MN} \ = \ \H^{MN}\V_{MN}^{(2)}  
    \ = \ 3\left(\frac{1}{8}\H^{MN}\p_M\H^{PQ}\p_N\H_{PQ}-\frac{1}{2}
  \H^{MN}\p_M\H^{KL}\p_L\H_{NK}\right)\;. 
 \ee  
Inserting now in (\ref{Honcemore}) and re-expressing $\phi=-2d$, we get  
 \be
  \begin{split}
  %\eta^{MN}\M_{MN}
  {\rm tr}({\cal T}) \ &= \ 3\, ( \,\frac{1}{8}{\cal H}^{MN}\p_M{\cal H}^{PQ}\p_N{\cal H}_{PQ}
   -\frac{1}{2}{\cal H}^{MN}\p_M{\cal H}^{KL}\p_L{\cal H}_{NK}
   -\p_M\p_N{\cal H}^{MN} \\[0.5ex]
   &\qquad \quad +4{\cal H}^{MN}\p_M\p_Nd+4\p_M{\cal H}^{MN}\p_Nd -4{\cal H}^{MN}\p_Md\,\p_Nd ~) \\
   \ &= \ 3\,{\cal R}(\H, d) \;.   
  \end{split}
 \ee  
Thus, we get exactly the scalar curvature and so the dilaton field equation. 

It is also instructive to verify that  
the equation tr$(\T)=0$ follows from the 
action (\ref{ACtionHERE}) upon using the $\T$ equation --- instead of using 
the first-order \textit{solution} of that equation. 
We thus vary  (\ref{3TermVersion}) with respect to $\phi$, 
and we are allowed to use $\M^2=1$ in the two-derivative 
terms as a consequence of the $\T$ equation. We obtain 
  \be
   \frac{1}{2}{\rm Tr}\big(\M-\frac{1}{3}\M^3\big)+{\cal R}(\M,\phi) \ = \ 0\;.
  \ee 
Inserting now  $\M^2=1+2{\cal V}^{(2)}$ this becomes   
 \be\label{STep2}
  {\rm Tr}\M-\M^{MN}{\cal V}_{MN}^{(2)}+3{\cal R} \ = \ 0\;.
 \ee 
In  $\M^{MN}{\cal V}_{MN}^{(2)}$ we may use $\M^2=1$, which with (\ref{simpHV}) yields 
 \be
  -\M^{MN}{\cal V}_{MN}^{(2)} \ = \ -3\left(\frac{1}{8}\M^{MN}\p_M\M^{PQ} \p_N\M_{PQ}
  -\frac{1}{2}\M^{MN} \p_N\M^{KL}\p_L\M_{KM}\right) \;.
 \ee
Back in (\ref{STep2}) this gives 
 \be
  \eta^{MN}\M_{MN}-6\M^{MN}\p_M\p_N\phi-3\p_M\p_N\M^{MN}
  -6\p_M\M^{MN}\p_N\phi-3\M^{MN}\p_M\phi \, \p_N\phi \ = \ 0 \;, 
 \ee
which is exactly the dilaton equation tr$(\T)=0$ as following from the OPE.

\sectiono{Prospects}

This paper provides a different approach to $\alpha'$ corrections
of low-energy string actions.
Traditionally these corrections have been gleaned from the string theory
S-matrix, and then terms are constructed for the low-energy theory that
reproduce such S-matrix results.  In this paper $\alpha'$ corrections
are seen as required by a modified gauge structure. They are predicted,
or at least constrained by a symmetry principle.  We want to emphasize
that our use of the double field theory approach does not mean that the
results are only valid for compactified theories.  The $\alpha'$ information
obtained is background independent. 

\noindent
\rgb{0 0 1}{{\bf\boldmath $\alpha'$-Geometry.}} In string theory $\alpha'$ corrections
are part of the (target space) 
classical theory. More precisely, classical closed string
field theory includes $\alpha'$ corrections of all orders.  For the massless
sector, the $\alpha'$ corrections parameterize the way in which the string theory
differs from the two-derivative Einstein action coupled to a two-form
and a dilaton. It therefore has been reasonable to expect
that the appropriate geometry of string theory should be an
$\alpha'$-deformation of Riemannian geometry.  The incorporation of T-duality
has forced on us a doubled geometry that can be viewed as a mild extension
of generalized geometry.  This is the case even for the two-derivative theory.
This geometry has an inner product, a C-bracket and generalized Lie derivatives,
that upon reduction from D+D to D dimensions give the inner product, the Courant
bracket and the Lie derivatives of generalized geometry.   
The $\alpha'$ corrections are nontrivial deformations of
the geometry.
 The inner product and C-bracket acquire a  correction that is in fact linear in $\alpha'$.
Gauge transformations or generalized Lie derivatives acquire a linear correction
for a vector field and a linear plus quadratic correction for a two-tensor.   
This is a ``complete" deformation: 
The bracket is fully consistent (has a trivial Jacobiator)
without higher-order corrections, and the commutator of 
generalized Lie derivatives gives precisely the Lie derivative along the
 C-bracket of the input gauge parameters.  We want to emphasize that
 the C-bracket
does not allow higher-order $\alpha'$ corrections consistent with linearity in its arguments,
so the correction we have is unique.  This indicates that the above represents
a first step in the construction of the $\alpha'$-geometry.  
Intriguingly, the corrections to all these
structures {\it do not} vanish when reduced from the doubled manifold to D dimensions.  
Therefore, they define an apparently unknown 
deformation of the Courant bracket and other
structures in generalized geometry.  

\noindent  
\rgb{0 0 1}{{\bf\boldmath Double metric $\M$.}}
The generalized metric $\H$ of the doubled manifold was a duality-covariant
gravitational field variable for the two-derivative theory.  
Surprisingly, the relevant OPE's indicated that the constraint $\H^2=1$
satisfied by this metric cannot be preserved when considering $\alpha'$ 
corrections.  We were thus led to consider
a double metric $\M$,  an unconstrained extension of the generalized
metric.   Just like ordinary metrics, $\M$ does not satisfy an algebraic 
constraint.  But even more is true:  We do not need to assume $\M$ is invertible 
to define the action, yet it
is invertible on-shell as a consequence of its field equation $\M^2 = 1 + \ldots$. 
The straightforward emergence of $\M$ and the simplicity of the action 
suggests that $\M$  is a natural variable for the fundamental
description of gravitational degrees of freedom in string theory.

\noindent
\rgb{0 0 1}{{\bf A new consistent truncation of string theory?}}  We have constructed an $\alpha'$
deformation of the low-energy effective action.  In terms of the gravitational
variable $\M$ and the dilaton, the action and field equations contain bounded
powers of $\alpha'$.  
In terms of 
$(g,b, \phi)$,  the equations of motion and
the action presumably contain terms to all orders in $\alpha'$. 
The obvious question is:
Is this the exact effective action of string theory for the massless sector?
It seems not:  The four-point and higher point amplitudes
in this theory are not expected to contain the poles
associated with the massive string states.  
The theory is, however, fully consistent: All $\alpha'$ dependent 
gauge symmetries are exact invariances. 
This indicates that this theory is a consistent
truncation of string theory in which some of the 
stringy non-locality has been eliminated.
The higher-derivative contributions that remain can be perhaps 
traced to those in the three-closed-string vertex.  
With a suitable off-shell definition of the vertex (not the one used
in closed string field theory, in fact) the massless field three-vertices
contain only finite number of derivatives (two, four, and 
six for the case of three gravitons).
The theory we have may be the consistent completion of such
cubic theory.

\noindent
\rgb{0 0 1}{{\bf A new worldsheet theory?}}
We extended double field theory consistently to higher order(s) in $Œ'$.  The method is a ``complete" result for a system related to the usual string theory, but employing free, chiral bosons.  Further investigation is required to determine how this chiral string relates to the usual string beyond the classical level.
It would be useful
to have a derivation of this theory obtained by gauge fixing
of the standard first-quantized action. Such gauge-fixing 
would teach us how to introduce ghost fields, which are needed
beyond the classical level discussed in this paper.  
In this theory the strong constraint ensures that the OPE of
fields is nonsingular.  Thus the derivation of the field equations
from conformal invariance is greatly simplified,  
as compared to the usual calculation of beta functions~\cite{Banks:1986fu}.
It also suggests a new string field theory based on the 
BRST operator for this chiral Virasoro algebra.

\noindent
\rgb{0 0 1}{{\bf Covariant derivatives, torsions, and curvatures.}}
For further clarification of the geometry, the inclusion of the Lorentz 
current algebra will allow for true covariant derivatives in a vielbein formalism (also required for supersymmetry) \cite{Siegel:2011sy}: $Œ'$ corrections to torsions, curvatures, and local Lorentz transformations will then automatically follow by the same methods used in this paper.
The corresponding expressions should exist in terms of the generalized metric used here and generalized Christoffel symbols as an extension of the methods
in~\cite{Hohm:2012mf} (eqns.¼(1.5) and (1.6)).
Some components
of Riemann would still be undetermined, since suitable
generalized constraints are still going to fail to fix the connection 
completely.   
But just as in the case of the two-derivative theory, the contractions
that give the scalar curvature may eliminate all undetermined components.
If this is so, the action density would simply be 
the ``scalar curvature" associated with the $\alpha'$-corrected Riemann.

\noindent  
\rgb{0 0 1}{{\bf Relation to conventional field theory.}} It is of great interest to see how the
theory given here is related to one that has a metric $g$ and a two-form field $b$ with conventional
gauge transformations --- of course, by sacrificing manifest T-duality.  This assumes that the
$\alpha'$-deformation of our gauge structures can be trivialized using T-duality violating
and gauge non-covariant field redefinitions. This seems very plausible, but should be investigated.  
%oh modified and extended 
%Identifying conventional
%$(g, b)$ fields from the generalized metric may be one possibility; 
Naively, one may try to identify the conventional fields
$g$ and $b$ via the generalized metric ${\cal H}$, as the latter is naturally parametrized in terms 
of these fields. As noted at the end of section 7.2,  however, writing 
${\cal M}$ in terms of  ${\cal H}$ was only 
possible \textit{on-shell}. Therefore, from this starting point one cannot derive 
off-shell gauge transformations of $g$ and $b$, nor  an off-shell 
action for these fields.   
Perhaps it will be possible to identify $g$ and $b$  directly as components of the double metric $\M$, which would also contain a number of
auxiliary field degrees of freedom, but this remains an open question.
% It remains as an open question whether the conventional 
%(background-independent) field variables can be identified directly in our $\alpha'$-deformed framework
Some progress may be possible in a perturbative analysis around
a constant background,  giving a relation between double metric fluctuations
and the (background-dependent) field variables in closed string field theory.  
Partial results along these lines have been obtained, and we hope to report on them in the near future. 
%oh end of additions 
A related question is the appearance of higher-order curvature terms in the action.  To this end we note that the cubic term of the action is essentially the same as the most singular ($1/z^6$) term in the OPE of three operators $\T$'s at three different $z$'s.  But this is the same calculation that gives  
the three-point function of   
the corresponding vertex operators (the operators minus their vacuum pieces, with the ghosts 
cancelling the $1/z^6$).  The result of the latter is the cubic pieces of $R+R^2+R^3$ (where ``$R$" stands for the Riemann tensor).  Our action is expected to yield the T-duality covariantization of this.

\noindent
\rgb{0 0 1}{{\bf Relation to conventional string theory.}}
We have dealt with genuine string theory, which 
is evident from our starting point,
where the equations of motion came from the closure of the Virasoro algebra
and the action was written to give such equations of motion.  The formalism,
however, allows one to define other gauge invariant terms that could be
added to the action, at the price of changing the field equations and perhaps
losing the connection to string theory.  Such alternative actions may be of some interest.
On the other hand, the existence of these higher-derivative gauge-invariant  
terms could allow the construction of those
$\alpha'$ contributions that turn the present theory into one that
reproduces the dual amplitudes of string theory.  

\noindent  
\rgb{0 0 1}{{\bf Other.}}  The action and field equations found here are unusual in that they contain both dynamics and algebraic constraints from the same field (no Lagrange multipliers or auxiliary fields).  This is similar to the decomposition of gauge fields into gauge, auxiliary, and dynamical components in a lightcone gauge, but here the decomposition is local and Lorentz covariant.  It would be interesting to see if this new concept can be extended to other systems.  An obvious avenue of extension of the current results is to superstrings, whose classical treatment was begun in  \cite{Siegel:1993xq,Siegel:1993th,Siegel:1993bj}. 
It may also be interesting to consider the inclusion of higher weight operators
describing higher spin fields.

\section*{Acknowledgments}

OH and BZ would like to thank Ashoke Sen for early discussions
on the problem of $\alpha'$ corrections and T-duality. 
WS was supported in part by National Science Foundation Grant No.\ PHY-0969739.
OH was supported by the
DFG Transregional Collaborative Research Centre TRR 33
and the DFG cluster of excellence ``Origin and Structure of the Universe".

\appendix

\sectiono{Quadratic identities}

\be \O_2 \circ_w \O_1 = (-1)^{w_1+w_2-w} e^{-\L} \O_1 \circ_w \O_2 \ee
\be
\begin{split}
\O_1' \circ_w \O_2 ¼=¼ & \ (w-w_1-w_2) \,\O_1 \circ_w \O_2\,,\\
( \O_1 \circ_w \O_2 )' ¼=¼& \  \O_1' \circ_{w+1} \O_2 + \O_1 \circ_{w+1} \O_2' 
\end{split}
\ee
\be
\label{saction}
\begin{split}
 \S \circ_{w_\O+1} \O \ = \ &\  \O'   \\
  \S \circ_{w_\O} \O\,  \ =\ & \  w_\O \O \\
\S \circ_{w_\O-1} \O \ ­ \ & \ \hbox{div}(\O)   \\
  \S \circ_{w_\O-2} \O\  ­ \ & \  \f12\hbox{tr}(\O) 
\end{split}
\ee

\subsection*{Examples}

\be
\begin{split}
 \hbox{tr}\, (\O')\  = \ &  (\hbox{tr}\,\O)' +6\,  
\hbox{div}\,\O , \\
\hbox{div}\, (\O') \ = \ & (\hbox{div}\,\O)' +2w_\O \O 
\end{split}\ee

\be \hbox{tr}\, V = \hbox{tr}\, f = \hbox{div}\, f = 0 \ee
\be \hbox{tr}(V') = 6 \,\hbox{div}ÊV ,â
\hbox{div}(V') = 2V + (\hbox{div}ÊV)' ,â
\hbox{div}(f') = 0 ,â
\hbox{div} (f'') = 2 f' ,â
\hbox{tr}\, (f'') = 0
\ee

\sectiono{Cubic identities}

\be\begin{split}
\O_1 \circ_{öw} ( \O_2 \circ_{w_2+w_3-w} \O_3 )  -
	\O_2 \circ_{öw} ( \O_1 \circ_{w_1+w_3-w'} \O_3 )  & ¼=¼
	Ý_{w''=1}^{w'} {w'-1\choose w''-1}( \O_1 \circ_{w_1+w_2-w''} \O_2 ) \circ_{öw} \O_3 \\
öw + w + w' & ¼=¼ w_1 + w_2 + w_3 \,, ~~ w' \geq 1   
\end{split}\ee

\subsection*{Examples}

{\boldmath$ w_1 = w_2 =w_3 =2:$ }
\medskip

\noindent
{\boldmath $w=2, w' = 4, \hat w =0$}
\be\begin{aligned}
 T_1 \circ_{0} ( T_2  \circ_{2} T_3 )  \ = \ & \ 
  T_2 \circ_{0} ( T_1 \circ_{0} T_3 )  \ +   ( T_1 \circ_{0} T_2 ) \circ_{0} T_3  \\[1.0ex] 
  &
   \ +3 ( T_1 \circ_{1} T_2 ) \circ_{0} T_3 
    \ +3 ( T_1 \circ_{2} T_2 ) \circ_{0} T_3 
     \ + ( T_1 \circ_{3} T_2 ) \circ_{0} T_3  ~~     
\end{aligned}\ee
{\boldmath $w=2, w'=3, \hat w =1$}
\be
 T_1 \circ_{1} ( T_2  \circ_{2} T_3 )  \ =  \ 
  T_2 \circ_{1} ( T_1 \circ_{1} T_3 )  \ +   ( T_1 \circ_{3} T_2 ) \circ_{1} T_3 
   \ +2 ( T_1 \circ_{2} T_2 ) \circ_{1} T_3 
    \ + ( T_1 \circ_{1} T_2 ) \circ_{1} T_3 
\ee
The two equations above with the first tensor taken to be $\S$ give
\be\begin{aligned}
 \f12 \hbox{tr}  ( T_1  \circ_{2} T_2 )  \ = \ & \ 
  \f12 \, T_1 \circ_{0} ( \hbox{tr} \, T_2 )  \ +  \f12 ( \hbox{tr} \,T_1 ) \circ_{0} T_2  
 \ +3 \, ( \hbox{div}\, T_1 ) \circ_{0} T_2 \ +6 \, T_1  \circ_{0} T_2 \ + T_1' \circ_{0} T_2  ~~   \\[1.0ex]  
 \hbox{div}  ( T_1  \circ_{2} T_2 )  \ = \ & \ 
  T_1 \circ_{1} (\hbox{div}\,  T_2 )  \ +   T_1' \circ_{1} T_2   \ +4 \, T_1  \circ_{1} T_2 \ + ( \hbox{div}\,  T_1 ) \circ_{1} T_2 
\end{aligned}\ee
{\boldmath $w=5, w'=1$}
\be 0 ¼= \  T_2 \circ_{0} ( T_1 \circ_{3} T_3 )  \ + \ ( T_1 \circ_{3} T_2 ) \circ_{0} T_3 \ee
{\boldmath $w=4, w'=1$}
\be  T_1 \circ_{1} ( T_2 \circ_{0} T_3 ) \ = \ T_2 \circ_{1} ( T_1 \circ_{3} T_3 ) \ + \  ( T_1 \circ_{3} T_2 ) \circ_{1} T_3 \ee
{\boldmath$w=4, w'=2$}
\be T_1 \circ_{0} ( T_2  \circ_{0} T_3 ) \ =  \ T_2 \circ_{0} ( T_1 \circ_{2} T_3 ) \ + ( T_1 \circ_{2} T_2 ) \circ_{0} T_3  \ + ( T_1 \circ_{3} T_2 ) \circ_{0} T_3  \ee
{\boldmath $w=2, w'=4$}
\be\begin{aligned}
 T_1 \circ_{0} ( T_2  \circ_{2} T_3 )  \ = \ & \ 
  T_2 \circ_{0} ( T_1 \circ_{0} T_3 )  \ +   ( T_1 \circ_{0} T_2 ) \circ_{0} T_3  \\[1.0ex] 
  &
   \ +3 ( T_1 \circ_{1} T_2 ) \circ_{0} T_3 
    \ +3 ( T_1 \circ_{2} T_2 ) \circ_{0} T_3 
     \ + ( T_1 \circ_{3} T_2 ) \circ_{0} T_3  ~~     
\end{aligned}\ee
{\boldmath $w=3, w'=3$}
\be
 T_1 \circ_{0} ( T_2  \circ_{1} T_3 ) \ = \ 
  T_2 \circ_{0} ( T_1 \circ_{1} T_3 ) \ +( T_1 \circ_{1} T_2 ) \circ_{0} T_3
   +2( T_1 \circ_{2} T_2 ) \circ_{0} T_3 +( T_1 \circ_{3} T_2 ) \circ_{0} T_3
\ee
\be\begin{aligned}
 \hbox{\rm div}   ( T_2  \circ_{1} T_3 ) \ = \ & \ 
  \langle T_2 | \hbox{\rm div}(T_3 )\rangle \ + \, \langle \hbox{\rm div} (T_2 ) |T_3\rangle
\end{aligned}\ee

\sectiono{Evaluated products}

\be
\begin{split}
\hbox{tr}(T) ¼=â & ú^{MN}T_{MN} -3 ( T^{MN}»_M »_N Ä + »ÉöT + öTÉ»Ä ) \\[0.5ex]
\hbox{div}(T)^M ¼=â &  »_N T^{MN} + T^{MN} »_N Ä -ü T^{NP}»_N »_P »^M Ä - öT^M  - ü »^M (» É öT + öTÉ»Ä ) \\[0.3ex]
\hbox{div}(V) ¼=â & »ÉV+VÉ»Ä
\end{split}\ee

\be
\label{scalar_tensor}\begin{split}
ÒT|fÔ¼=â& ü(T^{MN}»_M »_N f +öT^M »_M f) \\[0.5ex]
(T\circ_1 f)^M¼=â& T^{MN}»_N f +ü(»^M T^{NP})»_N »_P f \\[0.5ex]
(f\circ_1 T)^M¼=â& -T^{MN}»_N f +üT^{NP}»_N »_P »^M f +ü»^M(öT^N »_N f)
\end{split}\ee
\be\begin{split}
ÒV|TÔ¼=â& -ÒV|öTÔ -T^{MN}»_M V_N +ü(»^M T^{NP})»_N »_P V_M \\
(V\circ_1 T)^M¼=â& T^{MN}V_N -ü([V,ßT]_D^M +»^MÒV|öTÔ) +üT^{NP}»_N »_P V^M -T^{NP}»^M »_N V_P \\
	& -(»^N T^{MP})»_P V_N +ü(»^N T^{PQ})»^M »_P »_Q V_N
\end{split}\ee
($V\circ_2 T$ is in (\ref{gauge56}).  Here $ÒV|öTÔ$ means the inner product from treating $öT^M Z_M$ as if it were a vector and not a pseudovector.)
\be\begin{split}
T_1\bullet_1 T_2¼=â& - \f14 
 T_1{}^{PQ}\onª»{}_{\hskip-1pt M} T_2{}_{PQ} 
 +( T_1{}^{LK}  \p_LT_2{}_{KM} 
 -T_2{}^{LK} \p_LT_1{}_{KM} ) 
 + \f12 \p_PT_1{}^{LK}  
\onª»{}_{\hskip-1pt M}\p_LT_2{}_K{}^P \\
 & +\f12\, ( \p_KT_2{}^{PQ}\p_P\p_QT_1{}^K{}_M
-\, \p_KT_1{}^{PQ}  \p_P\p_QT_2{}^K{}_M )
- \f18 \p_P\p_Q T_1{}^{KL} \onª»{}_{\hskip-1pt M} \p_K \p_L T_2{}^{PQ}  \\[1.0ex]
& -\f12  ([ öT_1{} , öT_2{}]_{{}_C})_M
 + öT_1{}^K T_2{}_{KM} 
  - \p_PöT_1{}^K \p_K T_2{}^P{}_M
 -  \p_M\p_PöT_1{}^KT_2{}_K{}^P  
- öT_2{}^K T_1{}_{KM} \\
&   +\, \p_PöT_2{}^K \p_K T_1{}^P{}_M
 +  \p_M\p_PöT_2{}^KT_1{}_K{}^P  
 + \f12\p_P\p_Q öT_1{}_M T_2{}^{PQ}   \\[1.0ex]
& +\f12 \p_M\p_P\p_Q öT_1{}^K \p_KT_2{}^{PQ} 
 - \f12\p_P\p_Q öT_2{}_M T_1{}^{PQ}  
 -\f12 \p_M\p_P\p_Q öT_2{}^K \p_KT_1{}^{PQ}  \\[1.0ex]
 &   + \p_M  \bigl[ \f34 (\p_PöT_1{}^KT_2{}_K{}^P 
 -\p_PöT_2{}^KT_1{}_K{}^P )
 - \f38( \p_P\p_Q öT_1{}^K \p_KT_2{}^{PQ} 
 -\p_P\p_Q öT_2{}^K \p_KT_1{}^{PQ}) \bigr]  \\[1.0ex]
\end{split}\ee

\sectiono{Alternate projection}

We consider here a different, tilde projection from operators
$\O$ to operators $\tilde \O$.   There is also a different divergence operator $\widetilde{\hbox{Div}}$   
associated with this projection.  Although we
do not have a specific application in mind, this projection is in some ways
simpler than the overline projection.

The operator $\tilde \O$ is defined implicitly by the following relation 
\be
\label{jnpw} (÷\O)' ­ {1\over w_\O-1}Ê\O\circ_{w_\O+1}\S \,. \ee
The derivative identity shows that the above implies that
\be  \widetilde{(\O')} = 0\,.  \ee
We can evaluate $÷\O$ in (\ref{jnpw}) by use of the symmetry identity which
confirms that the right-hand side is a $z$-derivative.  We then get   
\be
\label{define_tilde_O}
 ÷\O = \O + {1\over w_\O-1}Ý_{w'=1}^{w_\O} {(-1)^{w'}\over (w'+1)!} ( \S\circ_{w_\O-w'}\O )^{(w')} ,âââ ÷{÷\O} = ÷\O \ee
For the new divergence we define
\be \widetilde{\hbox{Div}}
(\O) ­ 2 Ý_{w'=0}^{w_\O-1} {(-1)^{w'}\over (w'+2)!} ( \S\circ_{w_\O-w'-1}\O )^{(w')} = \hbox{div}(\O) -\f16[\hbox{tr}(\O)]'+ ... \ee
\be ÷\O = \O - {1\over 2 (w_\O -1)} (\widetilde{\hbox{Div}}Ê\O)' ,âââ \widetilde{\hbox{Div}}(÷\O) = 0 ,âââ \widetilde{\hbox{Div}}(\O') = 2 w_\O \O \ee

In particular,   
\be\begin{split}
\hbox{tr}Ê÷T &¼=\ \hbox{tr}ÊT -3Ê\hbox{div}^2ÊT = \hbox{tr}\, ÑT \\[0.5ex]
& ¼=\ ú^{MN}T_{MN} -3 »_M (»_N T^{MN} + 2T^{MN} »_N Ä) 
  -3 \,T^{MN} \p_M\phi \p_N\phi\\[0.8ex]
(\widetilde{\hbox{Div}}ÊT)^M &¼=\ [\, \hbox{div}ÊT- \f16 (\hbox{tr}ÊT)' ]^M \\[0.5ex]
& ¼=\  -öT^M + »_N T^{MN}  - \f16 »^M ú^{NP}T_{NP} + T^{MN} »_N Ä  + ü (»^M T^{NP}) »_N »_P Ä \\[0.8ex]
\widetilde T &¼=¼ T  - \, \f12 \,  (\widetilde{\hbox{Div}}ÊT )' =Ê ÑT +\f1{12}(trÊÑT)'' \\
& \ = \ \f12 T_{MN} Z^M Z^N \,  - ü Ó [ »_N T^{MN}  - \f16 »^M ú^{NP}T_{NP} + T^{MN} »_N Ä  + ü (»^M T^{NP}) »_N »_P Ä] Z_MÕ' 
\end{split}\ee
We also have the trivial cases
\be ÷\S = \S âÜâtrÊ÷\S = 2D,ââ  
\widetilde{\hbox{Div}}Ê\S = 0 \,.\ee
For the variation of an arbitrary projected operator, explicit evaluation yields
\be ¶÷\O = \widetilde{¶\O} -{1\over 4(w_\O-1)}(\O\circ_{w_\O-1}¶Ä)' \,.\ee


\begin{thebibliography}{99}


\small

%\cite{Siegel:1983es}
\bibitem{Siegel:1983es}
  W.~Siegel,
  ``Manifest Lorentz Invariance Sometimes Requires Nonlinearity,''
  Nucl.\ Phys.\ B {\bf 238} (1984) 307.
  %%CITATION = NUPHA,B238,307;%%

%\cite{Fradkin:1984pq}
\bibitem{Fradkin:1984pq}
  E.~S.~Fradkin and A.~A.~Tseytlin,
  ``Effective Field Theory from Quantized Strings,''
  Phys.\ Lett.\ B {\bf 158} (1985) 316.
  %%CITATION = PHLTA,B158,316;%%

%cite{Fradkin:1985ys}
\bibitem{Fradkin:1985ys}
  E.~S.~Fradkin and A.~A.~Tseytlin,
  ``Quantum String Theory Effective Action,''
  Nucl.\ Phys.\ B {\bf 261} (1985) 1.
  %%CITATION = NUPHA,B261,1;%%

%\cite{Banks:1986fu}
\bibitem{Banks:1986fu}
  T.~Banks, D.~Nemeschansky, and A.~Sen,
  ``Dilaton Coupling and BRST Quantization of Bosonic Strings,''
  Nucl.\ Phys.\ B {\bf 277} (1986) 67.
  %%CITATION = NUPHA,B277,67;%%

%\cite{Floreanini:1987as}
\bibitem{Floreanini:1987as}
  R.~Floreanini and R.~Jackiw,
  ``Selfdual Fields as Charge Density Solitons,''
  Phys.\ Rev.\ Lett.\  {\bf 59} (1987) 1873.
  %%CITATION = PRLTA,59,1873;%%

%\cite{Hull:1988dp}
\bibitem{Hull:1988dp}
  C.~M.~Hull,
  ``Covariant Quantization Of Chiral Bosons And Anomaly Cancellation,''
  Phys.\ Lett.\ B {\bf 206} (1988) 234.
  %%CITATION = PHLTA,B206,234;%%

%\cite{Duff:1989tf}
\bibitem{Duff:1989tf}
  M.~J.~Duff,
  ``Duality Rotations In String Theory,''
  Nucl.\ Phys.\ B {\bf 335} (1990) 610.
  %%CITATION = NUPHA,B335,610;%%

%\cite{Tseytlin:1990nb}
\bibitem{Tseytlin:1990nb}
  A.~A.~Tseytlin,
  ``Duality Symmetric Formulation Of String World Sheet Dynamics,''
  Phys.\ Lett.\ B {\bf 242} (1990) 163.
  %%CITATION = PHLTA,B242,163;%%
%\cite{Tseytlin:1990va}
\bibitem{Tseytlin:1990va}
  A.~A.~Tseytlin,
  ``Duality symmetric closed string theory and interacting chiral scalars,''
  Nucl.\ Phys.\ B {\bf 350} (1991) 395.
  %%CITATION = NUPHA,B350,395;%%

%\cite{Siegel:1993xq}
\bibitem{Siegel:1993xq} 
  W.~Siegel,
  ``Two vierbein formalism for string inspired axionic gravity,''
  Phys.\ Rev.\ D {\bf 47} (1993) 5453
  \xxxlink{hep-th/9302036}
  %%CITATION = HEP-TH/9302036;%%

%\cite{Siegel:1993th}
\bibitem{Siegel:1993th} 
  W.~Siegel,
  ``Superspace duality in low-energy superstrings,''
  Phys.\ Rev.\ D {\bf 48} (1993) 2826
  \xxxlink{hep-th/9305073}.
  %%CITATION = HEP-TH/9305073;%%

%\cite{Siegel:1993bj}
\bibitem{Siegel:1993bj} 
  W.~Siegel,
  ``Manifest duality in low-energy superstrings,''
  in *Berkeley 1993, Proceedings, Strings '93* 353-363, eds. M.B. Halpern, G. Rivlis, and A. Sevrin (World Scientific, 1995)
  \xxxlink{hep-th/9308133}.
  %%CITATION = HEP-TH/9308133;%%

%\cite{Siegel:2004dj}
\bibitem{Siegel:2004dj}
  W.~Siegel,
  ``Untwisting the twistor superstring,''
  \xxxlink{hep-th/0404255}.
  %%CITATION = HEP-TH/0404255;%%

%\cite{Hull:2009mi}
\bibitem{Hull:2009mi} 
  C.~Hull and B.~Zwiebach,
  ``Double Field Theory,''
  JHEP {\bf 0909} (2009) 099
  \xxxlink{0904.4664} [hep-th].
  %%CITATION = ARXIV:0904.4664;%%
  
%\cite{Zwiebach:1992ie}
\bibitem{Zwiebach:1992ie}
  B.~Zwiebach,
  ``Closed string field theory: Quantum action and the B-V master equation,''
  Nucl.\ Phys.\  B {\bf 390} (1993) 33
 \xxxlink{hep-th/9206084}.   % [arXiv:hep-th/9206084].
  %%CITATION = NUPHA,B390,33;%%

%\cite{Kugo:1992md}
\bibitem{Kugo:1992md}
  T.~Kugo and B.~Zwiebach,
  ``Target space duality as a symmetry of string field theory,''
  Prog.\ Theor.\ Phys.\  {\bf 87} (1992) 801
\xxxlink{hep-th/9201040}. % [arXiv:hep-th/9201040].
  %%CITATION = PTPKA,87,801;%%

%\cite{Hull:2009zb}
\bibitem{Hull:2009zb}
  C.~Hull and B.~Zwiebach,
  ``The Gauge algebra of double field theory and Courant brackets,''
  JHEP {\bf 0909} (2009) 090
  \xxxlink{0908.1792} [hep-th].
  %%CITATION = ARXIV:0908.1792;%%

%\cite{Hohm:2010pp}
\bibitem{Hohm:2010pp} 
  O.~Hohm, C.~Hull and B.~Zwiebach,
  ``Generalized metric formulation of double field theory,''
  JHEP {\bf 1008} (2010) 008
  \xxxlink{1006.4823} [hep-th].
  %%CITATION = ARXIV:1006.4823;%%
   
%\cite{Hohm:2010jy}
\bibitem{Hohm:2010jy} 
  O.~Hohm, C.~Hull and B.~Zwiebach,
  ``Background independent action for double field theory,''
  JHEP {\bf 1007} (2010) 016
 \xxxlink{1003.5027} [hep-th].
  %%CITATION = ARXIV:1003.5027;%%
 
 \bibitem{Hohm:2010xe}
  O.~Hohm, S.~K.~Kwak,
  ``Frame-like Geometry of Double Field Theory,''
  J.\ Phys.\ A {\bf A44} (2011) 085404
 \xxxlink{1011.4101} [hep-th].

  %\cite{Hohm:2011si}
\bibitem{Hohm:2011si}
  O.~Hohm and B.~Zwiebach,
  ``On the Riemann Tensor in Double Field Theory,''
  JHEP {\bf 1205} (2012) 126
 \xxxlink{1112.5296} [hep-th].
  %%CITATION = ARXIV:1112.5296;%%
  
 %\cite{Hohm:2012gk}
\bibitem{Hohm:2012gk} 
  O.~Hohm and B.~Zwiebach,
  ``Large Gauge Transformations in Double Field Theory,''
  JHEP {\bf 1302} (2013) 075
  \xxxlink{1207.4198} [hep-th].
  %%CITATION = ARXIV:1207.4198;%% 
  
 %\cite{Hohm:2012mf}
\bibitem{Hohm:2012mf} 
  O.~Hohm and B.~Zwiebach,
  ``Towards an invariant geometry of double field theory,''
  J.~Math.~Phys. {\bf 54} (2013) 032303
  \xxxlink{1212.1736} [hep-th].
  %%CITATION = ARXIV:1212.1736;%%

%\cite{Jeon:2010rw}
\bibitem{Jeon:2010rw}
  I.~Jeon, K.~Lee, J.-H.~Park,
  ``Differential geometry with a projection: Application to double field theory,''
  JHEP {\bf 1104} (2011) 014
  \xxxlink{1011.1324} [hep-th];\\
  ``Stringy differential geometry, beyond Riemann,''
  Phys.\ Rev.\  {\bf D84} (2011) 044022
  \xxxlink{1105.6294} [hep-th]. 

%\cite{Hitchin:2004ut,Gualtieri:2003dx,Gualtieri:2007bq}
\bibitem{Hitchin:2004ut} 
  N.~Hitchin,
  ``Generalized Calabi-Yau manifolds,''
  Quart.\ J.\ Math.\ Oxford Ser.\  {\bf 54} (2003) 281\\
  \xxxlink{math/0209099} [math-dg].
  %%CITATION = MATH/0209099;%%  
  
%\cite{Gualtieri:2003dx}
\bibitem{Gualtieri:2003dx} 
  M.~Gualtieri,
  ``Generalized complex geometry,''
  \xxxlink{math/0401221} [math-dg].
  %%CITATION = MATH/0401221;%%  
  
%\cite{Gualtieri:2007bq}
\bibitem{Gualtieri:2007bq} 
  M.~Gualtieri,
  ``Branes on Poisson varieties,''
  \xxxlink{0710.2719} [math.DG].
  %%CITATION = ARXIV:0710.2719;%%  

   %\cite{Hohm:2011cp}
\bibitem{Hohm:2011cp} 
  O.~Hohm and S.~K.~Kwak,
  ``Massive Type II in Double Field Theory,''
  JHEP {\bf 1111} (2011) 086
\xxxlink{1108.4937} [hep-th].
  %%CITATION = ARXIV:1108.4937;%%

\bibitem{Hohm:2011ex}
  O.~Hohm, S.~K.~Kwak,
  ``Double Field Theory Formulation of Heterotic Strings,''
  JHEP {\bf 1106} (2011) 096
  \xxxlink{1103.2136} [hep-th].

%\cite{Hohm:2011dv}
\bibitem{Hohm:2011dv} 
  O.~Hohm, S.~K.~Kwak and B.~Zwiebach,
  ``Double Field Theory of Type II Strings,''
  JHEP {\bf 1109} (2011) 013
\xxxlink{1107.0008} [hep-th];
  %%CITATION = ARXIV:1107.0008;%%
  %\cite{Hohm:2011zr}
%\bibitem{Hohm:2011zr} 
%  O.~Hohm, S.~K.~Kwak and B.~Zwiebach,
  ``Unification of Type II Strings and T-duality,''
  Phys.\ Rev.\ Lett.\  {\bf 107} (2011) 171603
 \xxxlink{1106.5452} [hep-th].
  %%CITATION = ARXIV:1106.5452;%%
  
  %\cite{Coimbra:2011nw}
\bibitem{Coimbra:2011nw} 
  A.~Coimbra, C.~Strickland-Constable and D.~Waldram,
  ``Supergravity as Generalised Geometry I: Type II Theories,''
  JHEP {\bf 1111} (2011) 091
   \xxxlink{1107.1733} [hep-th].
  %%CITATION = ARXIV:1107.1733;%%
  
%\cite{Hohm:2011nu}
\bibitem{Hohm:2011nu} 
  O.~Hohm and S.~K.~Kwak,
  ``N=1 Supersymmetric Double Field Theory,''
  JHEP {\bf 1203} (2012) 080
 \xxxlink{1111.7293} [hep-th].
  %%CITATION = ARXIV:1111.7293;%%
  
%\cite{Jeon:2011sq}
\bibitem{Jeon:2011sq}
  I.~Jeon, K.~Lee and J.-H.~Park,
  ``Supersymmetric Double Field Theory: Stringy Reformulation of Supergravity,''
  \xxxlink{1112.0069} [hep-th].   

%\cite{Siegel:2011sy}
\bibitem{Siegel:2011sy}
  W.~Siegel,
  ``New superspaces/algebras for superparticles/strings,''
  \xxxlink{1106.1585} [hep-th].
  %%CITATION = ARXIV:1106.1585;%%

 %\cite{Andriot:2011uh}
\bibitem{Andriot:2011uh}
  G.~Aldazabal, W.~Baron, D.~MarquŽs, C.~Nœ–ez,
  ``The effective action of Double Field Theory,''
  JHEP {\bf 1111} (2011) 052
  \xxxlink{1109.0290} [hep-th];\\
  D.~Geissb¬uhler,
  ``Double Field Theory and N=4 Gauged Supergravity,''
  \xxxlink{1109.4280} [hep-th].
  
 \bibitem{grana-marques}
   M.~Gra÷na and D.~MarquŽs,
  ``Gauged Double Field Theory,''
  JHEP {\bf 1204} (2012) 020
  \xxxlink{1201.2924} [hep-th].
  %%CITATION = ARXIV:1201.2924;%%
  
%\cite{Geissbuhler:2013uka}
\bibitem{Geissbuhler:2013uka} 
  D.~Geissb\"uhler, D.~MarquŽs, C.~Nœ–ez and V.~Penas,
  ``Exploring Double Field Theory,''
  \xxxlink{1304.1472} [hep-th].
  %%CITATION = ARXIV:1304.1472;%%  
  
%\cite{Andriot:2012wx}
\bibitem{Andriot:2012wx} 
  D.~Andriot, O.~Hohm, M.~Larfors, D.~L¬ust and P.~Patalong,
  ``A geometric action for non-geometric fluxes,''
  Phys.\ Rev.\ Lett.\  {\bf 108} (2012) 261602
  \xxxlink{1202.3060} [hep-th], 
  %%CITATION = ARXIV:1202.3060;%%
  ``Non-Geometric Fluxes in Supergravity and Double Field Theory,''
 Fortsch. Phys. {\bf 60} (2012) 1150 %-1186
%  Fortschritte der Physik, Volume 60, Issue 11-12, 11501186, 2012, 
 \xxxlink{1204.1979} [hep-th]. 
 
%\cite{Hillmann:2009ci,Berman:2010is}
\bibitem{Hillmann:2009ci} 
  C.~Hillmann,
  ``Generalized E(7(7)) coset dynamics and D=11 supergravity,''
  JHEP {\bf 0903} (2009) 135
 \xxxlink{0901.1581} [hep-th].
  %%CITATION = ARXIV:0901.1581;%% 
 
%\cite{Berman:2010is}
\bibitem{Berman:2010is} 
  D.~S.~Berman and M.~J.~Perry,
  ``Generalized Geometry and M theory,''
  JHEP {\bf 1106} (2011) 074
 \xxxlink{1008.1763} [hep-th].
  %%CITATION = ARXIV:1008.1763;%% 
  
  %\cite{Coimbra:2011ky}
\bibitem{Coimbra:2011ky} 
  A.~Coimbra, C.~Strickland-Constable and D.~Waldram,
  ``$E_{d(d)} \times \mathbb{R}^+$ Generalised Geometry, Connections and M theory,''
\xxxlink{1112.3989} [hep-th].
  %%CITATION = ARXIV:1112.3989;%%
  
  %\cite{Aldazabal:2013sca}
\bibitem{Aldazabal:2013sca} 
  G.~Aldazabal, D.~Marqu«es and C.~Nœ–ez,
  ``Double Field Theory: A Pedagogical Review,''
\xxxlink{1305.1907} [hep-th].
  %%CITATION = ARXIV:1305.1907;%% 
 
%\cite{Sen:1991zi}
\bibitem{Sen:1991zi} 
  A.~Sen,
  ``O(d) x O(d) symmetry of the space of cosmological solutions in string theory, scale factor duality and two-dimensional black holes,''
  Phys.\ Lett.\ B {\bf 271} (1991) 295.
  %%CITATION = PHLTA,B271,295;%%

%\cite{Meissner:1996sa}
\bibitem{Meissner:1996sa} 
  K.~A.~Meissner,
  ``Symmetries of higher order string gravity actions,''
  Phys.\ Lett.\ B {\bf 392} (1997) 298
  \xxxlink{hep-th/9610131}.
  %%CITATION = HEP-TH/9610131;%%

%\cite{Kaloper:1997ux}
\bibitem{Kaloper:1997ux} 
  N.~Kaloper and K.~A.~Meissner,
  ``Duality beyond the first loop,''
  Phys.\ Rev.\ D {\bf 56} (1997) 7940
  \xxxlink{hep-th/9705193}.
  %%CITATION = HEP-TH/9705193;%%


\end{thebibliography}
\end{document}